\documentclass[11pt]{article}
\usepackage{geometry}
\usepackage{graphicx}
\usepackage{cite}
\usepackage{booktabs} 
\usepackage{floatrow} 
\usepackage[labelfont=bf,labelsep=period]{caption} 
\usepackage{subfig}

\usepackage{hyperref}
\hypersetup{pdftitle={Coupling with opinion dynamics promotes prosocial behavior in multilayer networks}, pdfauthor={Jnanajyoti Bhaumik, Naoki Masuda}}
\usepackage{amsmath, amsfonts, amssymb}
\usepackage{bm}

\usepackage[normalem]{ulem} 
\usepackage{xcolor}

\usepackage{cancel}
\usepackage{array}
\newtheorem{axiom}{Axiom}
\newtheorem{lemma}{Lemma}

\begin{document}
	\title{Coupling with opinion dynamics promotes prosocial behavior in multilayer networks}
	\author{Jnanajyoti Bhaumik$^1$ \and Naoki Masuda$^{2,3,4,*}$}
	\date{%
$^1$Department of Mathematics, State University of New York at Buffalo, Buffalo, New York, United States of America\\
$^2$Gilbert S.\,Omenn Department of Computational Medicine and Bioinformatics, University of Michigan, Ann Arbor, Michigan, United States of America\\
$^3$Department of Mathematics, University of Michigan, Ann Arbor, Michigan, United States of America\\
$^4$Center for Computational Social Science, Kobe University, Kobe, Japan\\
$^*$naokimas@gmail.com\\[2ex]%
	}
	
\maketitle

\section*{Abstract}
Various theoretical and empirical studies have accounted for why humans cooperate in competitive environments. Although prior work has revealed that network structure and multiplex interactions can promote cooperation, most theory assumes that individuals play dilemma games in all social contexts. However, real-world agents may participate in a diversity of interactions, not all of which present dilemmas. We develop an evolutionary game model on multilayer networks in which one layer supports the prisoner's dilemma game, while the other follows constant-selection dynamics, representing biased but non-dilemmatic competition, akin to opinion or fad spreading. Our theoretical analysis reveals that coupling a social dilemma layer to a non-dilemmatic constant-selection layer robustly enhances cooperation in many cases, across different multilayer networks, updating rules, and payoff schemes. These findings suggest that embedding individuals within diverse networked settings---even those unrelated to direct social dilemmas---can be a principled approach to engineering cooperation in socio-ecological and organizational systems.

\section*{Author summary} 

Why do people cooperate even when selfish behavior looks rewarding? Network models show that who interacts with whom matters, but most assume the same kind of dilemma in every social setting. We develop a theory for populations embedded in two overlapping interaction networks: one in which individuals face a prisoner’s dilemma game, and another in which two opinions spread under competition with a built-in bias but no dilemma. We show that linking these layers can systematically increase the chance that cooperation takes over, across network structures, update rules, and how public goods are distributed. The work suggests new, principled ways to foster cooperation in dilemmatic situations by shaping non-dilemmatic social contexts.

\section{Introduction}

Altruistic behavior is a key feature of both humans and non-human organisms. Over past decades, studies of social dilemma games, most famously represented by the prisoner's dilemma game, have revealed various mechanisms with which individuals cooperate when prosocial behavior is desirable for the society but apparently irrational for individuals. Such mechanisms include kin selection, repeated interactions between the same individuals, reputations, and population structure~\cite{Nowak2006Science, Nowak2006book, Sigmund2010book}.

%
Networks, or population structure in general, promote prosocial behavior in multiple ways. For example, clustering of individuals, quantified by the abundance of short cycles composed of a few individuals, promotes cooperation because short cycles create locally connected clusters of cooperators that protect themselves against exploitation by non-cooperators~\cite{Axelrod1984book, Nowak1992Nature_spatial, Nowak1993IntJBifuChaos}. Any edge (i.e., dyad) forming a network in fact creates assortative connectivity between cooperators, promoting cooperation \cite{ohtsuki2006simple}. Heterogeneity in the degree (i.e., the number of neighbors that an individual has) across individuals also promotes cooperation under some assumptions~\cite{duran2005, Santos2005PhysRevLett, santos2008social}. Time-varying nature of networks can also promote cooperation~\cite{li2020evolution,li2023temporal,su2023strategy,meng2025promoting}, whereas the opposite holds true in some situations~\cite{cardillo2014evolutionary}. All these network features are shared by a vast majority of empirical social networks \cite{Easley2010book, Barabasi2016book, Newman2018book}.

In fact, humans, as well as animals and organizations, can interact with others in multiple social contexts and modalities. For example, we may interact with each other both in person and online. This situation can be modeled by a multilayer network, of which each network layer is defined by one type of edge (i.e., interaction)~\cite{Kivela2014JCompNetw, Boccaletti2014PhysRep, Bianconi2018book, de2023more}. 
Various numerical studies have shown that cooperation can be enhanced in multilayer networks
\cite{Gomezgardenes2012SciRep, wang2015evolutionary, Jusup2022PhysRep, basak2024evolution, zhu2025evolution}.
Su et al.\,pioneered a theoretical framework to understand cooperation in multilayer networks~\cite{su2022evolution}. They assumed that each player is involved in a distinct prisoner's dilemma game with peers in each network layer and that the payoff of each player that guides evolutionary dynamics in the different layers is the sum of the payoffs that the player obtains across all the network layers. They showed that multilayer
networks enhance cooperation under broad conditions, compared to each network layer separately considered.

Social dilemmas are not the only situation that humans are coping with.
Even if we face a social dilemma in our daily life, we would also be simultaneously involved in other types of social but non-dilemmatic interactions. A game formulation of this situation is to make individuals play different types of games in different network layers~\cite{santos2014biased,wang2020vaccination,raducha2023evolutionary,zhu2025evolution}.
%
%
Seminal studies also analyzed a related situation in which the social dilemma game and imitation of behavior occur in different network layers~\cite{ohtsuki2007breaking,ohtsuki2007evolutionary}.
%
%
How robust is enhanced cooperation in multilayer networks when individuals are not always participating in social dilemma games? Although there are some numerical results along this direction~\cite{zhu2025evolution}, can we provide theoretical underpinnings to elucidate how non-dilemmatic social dynamics modify the likelihood of prosocial behavior in dilemmatic situations? To answer these questions, we propose an evolutionary game model on two-layer networks (which can be readily extended to the case of more than two layers) in which players are involved in the prisoner's dilemma game in one network layer and constant-selection dynamics in the other layer. The constant-selection dynamics is a simple social dynamics in which two types (e.g., opinions) compete in a population. It is equivalent to the biased voter model, a model of consensus formation dynamics used in mathematics \cite{Bramson1981AnnProb, Durrett1988book} and interdisciplinary physics \cite{Antal2006PhysRevLett, Czaplicka2022ChaosSolitonsFractals} for many years.
By extending a recently developed theoretical framework~\cite{su2022evolution}, we show that the constant-selection dynamics can promote cooperation played in the opposite network layer in many cases. This study expands the set of scenarios that promote cooperation in multilayer populations and furnishes a theoretical framework to study agents' multifaceted payoff-seeking behavior when social-dilemma interactions account for only part of their payoffs.
	
\section{Model}

\begin{figure}[t]
	\centering
	\includegraphics[width=\textwidth]{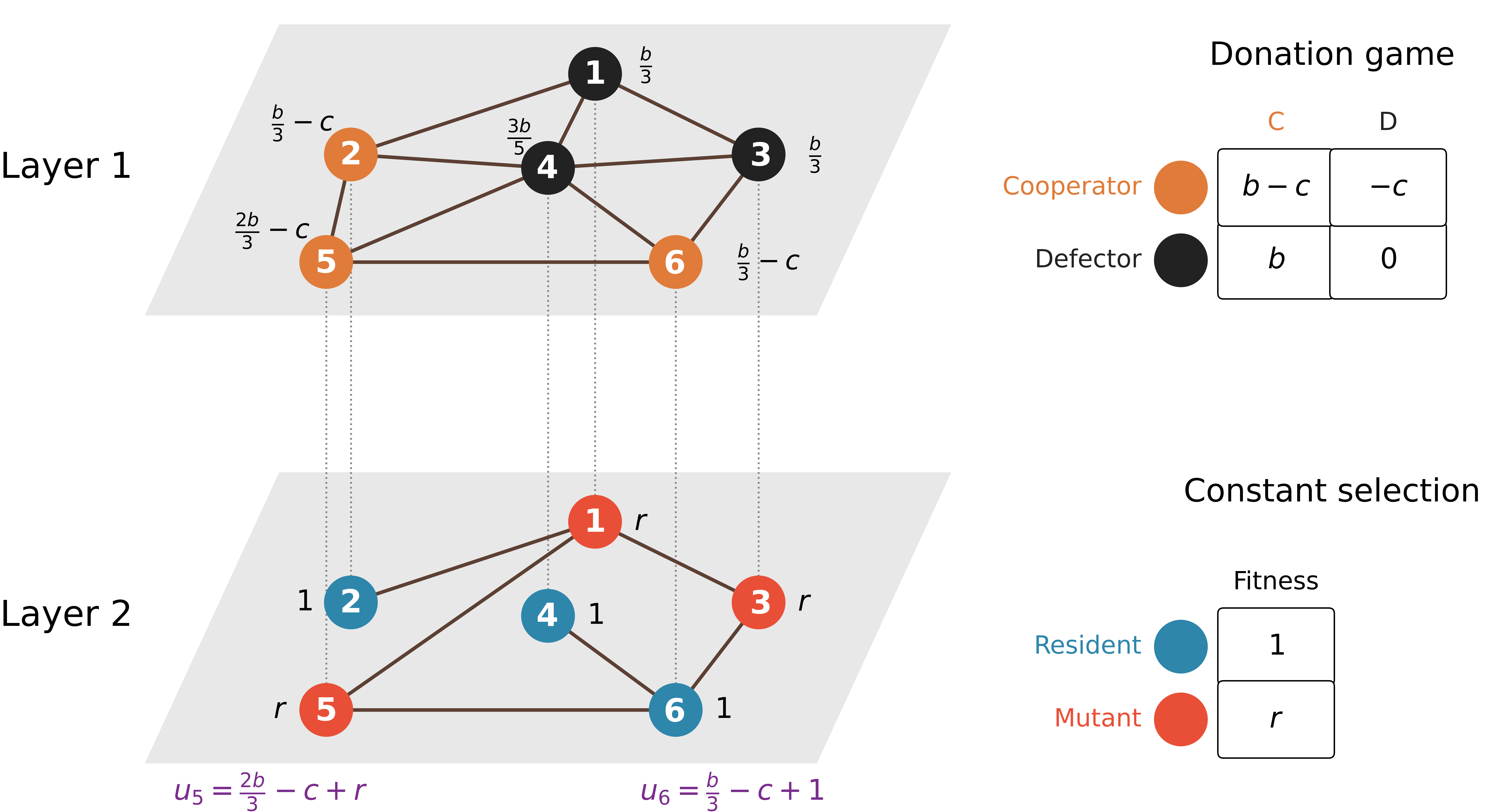}
	\caption{Schematic of the two-layer game. The donation game is played in layer 1. The constant-selection dynamics occurs in layer 2. The total payoff for each individual, which drives evolutionary dynamics in both layers, is the sum of the payoffs that the individual obtains from both layers. The figure shows the total payoffs for individuals 5 and 6 as examples. In both layers, replica nodes (shown by circles) that have a high total payoff tend to spread its type to its neighbors.}
	\label{fig:general_rule_multilayer}
\end{figure}

Fig.~\ref{fig:general_rule_multilayer} is a schematic of our evolutionary dynamics model on two-layer networks. We assume that there are $N$ individuals. The replica node refers to a node of the undirected network layer. Therefore, there are $2N$ replica nodes, and each individual is represented by two replica nodes. Each edge directly connects two replica nodes in the same layer and represents one type of pairwise interaction between individuals. We assume that the individuals play the donation game, a special case of the prisoner's dilemma game, with each of their neighbors in layer 1 and that the evolutionary dynamics are governed by constant selection in layer 2. 
	
In layer 1, each $i$th individual is either a cooperator or a defector, encoded into
$x_i^{[1]} = 1$ and $x_i^{[1]} = 0$, respectively. Each cooperator pays a cost $c$ and provides a benefit $b$ to each of its neighbors; this is called a pf goods scheme \cite{mcavoy2020social}; we will examine a different goods scheme later. The weight of an edge from the $i$th replica node to the $j$th replica node within layer $L \in \{1, 2 \}$ is denoted by $w_{ij}^{[L]}$. We denote the weighted degree (which we simply call the degree in the following text) of the $i$th replica node in layer $L$ by $s_{i}^{[L]} := \sum_{j=1}^N w_{ij}^{[L]}$. The payoff of the $i$th replica node in layer 1 is given by \cite{allen2017evolutionary, su2022evolution}
\begin{equation}
	u_i^{[1]} = -cx_{i}^{\left[1\right]}+\sum_{j=1}^{N} b p_{ij} x_j^{\left[1\right]},
\end{equation}
where $p_{ij} = w_{ij}^{[1]} / s_{i}^{[1]}$. 
This form of $p_{ij}$ models a variant of the ``pf goods'' scheme in \cite{mcavoy2020social}, which dictates that
the cooperator pays the cost $c$ per neighbor and each of its neighbor receives a benefit $b$. We divide the total payoff that each individual receives by the (weighted) degree of the node, $s_{i}^{[1]}$, for normalization. We examine a different payoff scheme
in the section ``Fixed benefits, fixed costs'' goods scheme.

In layer 2, each $i$th individual is assumed to be either the mutant or resident type, encoded into $x_i^{[2]} = 1$ and $x_i^{[2]} = 0$, respectively. The fitness of the mutant and resident type is denoted by $r$ and $1$, respectively. Therefore, the payoff of the $i$th replica node in layer 2 is given by
\begin{equation}
	u_i^{[2]} = x_i^{\left[2\right]} \left(r-1\right)+1.
	\label{eq:payoff-constant-selection_intro}
\end{equation}
The total payoff of the $i$th individual, denoted by $u_i$, is assumed to be the sum of payoff across both layers, i.e., $u_i = u_i^{[1]}+u_i^{[2]}$.

We assume that the evolutionary dynamics is governed by the death-Birth (dB) updating rule. (Capitalization of B indicates that selection occurs at birth \cite{hindersin2015most}; also see \cite{Masuda2009JTheorBiol, shakarian2012review, pattni2015evolutionary} for earlier studies that distinguish death-birth processes depending on whether selection occurs at death or birth.) At the beginning of every time step in the evolutionary dynamics, we choose a replica node $i$ in layer 1 uniformly at random (with probability $1/N$) to update its strategy. Then, the neighbors of  $i$ in layer 1 compete to disseminate its strategy to $i$. A neighbor of $i$ is chosen with the probability proportional to their fecundity, where fecundity of an individual $i$ is defined as $F_{i} = 1+\delta u_i$; note that $\delta$ ($\ge 0$) is the strength of selection. The type of $j$ (i.e., cooperation or defection) replaces that of $i$ in layer 1. Simultaneously, in layer 2, in a similar fashion, an individual $j$ is selected for death uniformly at random. Next, a neighbor of $j$ in layer 2 is selected for reproduction with the probability proportional to $u_j$. Then, the type of $j$ (i.e., mutant or resident) replaces that of $i$ in layer 2, concluding one time step of the evolutionary dynamics. This process is repeated until there are either only cooperators or only defectors in layer 1 and there are either only mutants or only residents in layer 2. Throughout the text, we assume weak selection, that is, $\delta\ll1$.
With the form of $F_i$ and reproduction rule described above, we are assuming that the selection strength is the same between the two layers.

We use the dB updating rule for layer 1 because it is often used for studying prosociality on networks~\cite{ohtsuki2006simple,fu2009evolutionary,nowak2010evolutionary,wang2015universal,su2019evolutionary,su2022evolution}.
We use the same updating rule in layer 2 for simplicity. We call the combined updating rule the dB-dB rule. In fact, a different updating rule called the Birth-death (Bd; selection occurs at birth) is more commonly used for constant-selection evolutionary dynamics \cite{lieberman2005evolutionary, hindersin2015most}. Therefore, we also investigate the Bd rule for layer 2. In this case, we call the combined updating rule the dB-Bd rule.

\section{Results}

\subsection{Conditions under which cooperators and mutants are favored}

We derived the condition under which the cooperator is favored in layer 1 and that under which the mutant is favored in layer 2 for arbitrary two-layer networks (see
Section~\ref{sec:SI-A}).
More specifically, we derived the fixation probability for a single cooperator introduced to the population of defectors in layer 1 under weak selection, and similar for a single mutant introduced in layer 2. The derivation expands that in \cite{su2022evolution}, where individuals are assumed to play the donation game in both networks. A key observation that we have exploited in our derivation is that constant-selection dynamics can be expressed as evolutionary game dynamics where the payoff matrix entries do not depend on the opponent's action. Nonetheless, the constant selection produces terms that are not present in the case of the two-layer donation game. After obtaining a general solution (i.e., last equation in
Section~\ref{sec:SI-A}),
we substitute the quantities for a given updating rule to obtain the conditions for the dB-dB and dB-Bd rules (see
Sections~\ref{sec:SI-B} and \ref{sec:SI-C},
respectively).

Under the dB-dB rule, we have found that the condition under which cooperation is favored in layer 1 is given by
 \begin{equation}
 \label{eqn:cooperation_criteria_reduced}
	c\theta_{2} + b\left(\theta_{1} - \theta_{3} \right) - \left(r-1\right) \phi_{2,0} > 0.
\end{equation}
Here, $\theta_{n}$ (denoted by $\theta_{n}^{\bm{\xi}\left[1\right]}$ in Section~\ref{sec:SI-B}) is the equilibrium probability that the initial and final nodes of a $n$-step random walk occurring within layer 1 have the same strategy, with a minor modification. It is calculated from a combination of the solution of a set of $O(N^2)$ linear equations,
 the $n$-step transition probability matrix of the random walk on the layer-1 network, and its stationary probability at each replica node. Quantity $\phi_{n,m}$ (denoted by $\phi_{n,m}^{\bm{\xi}\left[1,2\right]}$ in Section~\ref{sec:SI-B}) is 
the equilibrium probability that the initial node of the random walk is in layer 1 and the final node in layer 2 have the same strategy, with a minor modification, conditioned that the first $n$ steps of the random walk occur in layer 1 and the subsequent $m$ steps occur in layer 2. Quantity $\phi_{n,m}$ is
similarly obtained from a combination of the solution of a different set of $O(N^2)$ linear equations, the transition probability matrix of the aforementioned random walk with $n+m$ steps, and the stationary probability of the layer-1 random walk at each replica node.

Under the same dB-dB rule, the condition under which the mutant is favored in layer 2 is given by
\begin{equation}
\label{eqn:constant_drift_selection_criteria}
	-\left(r-1\right) \overline{\theta}_2 + c \phi_{ 0, 2} +
	b \left(\phi_{ 1, 0} - \phi_{ 1,2} \right) > 0,
\end{equation}
where $\overline{\theta}_2$ is the same as $\theta_2$ except that the random walk occurs within layer 2 instead of layer 1 (denoted by $\theta_{n}^{\bm{\xi}\left[2\right]}$ in Section~\ref{sec:SI-B}).
Equations~\eqref{eqn:cooperation_criteria_reduced} and \eqref{eqn:constant_drift_selection_criteria} are our main theoretical results. While they are outcomes of detailed mathematical analyses shown in Sections~\ref{sec:SI-A} and \ref{sec:SI-B}, we provide more intuitive explanation of these inequalities in Section~\ref{sec:SI-D}. We derive that explanation by imposing that a cooperator or mutant neighbor of an individual that is being updated, denoted by $i$, has a higher average payoff than a random neighbor of $i$. A similar explanation was provided for the two-layer network model in which individuals play the donation game in both layers~\cite{su2022evolution}.

These theoretical results hold for weighted two-layer networks. In the following demonstration, however, we focus on unweighted networks for simplicity.

\subsection{Classification of two-layer networks and exhaustive examination of networks with $N=6$ individuals}

\subsubsection{Evolution of cooperators}\label{sec:game-layer}

Let us more closely examine the conditions given by Eqs.~\eqref{eqn:cooperation_criteria_reduced} and \eqref{eqn:constant_drift_selection_criteria}. When $r=1$, Eq.~\eqref{eqn:cooperation_criteria_reduced} reduces to the condition for favoring cooperation in one-layer networks~\cite{allen2017evolutionary,su2022evolution}. This is because setting $r=1$ implies that the payoff from the constant-selection layer (i.e., layer 2) is equal to $1$ for all individuals and therefore only makes the selection weaker for the donation game layer (i.e., layer 1).

\begin{figure}[t]
	\centering
	\includegraphics[width=\textwidth]{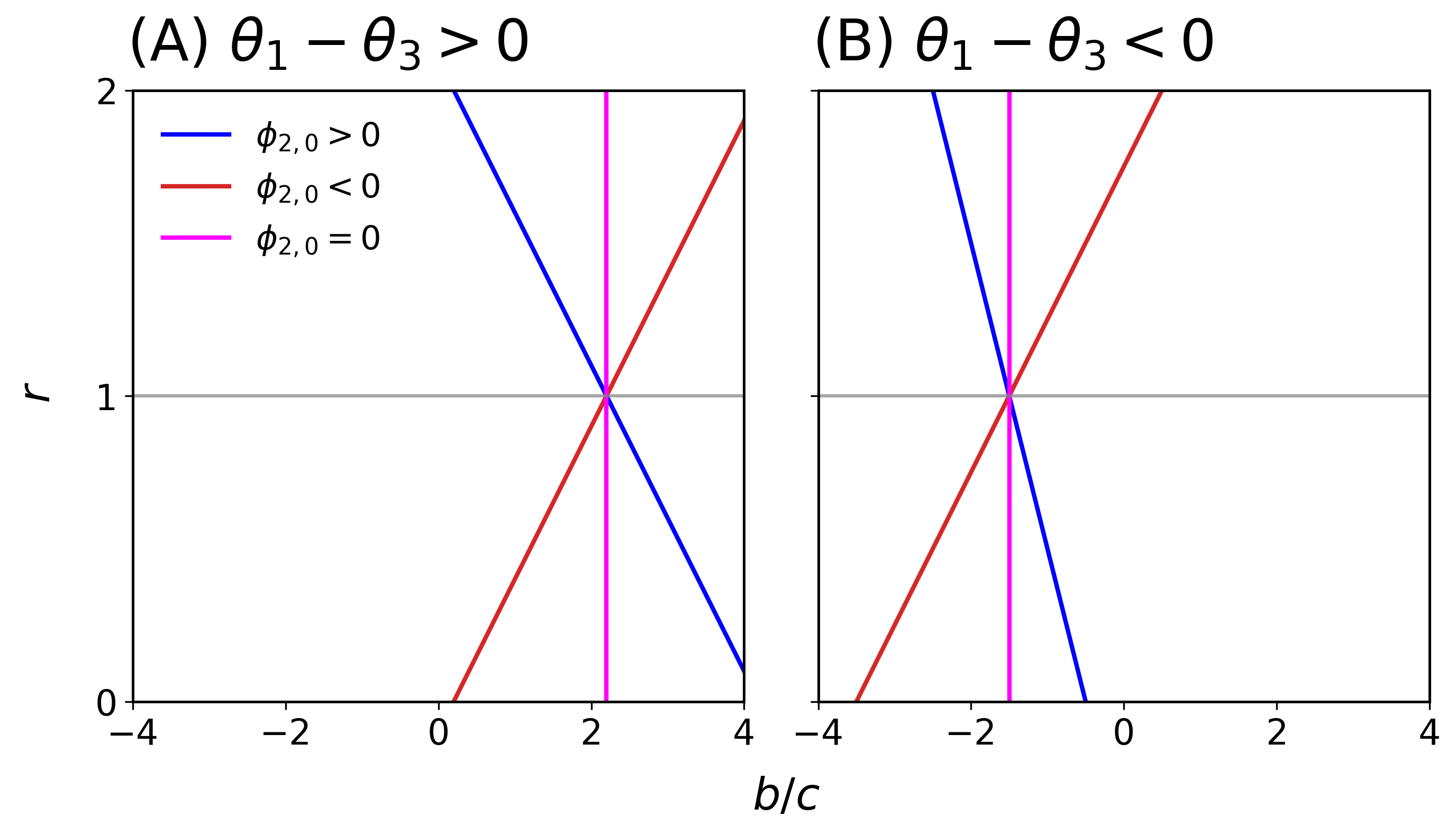}
	\caption{$(b/c)^*$ for various values of $r$. (A) $\theta_{1}-\theta_{3}>0$. (B) $\theta_{1}-\theta_{3}<0$. In (A), cooperation is favored when $b/c$ is larger than the value specified by the line, whose slope depends on $\phi_{2,0}$. In (B), spite is favored when $b/c$ is smaller than the value specified by the line depending on the $\phi_{2,0}$ value.}	
	\label{fig:schematic_b/c}
\end{figure}

To classify consequences of Eq.~\eqref{eqn:cooperation_criteria_reduced} depending on the parameter and variable values,
 we first assume that $\theta_{1} - \theta_{3} >0$ and rewrite Eq.~\eqref{eqn:cooperation_criteria_reduced} as follows:
\begin{equation}
	\frac{b}{c} > \left( \frac{b}{c} \right)^* \equiv \frac{- \theta_{2}}{\theta_{1} - \theta_{3}} +\frac{(r-1) \phi_{2,0} }{c \left( \theta_{1} - \theta_{3} \right)}.
\label{eqn:cooperation_criteria_reduced_rearranged-pos}
\end{equation}
The network-dependent constant $(b/c)^*$ is the value of $b/c$ above which cooperation is favored over defection in layer 1 of the two-layer network; however, we also discuss other interpretations of $(b/c)^*$ below.
When $r=1$, it is known that $(b/c)^*$ is positive because $\theta_{2} < 0$~\cite{allen2017evolutionary,su2022evolution,su2023strategy}, and Eq.~\eqref{eqn:cooperation_criteria_reduced_rearranged-pos} gives the condition for cooperation. Relative to this one-layer baseline case first shown in~\cite{allen2017evolutionary}, coupling with the constant-selection dynamics layer, i.e., $r\neq 1$, changes $(b/c)^*$ as follows.
Note that $(r-1)/c$ is the relative weight of the constant selection to the donation game. We distinguish between the following three cases (see Fig.~\ref{fig:schematic_b/c}A).
\begin{enumerate}
\item If $\phi_{2,0} > 0$, then $(b/c)^*$ is lower in the two-layer than the one-layer network if $0\le r < 1$, with the two-layer network easing cooperation. In contrast, the two-layer network makes cooperation harder if $r>1$. 
\item If $\phi_{2,0} < 0$, then $(b/c)^*$ is lower in the two-layer than the one-layer network if $1 < r < \overline{r}$, facilitating the cooperation. Constant $\overline{r}$ ($>1$) is given below. The two-layer network makes cooperation harder if $0\le r < 1$. 
\item If $\phi_{2,0} = 0$, then $(b/c)^*$ does not depend on $r$ such that $(b/c)^*$ remains the same as the case of $r=1$.
In this case, coupling two network layers does not change the condition for cooperation.
\end{enumerate}
In case 1, if $\phi_{2,0} > -c \theta_{2}$ $(> 0)$
and $0<r< \overline{r} \equiv 1 + \left(c \theta_{2} / \phi_{2,0} \right)$ $(<1)$, 
then cooperation is favored when $b/c > (b/c)^*$ with a negative threshold $(b/c)^*$.
It should be noted that, when the spite behavior (in place of cooperation) is favored over defection, the condition generally reads $b/c < (b/c)^* < 0$; this is how we interpret the $b/c < (b/c)^* < 0$ case in general~\cite{allen2017evolutionary}. In contrast, the inequality is in the opposite direction in the present case. When $b/c > (b/c)^*$ with a negative $(b/c)^*$, the interpretation of the result is difficult. Note that the same interpretability problem can occur when players are involved in different donation games in each network layer \cite{su2022evolution}. However, we confirmed in our numerical simulations described in the following text that 
there is no network yielding $b/c > (b/c)^*$ with a negative $(b/c)^*$
. Similarly, in case 2, if $r > \overline{r}$ $(>1)$, cooperation is favored when $b/c > (b/c)^*$ with a negative $(b/c)^*$. However, we have numerically confirmed that two-layer networks can show
$b/c > (b/c)^*$ with a negative $(b/c)^*$ only when $r$ is much larger than $1$, where the weak-selection assumption is compromised. 

Second, when $\theta_{1}-\theta_{3} <0$, we obtain Eq.~\eqref{eqn:cooperation_criteria_reduced_rearranged-pos} but with the inequality being flipped. Then, the first term on the right-hand side of Eq.~\eqref{eqn:cooperation_criteria_reduced_rearranged-pos} is negative.
Therefore, if $r=1$ or $\phi_{2,0} = 0$ (see Fig.~\ref{fig:schematic_b/c}B), then the condition reads $b/c < (b/c)^* < 0$
such that spite is favored with the same value of $(b/c)^*$ as that for the one-layer network. If $\phi_{2,0} \neq 0$, depending on whether $\phi_{2,0}$ is positive or negative, spite occurs more easily or less easily for different ranges of $r$, relative to $r=1$ (see Fig.~\ref{fig:schematic_b/c}B). Interpretation of the condition for the spite behavior $b/c<(b/c)^*$ would be difficult if $(b/c)^*$ is positive. However, we again numerically confirmed that this unphysical situation can only occur when $r$ is much larger than $1$.

Finally, when $\theta_{1}-\theta_{3} =0$, cooperation is favored if
$ \theta_{2}   > \frac{(r-1)\phi_{2,0}}{c}$.

In this manner, we can classify evolutionary outcomes with a linear analysis assuming weak selection. To assess which outcomes are frequent,
we exhaustively considered unique two-layer networks with six individuals and all possible unique initial conditions with just one cooperator in layer 1 and one mutant in layer 2. Following the convention, we use the same initial condition (i.e., single cooperator and single mutant) in all the following numerical simulations as well. We set $c=1$ in this and the following numerical simulations unless we state otherwise. We show the classification of the 2,763,739 unique pairs of two-layer network and initial condition in the left panel of Table~\ref{tab:thetapi_distribution_percentages}. We have several observations: First, more than 80\% of the unique pairs of network and initial condition yield negative $\theta_{1}-\theta_{3}$, implying spite behavior.
In fact, this result is not specific to two-layer networks because setting $r=1$ provides the condition for one-layer networks and the sign of 
$\theta_{1}-\theta_{3}$ is independent of $r$.
%
%
Second, the condition for cooperation when $\theta_{1}-\theta_{3} =0$ is hard to interpret, but this case rarely occurs.
Third, by additional numerical simulations, we confirmed that there is no case in which cooperation is selected for $b/c > (b/c)^*$ with a negative $(b/c)^*$ or the spite is selected for $b/c < (b/c)^*$ for a positive $(b/c)^*$ when $r$ is sufficiently close to $0$. The result that less than 20\% networks with $N=6$ individuals foster cooperation is underwhelming. However, we will later see that cooperation rather than spite is favored in a majority of larger networks.

\begin{table}[htbp]
	\centering
\caption{Classification of all unique pairs of the two-layer network with six individuals and the initial condition. (a) dB-dB. (b) dB-Bd. The marginal distribution of $\theta_1 - \theta_3$ is the same between the dB-dB and dB-Bd updating rules. This is because $\theta_1 - \theta_3$ only depends on the network structure and the updating rule used in layer 1, which are common between the dB-dB and dB-Bd rules.}
	\begingroup
	\setlength{\tabcolsep}{6pt}  
	\renewcommand{\arraystretch}{1.05}
	\resizebox{0.47\textwidth}{!}{
		\begin{tabular}{l cccc}
			\toprule
			dB-dB & \multicolumn{3}{c}{$\phi_{2,0}$ } & \\
			\cmidrule(lr){2-4}
			$\theta_{1}-\theta_{3}$  & $<0$ & $=0$ & $>0$ & Total \\
			\midrule
			$>0$ & 6.40\% & 0.00\% & 10.98\% & 17.38\% \\
			$=0$ & 0.54\% & 0.00\% & 1.08\%  & 1.61\% \\
			$<0$ & 21.46\% & 0.00\% & 59.54\% & 81.01\% \\
			\midrule
			Total & 28.40\% & 0.00\% & 71.60\% & 100.00\% \\
			\bottomrule
		\end{tabular}
	}
	\hspace{0.04\textwidth}
	\resizebox{0.47\textwidth}{!}{
		\begin{tabular}{l cccc}
			\toprule
			dB-Bd & \multicolumn{3}{c}{$\phi_{2,0}$} & \\
			\cmidrule(lr){2-4}
			$\theta_{1}-\theta_{3}$  & $<0$ & $=0$ & $>0$ & Total \\
			\midrule
			$>0$ & 8.23\% & 0.00\% & 9.15\% & 17.38\% \\
			$=0$ & 0.65\% & 0.00\% & 0.96\% & 1.61\% \\
			$<0$ & 38.29\% & 0.00\% & 42.72\% & 81.01\% \\
			\midrule
			Total & 47.17\% & 0.00\% & 52.83\% & 100.00\% \\
			\bottomrule
		\end{tabular}
	}
\label{tab:thetapi_distribution_percentages}
\endgroup
\end{table}

To explore which networks yield cooperation versus spite and enhancement of cooperation by coupling with the constant-selection layer, we first examined correlation between pairs of two outcome variables and three network metrics for single-layer networks with six nodes. The two outcome variables are $\theta_1 - \theta_3$ and $n_{\text{init}, \text{C}}$; the latter is the number of initial conditions out of the six initial conditions (i.e., choice of the single node which initially has the cooperator) for which $(b/c)^* > 0$. The three network metrics are the mean degree, clustering coefficient, and mean closeness centrality, where the mean is taken over the six nodes. Focusing on the dB-dB rule, we find that $\theta_1 - \theta_3$ and $n_{\text{init}, \text{C}}$ are strongly positively correlated with each other, that these two outcome variables are strongly negatively correlated with the three network metrics, and that the three network metrics are strongly positively correlated with each other (see Section~\ref{sec:SI-E} for details). Therefore, the small mean degree, small clustering coefficient, and small mean closeness tend to enhance cooperation. We remark that the small mean degree enhancing cooperation is an established result \cite{ohtsuki2006simple, allen2017evolutionary}. The clustering coefficient and mean closeness are negatively correlated with $\theta_1 - \theta_3$ and $n_{\text{init}, \text{C}}$ even after partialing out the mean degree (see Section~\ref{sec:SI-E}).

For single-layer networks with six nodes fostering cooperation (i.e., $(b/c)^* > 0$), we also find that the initial cooperator located at a node with a small degree, closeness, or local clustering coefficient promotes cooperation (i.e., by decreasing $(b/c)^*$). Finally, for two-layer networks with six nodes fostering cooperation, coupling with the constant-selection layer (i.e., layer 2) tends to allow enhancement of cooperation (i.e., large $\left| \text{d}(b/c)^* / \text{d}r \right|$) when the layer-1 network is dense (i.e., many edges). However, the edge density and other properties of the layer-2 network do not contribute to modulating the condition for cooperation. See Section~\ref{sec:SI-E} for details. While exploratory, these additional numerical results suggest that some network properties and initial conditions favor cooperation or enhance the effect of coupling with the constant-selection layer.

By numerical simulations of the stochastic evolutionary dynamics on some select two-layer networks with $N=6$ individuals, we also verified that 
correlation between the strategy in layer 1 (i.e., cooperation or defection) and the type in layer 2 (i.e., resident or mutant) does not particularly build up during the stochastic process (see Section~\ref{sec:SI-F} for results). For example, even if we start from a correlated initial condition in which the initial sole cooperator in layer 1 and the initial sole mutant in layer 2 are the same individual, the strategy-type correlation between the two layers decays regardless of whether cooperation or defection eventually fixates.

\subsubsection{Evolution of mutants}

We turn to the condition under which the mutant is favored in the two-layer network, Eq.~\eqref{eqn:constant_drift_selection_criteria}.
Because $\overline{\theta}_{2} < 0$, we can rewrite
Eq.~\eqref{eqn:constant_drift_selection_criteria} as
\begin{equation}
\label{eqn:constant_drift_selection_criteria_rearranged}
r > r^* \equiv 1+ \frac{
b \left(\phi_{ 1, 0} - \phi_{ 1, 2}\right) + c \phi_{ 0, 2 }}
{ \overline{\theta}_2 } \equiv 1 + A .
\end{equation}
The network-dependent constant $r^*$ is the value of $r$ above which the mutant is favored over the resident in layer 2 of the two-layer network.
This result is in stark contrast with that for one-layer networks because, in two-layer networks, both the payoff of the donation game and the structure of the two-layer networks in both layers influence the propensity that the mutant is favored through $A$. Remarkably, mutants whose fitness (i.e., $r$) is smaller than that of the resident type (i.e., $1$) can be favored if $A<0$. If $A>0$, then $r$ has to be sufficiently larger than $1$ for the mutant to be favored. Another observation is that
setting $b/c$ to the $(b/c)^*$ value for the one-layer network does not lead to $r^*=1$ in general. Therefore, in contrast to the condition for cooperation, for which setting $r=1$ recovered the one-layer network results, we do not have a mathematically solid baseline one-layer case for the constant selection layer. Furthermore, Eq.~\eqref{eqn:constant_drift_selection_criteria_rearranged} indicates that $b$ and $c$ independently affect $r^*$, and whether an increase in $b$ or $c$ increases or decreases $r^*$ depends on the network structure. 

\begin{figure}[t]
	\centering
	\includegraphics[width=0.6\textwidth]{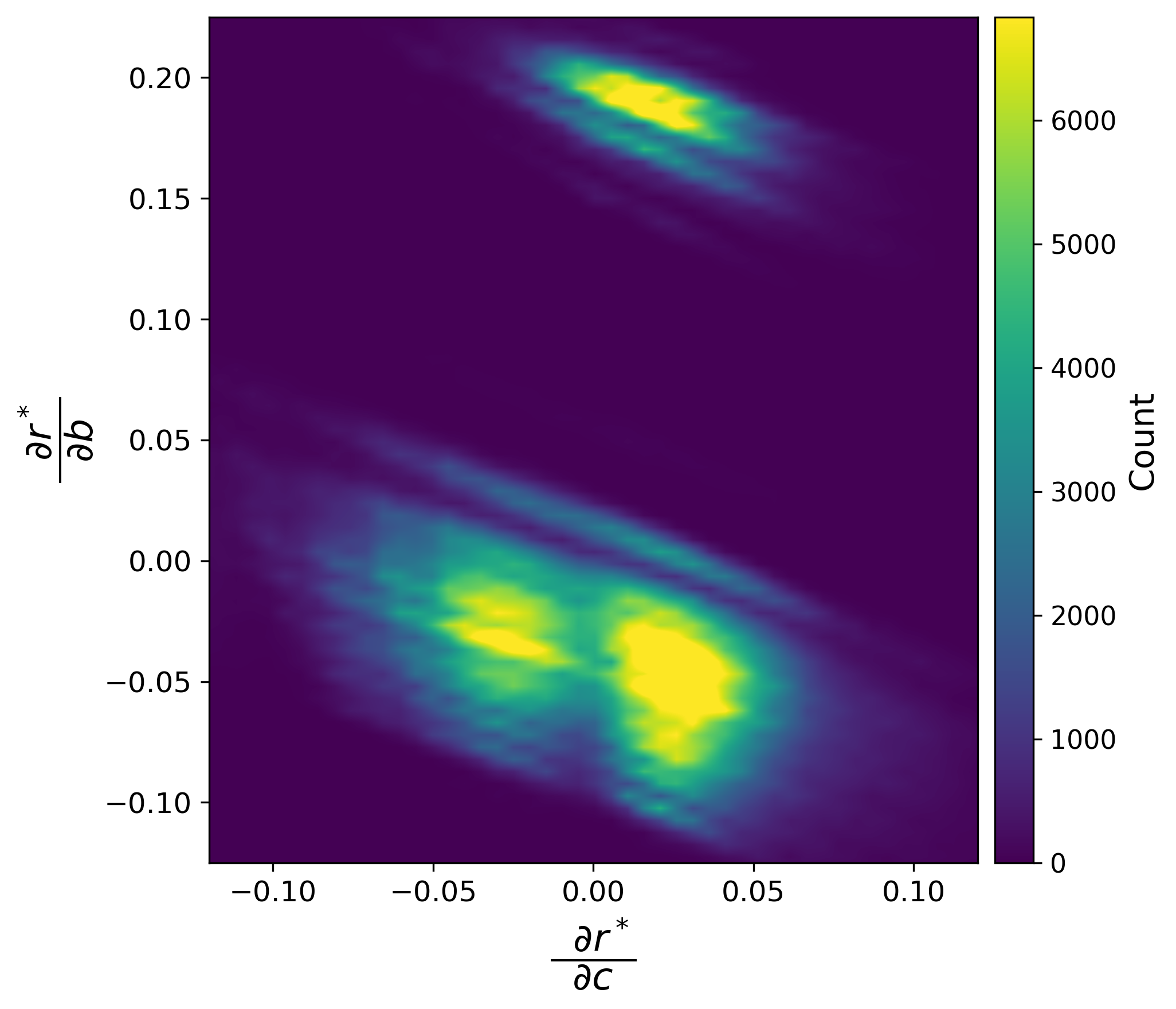}
\caption{Distribution of $dr^*/db$ and $dr^*/dc$ over the 2,763,739 unique pairs of the two-layer network with six individuals and initial condition. This is a non-smoothed two-dimensional histogram.}
\label{fig:dr}
\end{figure}

To examine responsiveness of $r^*$ to the coupling with the donation game network layer, we investigated the same 2,763,739 unique pairs of two-layer network with $N=6$ individuals and initial condition to calculate $dr^*/db = (\phi_{ 1,0} - \phi_{ 1,2}) / \overline{\theta}_2$ and $dr^*/dc = \phi_{ 0,2} / \overline{\theta}_2$. We show the distribution of ($dr^*/db$, $dr^*/dc$) in Fig.~\ref{fig:dr}. The figure suggests that $dr^*/db$ and $dr^*/dc$ can have either sign depending on the network and initial condition. The effect of $b$ and $c$ on $r^*$ is modest; the average of
$\left| dr^*/db \right|$ and $\left| dr^*/dc \right|$ over all the possible networks and initial condition is 0.0351 and 0.0665, respectively (with ranges
$-0.211 \le dr^*/db \le 0.294$ and $-0.262 \le dr^*/dc \le 0.247$; the density of unique pairs of two-layer network and initial condition is invisibly small outside the ($dr^*/db$, $dr^*/dc$) region shown in the figure). The different stripes present in Fig.~\ref{fig:dr} partially owe to the network structure and initial condition of layer 1 (see
Section~\ref{sec:SI-G}).
We conclude that the $b$ and $c$ values modestly, but consistently, affects $r^*$ in almost all cases.

\subsection{Sample two-layer networks}

\begin{figure}
	\centering
	\includegraphics[width=0.9\textwidth]{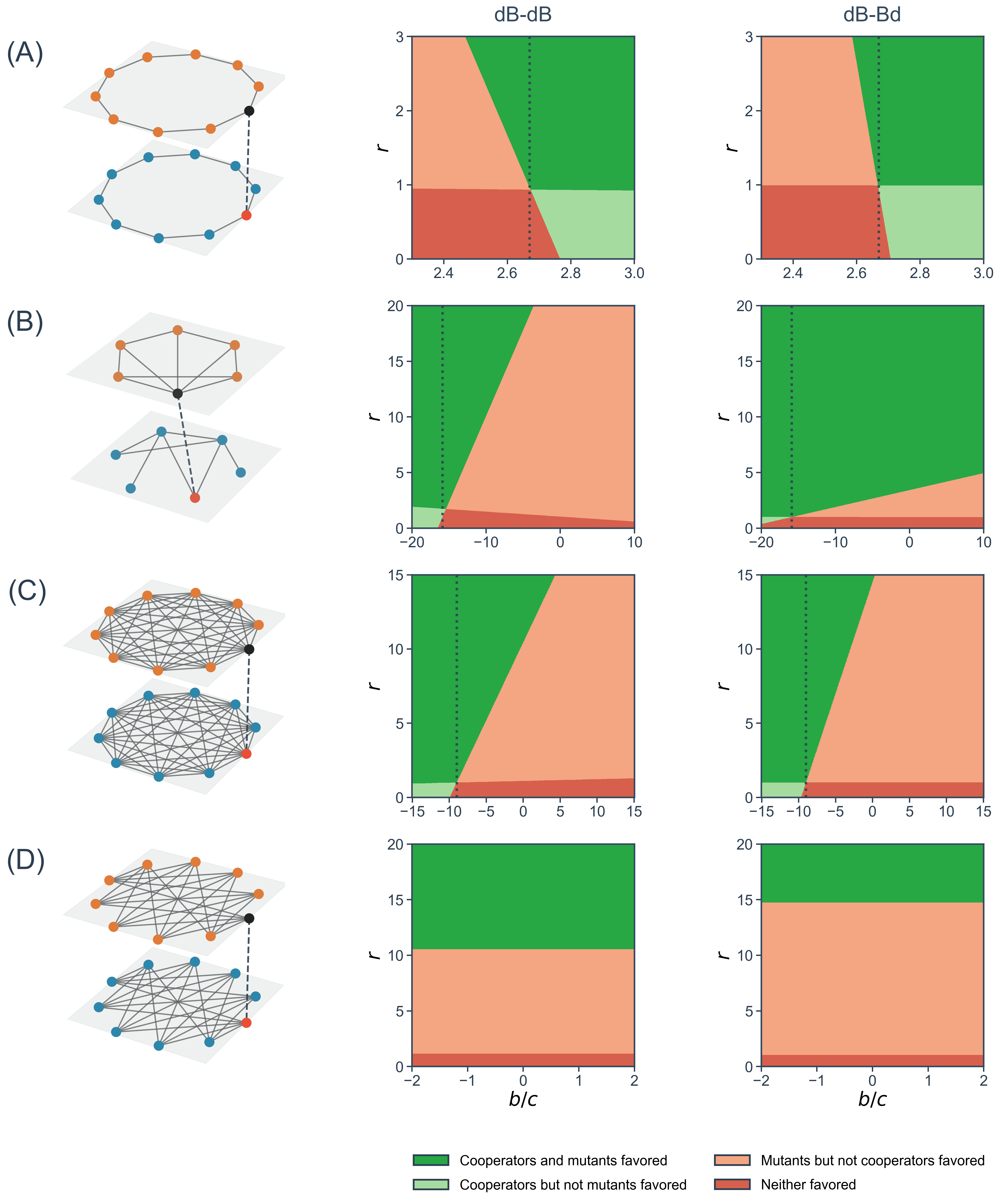}
	\caption{Selection of cooperators and mutants in four two-layer networks. Each row corresponds to the two-layer network and its initial condition visualized in the left panel. The colors of the replica nodes	are the same as those used in Fig.~\ref{fig:general_rule_multilayer}; orange, black, blue, and red represent cooperator, defector, resident, and mutant, respectively. The second column of each row of the figure shows the results under the dB-dB updating rule. The third column shows the results under the dB-Bd updating rule. Each colored region shows the $(b/c, r)$ region in which selection favors or disfavors cooperators in layer 1 or mutants in layer 2. (A) Coupled ring networks with $N=10$ individuals. We obtain $(b/c)^* = 8/3$ for the uncoupled one-layer ring, shown by the dotted lines.
(B) Coupled heterogeneous networks with $N=6$. We obtain $(b/c)^* = -15.89$ for the uncoupled layer-1 network. (C) Coupled complete graphs with $N=10$. The one-layer counterpart yields $(b/c)^*= -9$. (D) Coupled complete bipartite graphs with $N=10$. The one-layer counterpart yields $(b/c)^* = \infty$.} 
	\label{fig:examples}
\end{figure}

To examine whether various two-layer networks favor cooperators or mutants and how, we examine four larger synthetic networks. The exact structure of these networks and the initial condition are shown in the left panels of Fig.~\ref{fig:examples}. The first two of them were used in~\cite{su2022evolution}. First, consider a coupled ring network with $N=10$ individuals shown in Fig.~\ref{fig:examples}A. When each layer evolves independently, one obtains $(b/c)^*= 8/3$. When the two layers are coupled, $r>1$ causes cooperation to emerge in layer 1 under a more generous condition than in the case of a one-layer network, i.e., $(b/c)^* < 8/3$ (see the middle panel in Fig.~\ref{fig:examples}A). For example, we obtain $(b/c)^* = 2.57$ when $r = 2$.
Mathematically, in the limit of infinitely large two-layer ring network, $(b/c)^*$ decreases by $4(r-1)/c$ from the one-layer case (i.e., $(b/c)^* = 1/2$) if the individual that initially cooperates in layer 1 is of the mutant type in layer 2 (see
Section~\ref{sec:SI-H}).
In other cases, the effect of the constant-selection layer to modulate $(b/c)^*$ diminishes as the size of the ring increases.

Second, we investigated a heterogeneous two-layer network shown in Fig.~\ref{fig:examples}B. In this network, each layer promotes the evolution of spite (i.e.,  $(b/c)^* < 0$) when the two layers are uncoupled. We find for this two-layer network that spite is favored with a smaller punishment (i.e., negative $b/c$ values closer to $0$) when $r>1$ relative to the one-layer network and vice versa when $r<1$. 

Third, we investigated a two-layer network composed of two complete graph layers shown in Fig.~\ref{fig:examples}C. The results are qualitatively the same as those for the heterogeneous two-layer network investigated in Fig.~\ref{fig:examples}B. For this coupled complete graph, we analytically obtained the condition for favoring spite as 
$b/c < (b/c)^* = -(N-1) + \frac{r-1}{c} \cdot \frac{2(N-1)^2}{N(2N-3)}$; see
Section~\ref{sec:SI-I}
for the derivation. This result implies that spite can occur only in small coupled complete graphs because $(b/c)^* \to - \infty$ as $N \to\infty$. 

Fourth, we investigated a two-layer network composed of two complete bipartite networks shown in Fig.~\ref{fig:examples}D. We find that selection favors cooperation if $r$ is approximately larger than $10.5$, regardless of the value of $b/c$ (see the middle panel of Fig.~\ref{fig:examples}D). In contrast, a single complete bipartite network gives $(b/c)^*=\infty$, i.e., no cooperation.

Fifth, we also derived analytical results for the coupled star graph of arbitrary size 
(Section~\ref{sec:SI-J}).

These results are similar to those when the individuals play different donation games in two network layers of similar sizes \cite{su2022evolution}. Regardless of whether the second layer is subject to the social dilemma game or constant-selection dynamics, coupling of networks modulates the threshold value of $b/c$, i.e., $(b/c)^*$, for cooperation or spite. Quantitatively, in the coupled ring network (Fig.~\ref{fig:examples}A), 
$(b/c)^*$ decreases from $8/3$ in the case of the one-layer ring network to 
$1.74$ when the donation game with $b/c = 10$ is played on the other ring layer
\cite{su2022evolution}. In our model, reduction of $(b/c)^*$ to $1.74$ requires $r = 10.78$, which is by chance close to the aforementioned value of $b/c=10$.

\subsection{Random graphs}\label{sec:random-graphs} 

As a next test, we examined larger two-layer networks from two random network models. The first model is a two-layer Erd\H{o}s-R\'{e}nyi (ER) random graph.
For each layer, we generate an instance of the ER random graph with $N=15$ replica nodes with the probability of edge between each pair of nodes
$p_L \in\{0.1,0.2,0.3,0.4,0.5\}$ for layer $L \in \{1, 2 \}$. The two network layers are independently generated. 
We evaluated all the 225 possible initial conditions with one cooperator in layer 1 and one mutant in layer 2 for each generated two-layer network with given $p_1$ and $p_2$. 

\begin{figure}
\centering
\includegraphics[width=\textwidth]{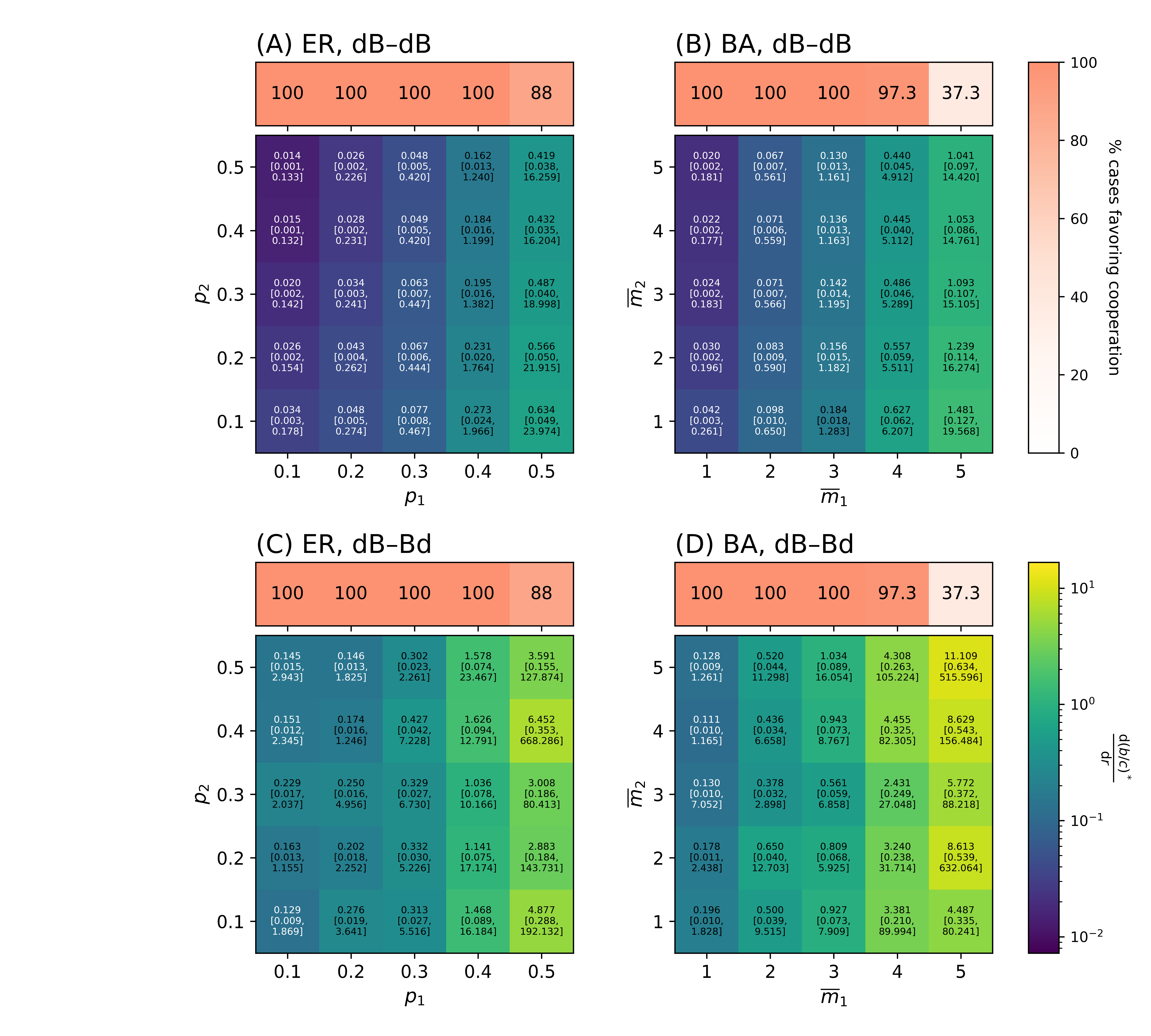}
\caption{Evolution of cooperation in coupled ER and BA networks. 
(A) ER, dB-dB rule. (B) BA, dB-dB rule. (C) ER, dB-Bd rule. (D) BA, dB-Bd rule. The upper part of each panel shows the fraction of pairs of one-layer network and its initial condition that yield cooperation when $b/c > (b/c)^*$ with a threshold value $(b/c)^* > 0$. The fraction values are the same between (A) and (C) and between (B) and (D) because this fraction only depends on the layer-1 network. The lower part of each panel shows the median along with the 5th and 95th percentiles (in square brackets) of $\left| \text{d}(b/c)^*/\text{d}r \right|$ for pairs of two-layer network and initial condition. Each two-layer network is composed of $N=15$ individuals.}
\label{fig:tables_random_graphs}
\end{figure}

We first analyzed the layer-1 ER random graphs as single-layer networks. Specifically, we counted the fraction of the layer-1 ER random graphs for which cooperation as opposed to spite is favored when $b/c > (b/c)^*$ with a positive value of $(b/c)^*$. We are interested in this case because, if this is the case, it is likely that one can choose a value of $r$ that lowers $(b/c)^*$ in the two-layer network such that cooperation is facilitated by the coupling with the constant-selection network layer (i.e., layer 2). 
The upper part of Fig.~\ref{fig:tables_random_graphs}A shows the fraction of pairs of one-layer ER random graph and initial condition yielding $(b/c)^* > 0$ as a single-layer network. Because this inquiry only concerns one-layer networks, we only show the result as a function of $p_1$. We find that the all instances of networks enable cooperation when $p_1 \le 0.4$ and that 88\% of them do so when $p_1=0.5$. These values are much larger than for smaller networks; compare these numbers with those for networks with $N=6$ individuals shown in Table~\ref{tab:thetapi_distribution_percentages}.

Next, we turned to two-layer networks. Specifically, we examined, among the runs yielding cooperation in layer 1, how much $(b/c)^*$ can be changed by the constant-selection dynamics. To quantify this effect, we measured $\left| \text{d}(b/c)^*/\text{d}r \right|$, i.e., how much $(b/c)^*$ moves by changing $r$ for each pair of two-layer network and initial condition. For each pair of $p_1$ and $p_2$, we show in the lower part of Fig.~\ref{fig:tables_random_graphs}A
the median along with the 5th and 95th percentiles of $\left| \text{d}(b/c)^*/\text{d}r \right|$. We show the median and percentiles instead of the mean and standard deviation because there are occasional outliers that yield a huge $\left| \text{d}(b/c)^*/\text{d}r \right|$ relative to typical pairs of two-layer network and initial condition. However, we have confirmed that the tendency that we are reporting with the median and percentiles remains similar when we measure the mean and standard deviation of $\left| \text{d}(b/c)^*/\text{d}r \right|$ (see
Section~\ref{sec:SI-K}).
In Fig.~\ref{fig:tables_random_graphs}A, we find that 
$\left| \text{d}(b/c)^*/\text{d}r \right|$ is larger for denser layer-1 networks (i.e., larger $p_1$) and sparser layer-2 networks (i.e., smaller $p_2$).
The $(b/c)^*$ value is reasonably responsive to the changes in $r$ (with median $\left| \text{d}(b/c)^*/\text{d}r \right| > 0.4$) when $p_1 = 0.5$ and modestly so (i.e., median $\left| \text{d}(b/c)^*/\text{d}r \right| > 0.16$) when $p_1 = 0.4$.

Given that most empirical contact networks are heterogeneous in terms of the node's degree, we ran the same analysis for two-layer Barab\'{a}si-Albert (BA) networks, where each new node added to the network layer $L$ has $\overline{m}_L$ nodes, making the average degree of each network layer approximately equal to $2\overline{m}_L$ (see Methods for the procedure for generating networks). Fig.~\ref{fig:tables_random_graphs}B shows the proportion of network layer 1 that facilitates cooperation as a one-layer network (upper part) and the median and percentiles of $\left| \text{d}(b/c)^*/\text{d}r \right|$ (lower part).
The results are qualitatively similar to those for the ER random graph. Specifically, one-layer BA networks favor cooperation in most cases if $\overline{m}_1 \le 4$. Note that $\overline{m}_1 \le 4$ does not imply particularly sparse networks because we are using networks with $N=15$ nodes. Furthermore, $\left| \text{d}(b/c)^*/\text{d}r \right|$ is overall larger for the two-layer BA than ER networks. In particular, at $\overline{m}_1 = 4$, a majority of BA networks support cooperation, and cooperation is substantially eased (i.e., median $\left| \text{d}(b/c)^*/\text{d}r \right| > 0.44$) by coupling with another BA network layer subject to the constant-selection dynamics.

These results reinforce our main finding that coupling with the constant-selection dynamics can facilitate cooperation on networks.

\subsection{Empirical networks}

\begin{figure}
\centering
\includegraphics[width=0.7\textwidth]{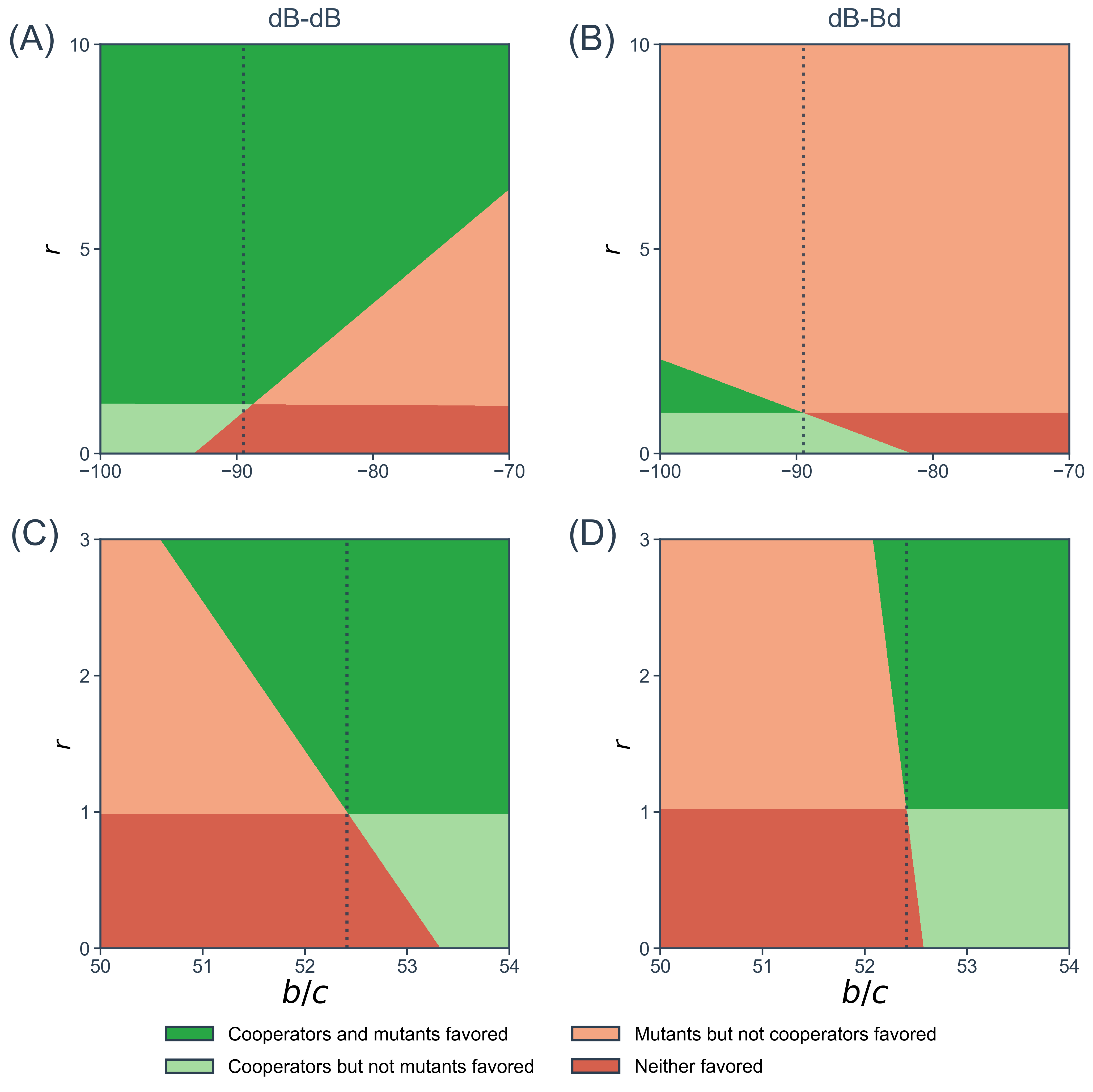}
\caption{Parameter regions in which the cooperator or mutant is selected in empirical two-layer networks. The vertical lines represent the $(b/c)^*$ value for the uncoupled layer-1 network. (A) VC7, dB-dB. (B) VC7, dB-Bd. (C) LLF, dB-dB. (D) LLF, dB-Bd.
%
%
}
\label{fig:empirical_networks}
\end{figure}

We examined two empirical two-layer networks, i.e., the Vickers–Chan 7th Graders (VC7) network with $N=29$ individuals~\cite{vickers1981representing} and the Lazega Law Firm (LLF) network with $N=71$ individuals~\cite{lazega2001collegial} (see Methods for the network description). We show in the left panels of Fig.~\ref{fig:empirical_networks} the ($b/c$, $r$) regions in which the cooperator or mutant is favored or disfavored. We obtain $(b/c)^* = -89.5$ for the VC7 two-layer network such that spite can evolve (see Fig.~\ref{fig:empirical_networks}A). By coupling with the constant-selection dynamics layer,
every change in the $r$ value by $1$ changes $(b/c)^*$ value by 3.64 (4.07\%). In the case of the LLF network, cooperation can be favored, and one obtains
$\left(\frac{b}{c}\right)^* = 52.4$ for the uncoupled layer-1 network. When the two layers are coupled, cooperation occurs more easily if $r>1$ and less easily if $r<1$, as shown in Fig.~\ref{fig:empirical_networks}C.
 
\subsection{Birth-death (Bd) rule in the constant-selection layer}

In the remainder of the Results section, including this section, we describe results of several robustness tests.

Constant selection on networks has most popularly been investigated under the Birth-death (Bd) updating rule, partly because the Bd rule amplifies the effect of selection in a majority of networks~\cite{hindersin2015most, tkadlec2020limits, Svoboda2024PlosComputBiol}. Therefore, we also derived the condition for favoring cooperators or mutants when layer 1 uses the dB rule and layer 2 uses the Bd rule, i.e., under the dB-Bd rule (see
Section~\ref{sec:SI-C}).

Under the dB-Bd rule, because the evolutionary dynamics of the donation game is still governed by the dB rule, the condition under which cooperation is favored in layer 1 is given by Eq.~\eqref{eqn:cooperation_criteria_reduced}. However, the value of $\phi_{2,0}^{\bm{\xi}[1,2]}$ changes from the case of the dB-dB rule, affecting the value of $(b/c)^*$. Note that $\theta_1^{\bm{\xi}[1]}$, $\theta_2^{\bm{\xi}[1]}$, and $\theta_3^{\bm{\xi}[1]}$ remain the same because they only depend on the network structure and updating rule in layer 1.

The results for the dB-Bd rule on two-layer networks with $N=6$ individuals (right panel of Table~\ref{tab:thetapi_distribution_percentages}), four sample networks (right panels of Fig.~\ref{fig:examples}), and two-layer ER and BA networks (Fig.~\ref{fig:tables_random_graphs}C and \ref{fig:tables_random_graphs}D) are qualitatively the same to those for the dB-dB rule (left panel of Table~\ref{tab:thetapi_distribution_percentages}, middle panels of Fig.~\ref{fig:examples}, and Fig.~\ref{fig:tables_random_graphs}A and \ref{fig:tables_random_graphs}B). Notable quantitative differences are: (i) $(b/c)^*$ under the dB-Bd rule is
%
%
$10.1$ times more sensitive to variation in $r$ than under the dB-dB rule in the two-layer network shown in Fig.~\ref{fig:examples}B. (ii) $\left| \text{d}(b/c)^*/\text{d}r \right|$ is substantially larger for the two-layer ER and BA networks under the dB-Bd than dB-dB rule across the network density parameters (i.e., $p_1$, $p_2$, $\overline{m}_1$, and $\overline{m}_2$). In the VC7 network, while $(b/c)^*$ increases as $r$ increases under the dB-dB rule, $(b/c)^*$ decreases as $r$ increases under the dB-Bd rule. Another observation on empirical two-layer networks is that $(b/c)^*$ is more responsive to the change in $r$ under the dB-Bd than dB-dB rule for the VC7 network and vice versa for the LLF network. We conclude that the effect of layer-to-layer coupling on evolution of cooperation is overall similar between the dB-Bd and dB-dB rules, broadening the generality of our results.

\subsection{Verification with stochastic numerical simulations}

To explore the validity of our theory, we computed the fixation probability for cooperation by direct numerical simulations of the evolutionary dynamics with the strength of selection being $\delta = 0.02$ and $\delta = 0.2$. 
We used the two-layer ring network with $N=10$ individuals shown in Fig.~\ref{fig:examples}A, two-layer BA networks with $N=15$ and $N=100$, and a two-layer ER random graph with $N=100$. See Section~\ref{sec:SI-L} for the methods and results in detail.

Across these networks and the two updating rules (i.e., dB-dB and dB-Bd), we found that, for $\delta = 0.02$, the $(b/c)^*$ values estimated by the direct numerical simulations agree well with our theoretical prediction. In all these networks, coupling with the constant-selection layer promoted cooperation with an appropriate value of $r$. For $\delta = 0.2$, the numerically estimated $(b/c)^*$ considerably deviated from the theoretical prediction. This deviation is expected because our theory strictly holds true in the weak selection limit (i.e., $\delta \to 0$). Nonetheless, the numerical results with $\delta = 0.2$ indicate that coupling with the constant-selection layer promotes cooperation as it does for $\delta = 0.02$.

It should be noted that the present numerical test includes larger networks (i.e., $N=100$) than those used in the previous sections. We conclude that our main theoretical results are consistent with numerically simulated results of the evolutionary dynamics, including on larger networks.

For ER, BA, and empirical networks of comparable sizes, we also investigated the extent to which coupling with the constant-selection network layer enhances cooperation depending on the network structure and the effects of initial condition on $(b/c)^*$ (see Section~\ref{sec:SI-E}). The results are more diverse than but roughly consistent with those for networks with $N=6$.

\subsection{``Fixed benefits, fixed costs'' goods scheme\label{sub:ff-scheme}}

We have considered the payoff scheme in which the cooperator pays the cost $c$ and any neighbor of the cooperator receives a benefit $b$. This payoff scheme is the ``pf goods'' scheme in \cite{mcavoy2020social}, except the difference regarding whether or not one divides the total payoff for each individual by the degree of the node, denoted by $k$ in this section, for normalization. To further explore generality of our results, here we consider the ``ff goods'' scheme \cite{mcavoy2020social}, with which a producer (i.e., cooperator) pays a fixed cost $c$ irrespectively of $k$, and each neighbor receives benefit $b/k$. Therefore, the benefit produced by a producer, $b$, does not scale with the producer's $k$, and the benefit $b$ is divided equally by all the neighbors of the producer.
%

We derived the condition for favoring cooperators and mutants under the ff goods scheme and then examined the derived condition for the model and empirical networks that we have used (see
Section~\ref{sec:SI-M}).
The results are qualitatively the same as those under the pf goods scheme shown in the previous sections. 

\subsection{Fixation time}

On one-layer networks, the cooperator's fixation probability and fixation time are often in a trade-off relationship~\cite{altrock2009fixation,sui2015speed}. The advantages of two-layer networks over one-layer networks in promoting cooperation would be compromised if fixation times are excessively longer for two-layer than one-layer networks. Therefore, we have numerically investigated fixation time in two-layer networks in some representative scenarios.
To make a fair comparison between the one-layer and two-layer cases, we declared the fixation of cooperation once it is attained in layer 1 even if the resident or mutant type has not fixated in layer 2. We computed the fixation time for cooperators as the average over the runs for which the cooperator has fixated.

We show the mean fixation time for cooperators for different networks and values of $b/c$, $r$, and $\delta$ (i.e., selection strength) in
Section~\ref{sec:SI-N}.
For four synthetic networks, we find that the fixation time does not substantially increase (i.e., the increase is at most $\approx 10\%$) in two-layer networks relative to one-layer networks including the case in which cooperation is facilitated by the two-layer networks. For the VC7 network, the fixation of cooperation needs at most 2.8 times more time when the two network layers are coupled than uncoupled.

\section{Discussion}

The donation game and constant selection are two widely studied models of evolutionary dynamics. Both models allow substantial theoretical analysis, in particular via fixation theory, providing general and quantitative understanding. These two models have mostly been studied in isolation. They capture different social processes that people and institutions may simultaneously encounter. In particular, constant-selection dynamics is equivalent to a biased voter model, a standard model of competing opinions studied for decades \cite{Bramson1981AnnProb, Durrett1988book, Antal2006PhysRevLett, Czaplicka2022ChaosSolitonsFractals}. We modeled this setting using a two-layer network in which each individual is involved in the donation game in one network layer and undergoes constant-selection dynamics in the other layer. By adapting the recently developed multilayer fixation theory \cite{su2022evolution}, we showed that coupling with the layer can enhance prosociality in the donation-game layer. This effect is analogous to the cooperation enhancement previously reported when both layers host donation games \cite{su2022evolution}. In other words, enhancing prosociality does not require coupling two dilemma layers. Coupling a dilemma layer to a non-dilemmatic layer can also produce a cross-layer spillover effect that shifts the effective condition for cooperation in the dilemma layer. In our model, the condition for favoring cooperation (i.e., $(b/c)^*$) is reasonably responsive to the bias in the strength of the two opinions competing in the other layer (i.e., $r$). Conversely, the donation-game parameters (i.e., $b$ and $c$) spill over to affect the condition for favoring the mutant type in the constant-selection layer. We tested robustness across different networks, updating rules, and goods (i.e., payoff) schemes. Extension to the case of more than two layers \cite{Gomezgardenes2012SciRep}, different types of non-dilemma or dilemma dynamics in layer 2 \cite{santos2014biased,wang2020vaccination,raducha2023evolutionary,basak2024evolution, zhu2025evolution}, layers of hypergraphs modeling group interactions~\cite{santos2008social, Jusup2022PhysRep, Alvarezrodriguez2021NatHumanBehav,%
%
%
de2023more,sheng2024strategy}, different updating rules such as a pairwise comparison rule \cite{SzaboToke1998PhysRevE, ohtsuki2006simple, TraulsenNowakPacheco2006PhysRevE}, and different initial conditions may be fruitful.

 Engineering the structure of one network layer to enhance cooperation in the other network layer was briefly examined in \cite{su2022evolution} for the case of donation games in different network layers. 
Such an intervention method may be similarly effective at enhancing cooperation in our model. However, clarifying how one should manipulate layer 2, in which the constant-selection dynamics occurs, awaits investigation. In addition, the correspondence between the constant-selection dynamics and the biased voter model opens further avenues for engineering cooperation. Our results suggest that deploying a new opinion or fad currently dominated by a resident opinion may enhance cooperation in the other network layer. One may be able to target the initial adopters of the new opinion to further facilitate cooperation. 
Furthermore, cooperation may be better enhanced by introducing zealots, i.e., a small number of individuals who stick to their opinion regardless of social influence ~\cite{Mobilia2003PhysRevLett,Mobilia2007JStatMech,XieSreenivasan2011PhysRevE}. Introducing zealot cooperators \cite{Masuda2012SciRep} guarantees fixation of cooperation in finite populations~\cite{NakajimaMasuda2015JMathBiol}. In practice, however, it may be less costly to introduce zealots in the opinion layer to indirectly promote cooperation than to introduce zealot cooperators directly.

We assumed that all the individuals generate payoffs in both layers and that the payoff summed over the two layers, $F_i$, determines fecundity of the $i$th individual in both layers. This corresponds to a simple type of multilayer networks called multiplex networks. In fact, multilayer network research has advocated various ways in which different network layers depend on each other, driven by real-world examples~\cite{Kivela2014JCompNetw, Boccaletti2014PhysRep, Bianconi2018book, de2023more}. 
Extending the present framework, as well as prior analyses of donation games in both layers~\cite{su2022evolution} and constant-selection dynamics in both layers~\cite{ruodan2023multilayer}, to various types of multilayer network setting with analytical rigor warrants future work. First, inter-layer coupling may be weaker or unidirectional. A weaker inter-layer coupling implies that the fecundity in one layer more weakly depends on the payoff in the other layer \cite{wang2015evolutionary, ShenChu2018EPL}. The unidirectional coupling implies that the payoff in one layer influences the fecundity in the other layer, but not vice versa. Second, some nodes may be missing in one layer. 
Third, a ``network of networks'' setting would include more than two layers and specify which layer pairs interact. A network of networks may better reflect human behavior when individuals playing a social dilemma game use information from selected other social contexts, rather than from all layers equally.

The present work has two main limitations. First, as in fixation-based analyses of multilayer networks
~\cite{su2022evolution, ruodan2023multilayer}, it is computationally demanding to check conditions for favoring cooperators and mutants. Therefore, our present analysis is limited to small two-layer networks. However, we analyzed networks of different sizes to the best of our effort and also carried out robustness tests along additional dimensions. Second, as for a plethora of work in mathematically founded studies of evolutionary game and graph theory, we assumed weak selection. Focusing on simple network structure such as the coupled complete graph and the coupled ring network to analyze the case of strong selection may be a useful analytical approach. Our main message is that coupling a dilemma layer with the constant-selection dynamics, or the biased voter model dynamics, can enhance prosociality in the dilemma layer, similar to the case of coupling two dilemma layers. An important direction for future research is to test this prediction by controlled experiments and data-driven multilayer network modeling.

\section*{Methods}

\subsection*{Generation of two-layer BA networks}

For each layer $L \in \{1, 2 \}$, the initial condition of the network growth process was the star graph on $\overline{m}_L +1$ nodes. Then, we grew the network by sequentially adding nodes with $\overline{m}_L $ edges according to the preferential attachment rule~\cite{Barabasi1999Science}. For each $\overline{m}_L \in\{1,2,3,4,5\}$, we generated 5 instances of BA networks with $N=15$ nodes for each layer, yielding $25$ two-layer BA networks. We then uniformly randomly shuffled the node label of layer 2 to avoid the tendency that individuals with small indices (i.e., joining the network early in the growth process) are hubs in both layers such that the node's degree is positively correlated between the two layers.

\subsection*{Empirical two-layer networks}

We have used two empirical two-layer networks.
The first is the Vickers–Chan 7th Graders (VC7) network, which comprises two layers representing academic and friendship connections among 29 seventh-grade students from a school in Victoria, Australia~\cite{vickers1981representing}. We arbitrarily assigned the academic connection network to layer 1. The VC7 network has 126 and 152 edges in layer 1 and 2, respectively. The second network is the Lazega Law Firm (LLF) network, in which the two network layers represent professional and cooperative relationships among 71 partners within the firm~\cite{lazega2001collegial}. We arbitrarily assigned the professional relation network to layer 1. The LLF network has 717 and 726 edges in layer 1 and 2, respectively.

\section*{Acknowledgments}

We thank Alex McAvoy for discussion.
During the preparation of the revised manuscript, the authors used Claude Opus 4.7 for code development assistance. All numerical results and code were manually verified by the authors.

\section*{Data availability}

All other data was produced by numerical analysis and simulation.
The code to produce the data in this study is available at GitHub \cite{Bhaumik2026Github}.

\section*{Funding}

N.M. acknowledges support from the Japan Science and Technology Agency (JST) Moonshot R\&D (under grant no.\,JPMJMS2021), the National Science Foundation (under grant no.\,2204936), and JSPS KAKENHI (under grant nos.\,JP 23H03414, 24K14840, and 24K03013). The funders had no role in study design, data collection and analysis, decision to publish, or preparation of the manuscript.


\clearpage
\setcounter{section}{0}
\setcounter{subsection}{0}
\setcounter{table}{0}
\setcounter{figure}{0}
\setcounter{equation}{0}
\renewcommand{\thesection}{S\arabic{section}}
\renewcommand{\thesubsection}{\thesection.\Alph{subsection}}
\renewcommand{\thesubsubsection}{\thesubsection.\arabic{subsubsection}}
\renewcommand{\thetable}{S\arabic{table}}
\renewcommand{\thefigure}{S\arabic{figure}}
\renewcommand{\theequation}{S\arabic{equation}}
\renewcommand{\theHsection}{S\arabic{section}}
\renewcommand{\theHsubsection}{S\arabic{section}.\Alph{subsection}}
\renewcommand{\theHsubsubsection}{S\arabic{section}.\Alph{subsection}.\arabic{subsubsection}}
\renewcommand{\theHtable}{S\arabic{table}}
\renewcommand{\theHfigure}{S\arabic{figure}}
\renewcommand{\theHequation}{S\arabic{equation}}

\begin{center}
\begin{LARGE}{\textbf{Supplementary Information for: \\
\vspace{12pt}
Coupling with opinion dynamics promotes prosocial behavior in multilayer networks}}
\vspace{12pt}
\end{LARGE}

\vspace{12pt}

Jnanajyoti Bhaumik, Naoki Masuda
\end{center}
\section{Derivation of the condition under which the cooperator or mutant is favored}\label{sec:SI-A}\label{sec:methods}

\subsection{Replacement events and the fixation axiom}\label{sec:replacement-events-and-the-fixation-axiom}
	
We use the notations and assumptions, particularly regarding the replacement events and fixation axiom, in Ref.~\cite{su2022evolution}, which was originally introduced in Ref.~\cite{allen2019mathematical} and also used in Ref.~\cite{mcavoy2021fixation}. We consider a population with two layers. In each time step, we choose a replacement event, $(R, \alpha)$, which consists of a pair $\left(R^{[L]}, \alpha^{[L]}\right)$ for each layer, $L \in \{1, 2 \}$, where $R^{[L]}$ is the set of `offspring' nodes that are chosen for replacement, and $\alpha^{[L]} : R^{[L]} \rightarrow \{1, \ldots, N\}$ is the offspring-to-parent map in layer $L$. We refer to cooperators and mutants as type A, and defectors and residents as type B. We denote by $p_{(R, \alpha)} (\bm{x})$ the probability of choosing $(R, \alpha)$ in state $\bm{x} \in \{0,1\}^{N} \times \{0,1\}^{N}$, where each entry of $\bm{x}$, denoted by $x_{i}^{[L]}$, is 1 and 0 if individual $i$ has type A and B, respectively, in layer $L$. 
In each layer $L$, the map $\alpha^{[L]} : R^{[L]} \rightarrow \{1, \ldots, N\}$ extends to a map $\tilde{\alpha}^{[L]} : \{1, \ldots, N\} \rightarrow \{1, \ldots, N\}$ defined by $\tilde{\alpha}^{[L]}(i) = \alpha^{[L]}(i)$ if $i \in R^{[L]}$ and $\tilde{\alpha}^{[L]}(i) = i$ if $i \notin R^{[L]}$. For any state $\bm{x} \in \{0,1\}^{N} \times \{0,1\}^{N}$, we write $\bm{x}^{[L]} \in \{0,1\}^{N}$ for the state of the population in layer $L \in \{1, 2\}$. This extension of $\alpha$, denoted $\tilde{\alpha}$, gives an updated state $\bm{x}_{\tilde{\alpha}} \in \{0,1\}^{N} \times \{0,1\}^{N}$ defined by $(\bm{x}_{\tilde{\alpha}})_{i}^{[L]} = \bm{x}_{\tilde{\alpha}^{[L]}(i)}^{[L]}$. In the case of the dB and Bd processes, $R^{[L]}$ is a singleton set containing the individual $i_0$ to be replaced by its parent $j$, and we obtain $\tilde{\alpha}^{[L]}(i) = i$ if $i \neq i_0$ and $\tilde{\alpha}^{[L]}(i_0) = j$.

For $\bm{x}, \bm{y} \in \{0,1\}^{N} \times \{0,1\}^{N}$, we then have the state transition probability as follows:
\begin{equation}
	P_{\bm{x} \rightarrow \bm{y}} = \sum_{(R, \alpha) \; \text{s.t.} \; \bm{x}_{\tilde{\alpha}} = \bm{y}} p_{(R, \alpha)} (\bm{x}).
\end{equation}
In addition to being smooth, we assume that the replacement rule satisfies the following fixation axiom \cite{su2022evolution}:
\begin{axiom}[Fixation axiom]
	There exists $(i_1, i_2) \in \{1, \ldots, N\} \times \{1, \ldots, N\}$, an integer $m \geq 1$, and a sequence of replacement events $\{(R_k, \alpha_k)\}_{k=1}^{m}$ such that:
	\begin{itemize}
		\item[(i)] $p_{(R_k, \alpha_k)} (\bm{x}) > 0$ for every $k \in \{1, \ldots, m\}$ and $\bm{x} \in \{0,1\}^{N} \times \{0,1\}^{N}$.
		\item[(ii)] For each $L \in \{1, 2\}$, there exists $k_L$ such that $i_L \in R_{k_L}^{[L]}$.
		\item[(iii)] For $L \in \{1, 2\}$, we have $\tilde{\alpha}_{1}^{[L]} \circ \tilde{\alpha}_{2}^{[L]} \circ \cdots \circ \tilde{\alpha}_{m}^{[L]}(j) = i_L$ for every $j \in \{1, \ldots, N\}$.
	\end{itemize}
\end{axiom}

\subsection{Weak selection and a mutation-modified Markov chain}\label{sec:weak-selection}

We assume weak selection. In Ref.~\cite{su2022evolution}, they considered a mutation-modified Markov chain obtained by sending each absorbing state to a fixed, transient state $\bm{\xi}$ $\in \{0,1\}^{N} \times \{0,1\}^{N}$ with mutation probability $u > 0$ in each time step. With probability $1-u$, the chain remains in the same absorbing state in each time step. The benefit of this approach is that the mutation-modified Markov chain has a unique stationary distribution that facilitates further calculations. This chain has transition probabilities given by
	\begin{equation}
		P_{\bm{x}\rightarrow\bm{y}}^{\circlearrowright\{\bm{\xi}\}}=\begin{cases}
			u & \text{if } \bm{x} \in \{\bm{AA}, \bm{AB}, \bm{BA}, \bm{BB}\}, \bm{y} = \bm{\xi}, \\
			(1 - u) P_{\bm{x} \to \bm{y}} & \text{if } \bm{x} \in \{\bm{AA}, \bm{AB}, \bm{BA}, \bm{BB}\}, \bm{y} \neq \bm{\xi}, \\
			P_{\bm{x} \to \bm{y}} & \text{if } \bm{x} \notin \{\bm{AA}, \bm{AB}, \bm{BA}, \bm{BB}\}.
		\end{cases}
\label{eq:chain-transition-probability}
	\end{equation}
In Eq.~\eqref{eq:chain-transition-probability}, $\bm{AA}$ is the state in which all the individuals are of type A in both layers, and $\bm{AB}$ is the state in which all individuals are of type A in layer 1 and type B in layer 2. States $ \bm{BA}$ and $\bm{BB}$ are similarly defined.
The probability that $i$ transmits its offspring to $j$ in layer $L$ and in state $\bm{x}$
%
%
is
\begin{equation}
e_{ij}^{[L]}(\bm{x}) := \sum_{(R, \alpha) \; \text{s.t.} \; \alpha^{[L]}(j) = i} p_{(R, \alpha)} \, \left(\bm{x}\right).
\end{equation}
	
We recall that $\delta$ represents the strength of selection. If $\delta=0$, then we have neutral drift, which is denoted by $\circ$. We denote by $e_{ij}^{\circ [L]}$ the probability that $i$ transmits its offspring to $j$ under neutral drift in layer $L$. We also denote by $\pi_i^{[L]}$ the reproductive value (RV) of individual $i$~\cite{fisher1999genetical,taylor1990allele,taylor1996inclusive,grafen2006theory,maciejewski2014reproductive} in layer $L$ under neutral drift. We obtain
\begin{equation}\label{eqn:reproductive-value-1}
	\sum_{j=1}^N e_{ij}^{\circ [L]} \pi_j^{[L]} = \pi_i^{[L]} \sum_{j=1}^N e_{ji}^{\circ [L]}
\end{equation}
and	
\begin{equation}\label{eqn:reproductive-value-2}
	\sum_{i=1}^N \pi_i^{[L]} = 1.
\end{equation}
	
The RV-weighted frequency of state $\bm{x}$ in layer $L$ is denoted by
\begin{equation}
\widehat{x}^{[L]} := \sum_{i=1}^{N}\pi_i^{[L]} x_i^{[L]}.
\end{equation}
The change in the RV-weighted frequency in layer $L$ due to selection in one step of the evolutionary dynamics is
	\begin{equation}\label{eqn:RV-selection}
		\widehat{\Delta}^{\left[L\right]}_{\text{sel}}(\bm{x}) = \sum_{i=1}^{N} \sum_{j=1}^{N} \pi_{i}^{[L]} (x_j - x_i) e_{ji}^{[L]}(\bm{x}).
	\end{equation}
The change in the RV-weighted frequency in one time step due to both mutation and selection is given by
	\begin{equation}\label{eqn:rvchange}
		\widehat{\Delta}^{[L]}(\bm{x}) =
		\begin{cases}
			u \left( \widehat{\xi}^{[L]}-1\right) & \text{if } \bm{x}^{[L]} = \bm{A}, \, \bm{x}^{[-L]} \in \{\bm{A}, \bm{B}\}, \\
			u \widehat{\xi}^{[L]} & \text{if } \bm{x}^{[L]} = \bm{B}, \, \bm{x}^{[-L]} \in \{\bm{A}, \bm{B}\}, \\
			\widehat{\Delta}^{[L]}_{\text{sel}}(\bm{x}) & \text{if } \bm{x} \notin \{\bm{AA}, \bm{AB}, \bm{BA}, \bm{BB}\},
		\end{cases}
	\end{equation}
where $-L$ means the other layer (i.e., layer 2 if $L=1$ and layer 1 if $L=2$); $\widehat{\xi}^{[L]}=\sum_{i=1}^{N}\pi_i^{[L]} \xi_i^{[L]}$ is the RV-weighted frequency of type A in state $\bm{\xi}$ in layer $L$.
	
Next, we compute the expectation of the change in the RV-weighted frequency, which will enable us to compute the fixation probability of the mutation-modified Markov chain under the rare mutation condition. Because we are computing the expectation with respect to the stationary probability distribution, we obtain
	\begin{equation}
		\mathbb{E}_{\circlearrowright\left(\bm{\xi}\right)} [\widehat{\Delta}^{[L]}] = 0.
\label{eq:RV-weighted-frequency-martingale}
	\end{equation}
Equation~\eqref{eq:RV-weighted-frequency-martingale} yields
\begin{equation}
	\mathbb{E}_{\circlearrowright(\bm{\xi})}\left[\widehat{\Delta}_{\text{sel}}^{[L]}\right] = 
	\begin{cases} 
		u \pi_{\circlearrowright(\bm{\xi})} \left( \{ \bm{A} \} \times \{ \bm{A},\bm{B} \} \right) \left(1 - \widehat{\xi}^{[L]} \right) - u \pi_{\circlearrowright(\bm{\xi})} \left( \{ \bm{B} \} \times \{ \bm{A},\bm{B} \} \right) \widehat{\xi}^{[L]} & \text{if } L = 1, \\
		u \pi_{\circlearrowright(\bm{\bm{\xi}})} \left( \{ \bm{A},\bm{B} \} \times \{ \bm{A} \} \right) \left(1 - \widehat{\xi}^{[L]} \right) - u \pi_{\circlearrowright(\bm{\xi})} \left( \{ \bm{A},\bm{B} \} \times \{ \bm{B} \} \right) \widehat{\xi}^{[L]} & \text{if } L = 2.
	\end{cases}
\end{equation}

We define $\rho_{\bm{A}}^{[L]}\left(\bm{\xi}\right)$ as the fixation probability of the mutation-modified Markov chain reaching a state in which all individuals have type A in layer $L$ when the initial state of the network is $\bm{\xi}$. By the known rare-mutation results \cite{allen2019mathematical,mcavoy2021fixation}, we have
\begin{align}
	\lim_{u \to 0} \pi_{\circlearrowright(\bm{\xi})} \left( \{\bm{A}\} \times \{\bm{A},\bm{B}\} \right) &= \rho_{\bm{A}}^{[1]}(\bm{\xi}), \\
	\lim_{u \to 0} \pi_{\circlearrowright(\bm{\xi})} \left( \{\bm{B}\} \times \{\bm{A},\bm{B}\} \right) &= 1 - \rho_{\bm{A}}^{[1]}(\bm{\xi}), \\
	\lim_{u \to 0} \pi_{\circlearrowright(\bm{\xi})} \left( \{\bm{A},\bm{B}\} \times \{\bm{A}\} \right) &= \rho_{\bm{A}}^{[2]}(\bm{\xi}), \\
	\lim_{u \to 0} \pi_{\circlearrowright(\bm{\xi})} \left( \{\bm{A},\bm{B}\} \times \{\bm{B}\} \right) &= 1 - \rho_{\bm{A}}^{[2]}(\bm{\xi}).
\end{align}
Then, we can write down the fixation probability~\cite{su2022evolution} as
\begin{equation}\label{eqn:defn_fixation_prob}
	\rho_{\bm{A}}^{\left[L\right]}\left(\bm{\xi}\right)=\widehat{\xi}^{\left[L\right]}+\left.\frac{\text{d}}{\text{d}u}\right|_{u=0} \mathbb{E}_{\circlearrowright(\bm{\xi})}\left[\widehat{\Delta}_{\text{sel}}^{[L]}\right] .
\end{equation}
Because  $\pi_{\circlearrowright(\bm{\xi})}, \widehat{\xi}^{[L]}$, and $u$ are smooth in terms of $\delta$ and $u$, we have \cite{su2022evolution,allen2019mathematical}
\begin{equation}\label{eqn:derivative_fixation_prob}
	\left.\frac{\text{d}}{\text{d}\delta}\rho_{\bm{A}}^{\left[L\right]}\left(\bm{\xi}\right)\right|_{\delta=0} = \left. \frac{\text{d}}{\text{d}u} \right|_{u=0} \mathbb{E}^{\circ}_{\circlearrowright(\bm{\xi})} \left[ \left. \frac{\text{d}}{\text{d}\delta} \right|_{\delta=0} \widehat{\Delta}_{\text{sel}}^{[L]} \right].
\end{equation}
Finally, for $\bm{x} \notin \{\bm{AA}, \bm{AB}, \bm{BA}, \bm{BB}\}$, consider the rare-mutation conditional (RMC) distribution \cite{su2022evolution} given by
\begin{equation}\label{eqn:pi_rmc}
	\pi_{\text{RMC}(\bm{\xi})} (\bm{x}) := K \left. \frac{\text{d}}{\text{d}u} \right|_{u=0} \pi_{\circlearrowright(\bm{\xi})} (\bm{x}),
\end{equation}
where
\begin{equation}
K := \left(\sum_{\bm{y}\notin\{\bm{AA}, \bm{AB}, \bm{BA}, \bm{BB}\}}\left.\frac{\text{d}}{\text{d}u}\right|_{u=0}\pi_{\circlearrowright(\bm{\xi})} (\bm{y})\right)^{-1}.
\label{eq:def-K}
\end{equation}
The following lemma holds true:
\begin{lemma}
	For any function $\varphi : \{0,1\}^{N} \times \{0,1\}^{N} \rightarrow \mathbb{R}$,
	\begin{equation}\label{eqn:ExpRMC}
		\begin{aligned}
			\mathbb{E}^{\circ}_{\text{RMC}(\bm{\xi})} [\varphi] = K^{\circ} \left[ \right.
			& \varphi(\bm{\xi}) - \widehat{\xi}^{[1]} \widehat{\xi}^{[2]} \varphi(\bm{A}, \bm{A}) - \widehat{\xi}^{[1]} (1 - \widehat{\xi}^{[2]}) \varphi(\bm{A}, \bm{B}) \\
			& \left. - (1 - \widehat{\xi}^{[1]}) \widehat{\xi}^{[2]} \varphi(\bm{B}, \bm{A}) - (1 - \widehat{\xi}^{[1]}) (1 - \widehat{\xi}^{[2]}) \varphi(\bm{B}, \bm{B})
			\right] \\
			& + \sum_{(R, \alpha)} p^{\circ}_{(R, \alpha)} \mathbb{E}^{\circ}_{\text{RMC}(\bm{\xi})} [\varphi _{\tilde{\alpha}}].
		\end{aligned}
	\end{equation}
\end{lemma}
\noindent We recall that $\varphi_{\tilde{\alpha}}$ is the extension of the off-spring to parent map that maps the nodes in $R$ to their parents  and the nodes not in $R$ to themselves. As indicated in Ref.~\cite{su2022evolution}, the proof of Eq.~\eqref{eqn:ExpRMC} is an adaptation of the proof of Lemma 1 in Ref.~\cite{mcavoy2021fixation}. Note that $\varphi(\bm{x})$ in Lemma 1 in Ref.~\cite{mcavoy2021fixation} is replaced with function 
$\varphi(\bm{\xi}) - \widehat{\xi}^{[1]} \widehat{\xi}^{[2]} \varphi(\bm{A}, \bm{A}) - \widehat{\xi}^{[1]} (1 - \widehat{\xi}^{[2]}) \varphi(\bm{A}, \bm{B}) - (1 - \widehat{\xi}^{[1]}) \widehat{\xi}^{[2]} \varphi(\bm{B}, \bm{A}) - (1 - \widehat{\xi}^{[1]}) (1 - \widehat{\xi}^{[2]}) \varphi(\bm{B}, \bm{B})$ in Eq.~\eqref{eqn:ExpRMC}, as mentioned in Ref.~\cite{su2022evolution}. For any of the four absorbing states, $ \varphi(\bm{\xi}) - \widehat{\xi}^{[1]} \widehat{\xi}^{[2]} \varphi(\bm{A}, \bm{A}) - \widehat{\xi}^{[1]} (1 - \widehat{\xi}^{[2]}) \varphi(\bm{A}, \bm{B}) - (1 - \widehat{\xi}^{[1]}) \widehat{\xi}^{[2]} \varphi(\bm{B}, \bm{A}) - (1 - \widehat{\xi}^{[1]}) (1 - \widehat{\xi}^{[2]}) \varphi(\bm{B}, \bm{B})$  is equal to $0$, which is the requirement for using Lemma 1 in Ref.~\cite{mcavoy2021fixation}. We also use the fact that $\mathbb{E}\left[1\right]=1$ to derive the last term on the right-hand side of Eq.~\eqref{eqn:ExpRMC}.

\subsection{Condition for the cooperator and mutant to be selected in a general form}\label{sec:evolutionary-games}

In this section, we derive recurrence relations for deducing conditions for selecting cooperation in the game layer (i.e., layer 1) and mutant in the constant-selection layer (i.e., layer 2). The payoff matrix of the donation game, which is used in layer 1, is given by
\begin{equation}\label{payoff_matrix}
\bordermatrix{
	& {\rm C} & {\rm D} \cr
	{\rm C} & b-c & -c \cr
	{\rm D} & b & 0 \cr}, \;
\end{equation}
where the entries of the matrix represent the payoff to the row player.
Recall that $p_{ij} = \frac{w_{ij}}{s_i}$, where $w_{ij}$ is the $(i, j)$ entry of the adjacency matrix of the undirected network layer 1, and $s_i$ is the degree of the $i$th node in layer 1. The payoff for the $k$th replica node in layer 1 is given by
\begin{equation}
u_k^{\left[1\right]}(\bm{x}) = \sum_{\ell=1}^{N}\left(-c \cdot p_{k\ell}\cdot x_k^{\left[1\right]} + b\cdot  p_{k\ell }\cdot x_{\ell}^{\left[1\right]}\right)= 
\sum_{\ell=1}^{N}\left(-C_{k\ell} \cdot x_k^{\left[1\right]} + B_{\ell k}\cdot x_{\ell}^{\left[1\right]}\right),
\end{equation}
where $C_{kl}=c\cdot p_{kl}$ and $B_{lk}=b\cdot p_{k\ell}$. We write the payoff matrix in the case of constant selection as
\begin{equation}\label{payoff_matrix_mutant}
\bordermatrix{
	& {\rm M} & {\rm R} \cr
	{\rm M} & r & r \cr
	{\rm R} & 1 & 1 \cr}, \;
\end{equation}
where M and R stand for mutant and resident types, respectively. The payoff for the $k$th replica node in layer 2 is given by
\begin{equation}
	u_k^{[2]}(\bm{x}) = x_k^{\left[2\right]} \left(r-1\right)+1.
\label{eq:payoff-constant-selection}
\end{equation}
The total payoff for the $k$th individual is given by
\begin{equation}
	u_k(\bm{x}) = u_k^{[1]}(\bm{x}) +u_k^{[2]}(\bm{x}) = \sum_{\ell=1}^{N}\left(-C_{k\ell}\cdot x_k^{\left[1\right]} + B_{\ell k}\cdot x_{\ell}^{\left[1\right]}\right)
+	x_k^{\left[2\right]} \left(r-1\right)+1.
\end{equation}

We define $\mathbf{F} = (F_1, \ldots, F_N)$, where we remind that $F_i$ is the fecundity of the $i$th node. The marginal effect of the individual $k$'s fitness on the probability that $i$ replaces $j$ in layer $L$~\cite{mcavoy2020social} is denoted by
\begin{equation}
	m_{k;ij}^{[L]} = \left. \frac{\partial e_{ij}^{[L]}}{\partial F_k} \right|_{\mathbf{F}=\bm{1}},
\label{eq:def-m}	
\end{equation}
where $\bm{1} = (1, \ldots, 1)$. We find that
\begin{equation}
	\left. \frac{\text{d}}{\text{d}\delta} e_{ij}^{[L]}(\bm{x}) \right|_{\delta=0} = 
	\sum_{k=1}^{N} \left. \frac{\partial e_{ij}^{[L]}}{\partial F_k} \right|_{\mathbf{F}=1}\cdot \left. \frac{\text{d}F_k}{\text{d}\delta}\right|_{\delta=0}=
	\sum_{k=1}^{N} m_{k; ij}^{[L]} u_k(\bm{x}) = 
	\sum_{k=1}^{N} m_{k; ij}^{[L]} \left[ u_k^{[1]}(\bm{x}) + u_k^{[2]}(\bm{x}) \right].
\end{equation}
Thus, it follows from the definition of $\widehat{\Delta}_{\text{sel}}^{[L]}(\bm{x})$ (i.e., Eq.~\eqref{eqn:RV-selection}) that
\begin{align}
	\left. \frac{\text{d}}{\text{d}\delta} \widehat{\Delta}_{\text{sel}}^{[L]}(\bm{x}) \right|_{\delta=0} &= \sum_{i,j,k=1}^{N} \pi_{i}^{[L]} m_{k;ji}^{[L]} \left( x_j^{[L]} - x_i^{[L]} \right) \left[ u_k^{[1]}(\bm{x}) + u_k^{[2]}(\bm{x}) \right] \notag\\
	&= \sum_{i,j,k=1}^{N} \pi_{i}^{[L]} m_{k;ji}^{[L]} \left( x_j^{[L]} - x_i^{[L]} \right) \left\{\left[\sum_{\ell=1}^{N}\left(-C_{k\ell}\cdot x_k^{\left[1\right]} + B_{\ell k}\cdot x_\ell^{\left[1\right]}\right)\right]
	+	\left[x_k^{\left[2\right]} \left(r-1\right)+1\right]\right\} \notag\\
	&=  \sum_{i,j,k=1}^{N} \pi_{i}^{[L]} m_{k;ji}^{[L]} \left\{-\left[\sum_{\ell=1}^{N} \left(x_{j}^{\left[L\right]} x_k^{\left[1\right]} - x_{i}^{\left[L\right]} x_k^{\left[1\right]}\right)C_{k\ell} - \left(x_{j}^{\left[L\right]} x_\ell^{\left[1\right]} - x_{i}^{\left[L\right]} x_\ell^{\left[1\right]}\right)B_{\ell k}\right] \right. \notag \\
	&\quad + \left. \left[\left(x_{j}^{\left[L\right]} x_k^{\left[2\right]} - x_{i}^{\left[L\right]} x_k^{\left[2\right]}\right)\left(r-1\right)+\left(x_{j}^{\left[L\right]}  - x_{i}^{\left[L\right]} \right)\right] \right\}.
\label{eqn:explicit_form_delta_derivative_change_in_selection}
\end{align}

We denote by $x_{ij}^{\bm{\xi}[L, L']} \equiv \mathbb{E}_{\text{RMC}(\bm{\xi})} \left[ x_i^{[L]} x_j^{[L']} \right]$ the probability that individual $i$ in layer $L$ and individual $j$ in layer $L'$ are both of type A in the neutral RMC distribution. By taking the expectation of both sides of Eq.~\eqref{eqn:explicit_form_delta_derivative_change_in_selection} and combining the result with Eq.~\eqref{eqn:derivative_fixation_prob} and the definition of the RMC distribution, we obtain
\begin{align}
\left. \frac{\text{d}}{\text{d}\delta} \rho_{\bm{A}}^{[L]}(\bm{\xi}) \right|_{\delta=0}
=& \frac{1}{K^{\circ}} \mathbb{E}^{\circ}_{\text{RMC}(\bm{\xi})} \left[ \frac{\text{d}}{\text{d}\delta} \widehat{\Delta}_{\text{sel}}^{[L]} \right]_{\delta=0} \notag\\
=&  \frac{1}{K^{\circ}}\sum_{i,j,k=1}^{N} \pi_{i}^{[L]} m_{k;ji}^{[L]} \left\{ -\left[\sum_{\ell=1}^{N} \left(x_{jk}^{\bm{\xi}[L, 1]}-x_{ik}^{\bm{\xi}[L, 1]}\right)C_{k\ell} - \left(x_{j\ell}^{\bm{\xi}[L, 1]}-x_{i\ell}^{\bm{\xi}[L, 1]}\right)B_{\ell k}\right] \right. \notag \\
&+ \left. \left(x_{jk}^{\bm{\xi}[L, 2]}-x_{ik}^{\bm{\xi}[L, 2]}\right)\left(r-1\right)+
	\mathbb{E}_{\text{RMC}(\bm{\xi})} \left[x_{j}^{\left[L\right]}  - x_{i}^{\left[L\right]} \right]\right\}.
\label{eqn:weak_fixn_prob}
\end{align}
Let
\begin{equation}
\label{eqn:beta_ij_defn}
\beta_{ij}^{\bm{\xi}[L]} := \frac{x_{ij}^{\bm{\xi}[LL]}}{K^\circ}, \quad L \in \{1, 2\}.
\end{equation}
Owing to Lemma 1 in Ref.~\cite{su2022evolution}, $\beta_{ij}^{\bm{\xi}[1]}$ satisfies the following recurrence relation including when $i=j$:
\begin{equation}\label{eqn:beta_reccurence_soln}
	\beta_{ij}^{\bm{\xi}[1]} = \xi_{i}^{[1]} \xi_{j}^{[1]} - \widehat{\xi}^{[1]} + \sum_{(R^{[1]},\alpha^{[1]})} p^{\circ}_{(R^{[1]},\alpha^{[1]})} \beta_{\tilde{\alpha}^{[1]}(i) \tilde{\alpha}^{[1]}(j)}^{\bm{\xi}[1]}.
\end{equation}
For the ``cross terms'' which are associated to the two layers jointly, we let
\begin{equation}
\label{eqn:gamma_ij_defn}
\gamma_{ij}^{\bm{\xi}[12]} := \frac{x_{ij}^{\bm{\xi}[12]}}{K^\circ}
\end{equation}
and
\begin{equation}
\gamma_{ij}^{\bm{\xi}[21]} := \frac{x_{ij}^{\bm{\xi}[21]}}{K^\circ}.
\end{equation}
Owing to Lemma 1 in Ref.~\cite{su2022evolution}, $\gamma_{ij}^{\bm{\xi}[12]}$
satisfies the recurrence relation given by
\begin{equation}\label{eqn:gamma_reccurence}
\gamma_{ij}^{\bm{\xi}[12]} = \xi_{i}^{[1]} \xi_{j}^{[2]} - \widehat{\xi}^{[1]} \widehat{\xi}^{[2]} + \sum_{(R^{[1]},\alpha^{[1]}), (R^{[2]},\alpha^{[2]})} p^{\circ}_{(R^{[1]},\alpha^{[1]})} p^{\circ}_{(R^{[2]},\alpha^{[2]})} \gamma_{\tilde{\alpha}^{[1]}(i) \tilde{\alpha}^{[2]}(j)}^{\bm{\xi}[1,2]}
\end{equation}
including when $i=j$. Equation~\eqref{eqn:gamma_reccurence} implies that $\gamma_{ij}^{\bm{\xi}[12]} =\gamma_{ij}^{\bm{\xi}[21]}$.
To compute $\mathbb{E}_{\text{RMC}(\bm{\xi})} \left[x_{j}^{\left[L\right]}  - x_{i}^{\left[L\right]} \right]$ on the right-hand side of Eq.~\eqref{eqn:weak_fixn_prob}, we let
	\begin{equation}\label{eqn:eta_definition}
		\eta_{i}^{\bm{\xi}[1]} := \mathbb{E}_{\text{RMC}(\bm{\xi})} \left[\frac{x_i^{\left[1\right]}}{K^\circ}\right].
	\end{equation}
Owing to Lemma 1 in Ref.~\cite{su2022evolution}, $\eta_{1}^{\bm{\xi}[1]}$, $\ldots$, $\eta_{N}^{\bm{\xi}[1]}$ satisfy the recurrence relation given by
\begin{equation}\label{eqn:eta_recurrence}
	\eta_{i}^{\bm{\xi}[1]} = \xi_{i}^{\left[1\right]} - \widehat{\xi}^{\left[1\right]} \widehat{\xi}^{\left[2\right]} - \widehat{\xi}^{\left[1\right]} \left(1 - \widehat{\xi}^{\left[2\right]}\right) + \sum_{(R^{[1]},\alpha^{[1]})} p^{\circ}_{(R^{[1]},\alpha^{[1]})} \eta_{\tilde{\alpha}^{[1]}(i)}^{\bm{\xi}[1]}.
\end{equation}
By setting $L=1$ in Eq.~\eqref{eqn:weak_fixn_prob} and using Eqs.~\eqref{eqn:beta_ij_defn}, \eqref{eqn:gamma_ij_defn}, and \eqref{eqn:eta_definition}, we obtain
\begin{align}
\label{eqn:fixn_beta_gamma}
	\left.\frac{\text{d}}{\text{d}\delta}\rho_{\bm{A}}^{\left[1\right]}\left(\bm{\xi}\right)\right|_{\delta=0}= 
	\sum_{i,j,k=1}^{N} \pi_{i}^{[1]} m_{k;ji}^{[1]} 
	\left\{ -\left[\sum_{\ell=1}^{N} \left(\beta _{jk}^{\bm{\xi}[1]}-\beta_{ik}^{\bm{\xi}[1]}\right)C_{k\ell} - \left(\beta_{jl}^{\bm{\xi}[ 1]}-\beta_{il}^{\bm{\xi}[ 1]}\right)B_{\ell k}\right] \right. \notag \\
	+ \left. \left[\left(\gamma_{jk}^{\bm{\xi}[1,2]}-\gamma_{ik}^{\bm{\xi}[1,2]}\right)\left(r-1\right)+\left(
\eta_{j}^{\bm{\xi}[1]}-\eta_{i}^{\bm{\xi}[1]}\right)\right]\right\}.
\end{align}
If $\beta_{ij}^{\bm{\xi}\left[1\right]}$ is a solution to Eq.~\eqref{eqn:beta_reccurence_soln}, then $\beta_{ij}^{\bm{\xi}\left[1\right]} + C'$ is also a solution to Eq.~\eqref{eqn:beta_reccurence_soln} for any constant $C'$. Similarly, if $\gamma_{ij}^{\bm{\xi}\left[1,2\right]}$ is a solution to Eq.~\eqref{eqn:gamma_reccurence},  then $\gamma_{ij}^{\bm{\xi}\left[1,2\right]} + C''$ is also a solution to Eq.~\eqref{eqn:gamma_reccurence} for any constant $C''$. The authors of Ref.~\cite{su2022evolution} also pointed out that it does not matter which solution is used because only the differences of $\beta_{ij}^{\bm{\xi}\left[1\right]}$ and $\gamma_{ij}^{\bm{\xi}\left[1,2\right]}$ are used in Eq.~\eqref{eqn:fixn_beta_gamma}. By the fixation axiom in Refs.~\cite{allen2019mathematical,mcavoy2021fixation,su2022evolution}, the space of solutions is two-dimensional. In Ref.~\cite{su2022evolution}, they enforced two additional arbitrary constraints, which altogether ensure that the solution is unique. We impose these conditions, which are
\begin{equation}
\label{eqn:sum_beta_ii=0}
	\sum_{i=1}^{N}\pi _i ^{\left[1\right]}\beta_{ii}^{\bm{\xi}\left[1\right]}=0
\end{equation}
and
\begin{equation}
\label{eqn:sum_gamma_ii=0}
\sum_{i=1}^{N}\pi _i ^{\left[1\right]}\gamma_{ii}^{\bm{\xi}\left[1,2\right]}=0.
\end{equation}
In our case, we have the term $\eta_{j}^{\bm{\xi}\left[1\right]}-\eta_{i}^{\bm{\xi}\left[1\right]}$, which is not present in Ref.~\cite{su2022evolution}. Therefore, we also impose that $\sum_{i=1}^{N}\eta_{i}^{\bm{\xi}\left[1\right]}=0$ to ensure uniqueness of the solution, which is again justified because only the pairwise difference $\eta_{j}^{\bm{\xi}\left[1\right]}-\eta_{i}^{\bm{\xi}\left[1\right]}$ matters.

Equation~\eqref{eqn:defn_fixation_prob} implies that selection favors cooperation in the limit of weak selection when starting from state $\bm{\xi}$ if and only if
\begin{equation}
	\left.\frac{\text{d}}{\text{d}\delta}\rho_{\bm{A}}^{\left[1\right]}\left(\bm{\xi}\right)\right|_{\delta=0} >0.
\label{eq:condition-for-cooperation-original}
\end{equation}
By substituting Eq.~\eqref{eqn:fixn_beta_gamma} in Eq.~\eqref{eq:condition-for-cooperation-original}, we obtain
\begin{align}	
	\sum_{i,j,k=1}^{N} \pi_{i}^{[1]} m_{k,ji}^{[1]} 
	\left\{- \left[\sum_{\ell=1}^{N} \left(\beta_{jk}^{\bm{\xi}[1]} - \beta_{ik}^{\bm{\xi}[1]}\right)C_{k\ell} - \left(\beta_{jl}^{\bm{\xi}[1]} - \beta_{il}^{\bm{\xi}[1]}\right)B_{\ell k}\right] \right. \notag \\
	+ \left. \left[\left(\gamma_{jk}^{\bm{\xi}[1,2]} - \gamma_{ik}^{\bm{\xi}[1,2]}\right)\left(r-1\right) + \left(\eta_{j}^{\bm{\xi}[1]} - \eta_{i}^{\bm{\xi}[1]}\right)\right]\right\} > 0.
\label{eq:condition-for-cooperation}
\end{align}

\newpage
\clearpage

\section{Evolution of the cooperator and mutant under the dB-dB rule\label{sec:application-to-the-death-birth-db-rule}}\label{sec:SI-B}

In this section, we derive the condition under which the cooperator is favored in layer 1 and that under which the mutant is favored in layer 2, assuming the dB rule in both layers, i.e., dB-dB rule. The derivation amounts to calculating individual quantities in Eq.~\eqref{eq:condition-for-cooperation} and simplifying the expressions to facilitate interpretation and numerical computations.

\subsection{Preliminaries}

For the dB rule, we obtain
\begin{equation}
e_{ji}^{\left[1\right]}=\frac{1}{N}\frac{w_{ij}^{[1]}F_i\left(x\right)}{\sum_{\ell=1}^{N}w_{\ell i}F_{\ell}\left(x\right)}.
\label{eq:e-dB}
\end{equation}
By substituting Eq.~\eqref{eq:e-dB} in Eq.~\eqref{eq:def-m}, we obtain
\begin{align}
\label{eqn:marginalderivation}
m_{k;ji}=\left. \frac{\partial e_{ji}}{\partial F_k} \right|_{\mathbf{F}=\bm{1}}
&=	\left. \frac{\partial }{\partial F_k} \frac{1}{N}\frac{w_{ji}F_j\left(x\right)}{\sum_{\ell=1}^{N}w_{\ell i}F_{\ell}\left(x\right)}\right|_{\mathbf{F=\bm{1}}} \notag\\
&= \begin{cases}\left. - \frac{1} {N}w_{ji}F_{j}\left(x\right) \frac{w_{ki}}{\left[\sum_{\ell=1}^{N}w_{\ell i}F_{\ell}\left(x\right)\right]^2} \right|_{\mathbf{F}=\bm{1}}& \text{if }  j\neq k \notag\\
	\left.\frac{1}{N}w_{ki} \frac{\left[\sum_{\ell =1}^{N}w_{\ell i}F_{\ell}\left(x\right)\right] - F_{k}\left(x\right) w_{ki}}{\left[\sum_{\ell=1}^{N}w_{\ell i}F_{\ell}\left(x\right)\right]^2} \right|_{\mathbf{F}=\bm{1}} & \text{ if } j=k
\end{cases} \notag\\
&= \begin{cases} - \frac{1} {N}w_{ji} \frac{w_{ki}}{\left(\sum_{\ell=1}^{N}w_{\ell i}\right)^2} & \text{if }  j\neq k\\
	\frac{1}{N}w_{ki} \frac{\left(\sum_{\ell =1}^{N}w_{\ell i}\right)- w_{ki}}{\left[\sum_{\ell=1}^{N}w_{\ell i}F_{\ell}\left(x\right)\right]^2} & \text{ if } j=k
\end{cases} \notag\\
&= \begin{cases}-\frac{1}{N} p_{ij} p_{ik}& \text{if }  j\neq k\\
	\frac{1}{N}p_{ij}\left(1-p_{ik}\right)& \text{ if } j=k
\end{cases} \notag\\
&=\frac{1}{N}p_{ij}\left(\delta_{j,k}-p_{ik}\right).
\end{align}
We have dropped the superscript $[L]$ for $m_{k;ji}^{[L]}$ and $w_{ij}^{[L]}$ in Eq.~\eqref{eqn:marginalderivation} because the calculations are identical for any $L$.

\subsection{Conditions for the cooperator to be selected}

To compute the condition under which the cooperator is favored, Eq.~\eqref{eq:condition-for-cooperation}, we introduce a short-hand notation
\begin{equation}
\label{eqn:substitution_f_ijk}
f_{ijk} \equiv -\beta_{ij}^{\bm{\xi}\left[1\right]}C_{jk}+\beta_{ik}^{\bm{\xi}\left[1\right]}B_{kj}+\gamma_{ij}^{\bm{\xi}\left[1,2\right]}\left(r-1\right)+\eta_{i}^{\bm{\xi}\left[1\right]}.
\end{equation}
By substituting Eq.~\eqref{eqn:substitution_f_ijk} in Eq.~\eqref{eqn:fixn_beta_gamma}, we obtain

\begin{align}\label{eqn:derivative_rho_cooperation}
	\left.\frac{\text{d}}{\text{d}\delta}\rho_{\bm{A}}^{\left[1\right]}\left(\bm{\xi}\right)\right|_{\delta=0} &= \sum_{i,j,k,\ell=1}^{N} \pi_{i}^{[1]} m_{k;ji}^{[1]} (f_{jkl} - f_{ikl}) \notag\\
	&= \sum_{i,j,k,\ell=1}^{N} \pi_{i}^{[1]} \frac{1}{N} p_{ij}^{[1]} (\delta_{j,k} - p_{ik}^{[1]}) (f_{jkl} - f_{ikl}) \notag\\
	&= \frac{1}{N} \sum_{i,j,k,\ell=1}^{N} \pi_{i}^{[1]} p_{ij}^{[1]} \delta_{j,k} f_{jkl} - \frac{1}{N} \sum_{i,j,k,\ell=1}^{N} \pi_{i}^{[1]} p_{ij}^{[1]} p_{ik}^{[1]} f_{jkl} \notag\\
	& \quad - \frac{1}{N} \sum_{i,j,k,\ell=1}^{N} \pi_{i}^{[1]} p_{ij}^{[1]} \delta_{j,k} f_{ikl} + \frac{1}{N} \sum_{i,j,k,\ell=1}^{N} \pi_{i}^{[1]} p_{ij}^{[1]} p_{ik}^{[1]} f_{ikl} \notag\\
	&= \frac{1}{N} \sum_{i,j,\ell=1}^{N} \pi_{i}^{[1]} p_{ij}^{[1]} f_{jjl} - \frac{1}{N} \sum_{i,j,k,\ell=1}^{N} \pi_{i}^{[1]} p_{ji}^{[1]} p_{ik}^{[1]} f_{jkl} \notag\\
	& \quad - \frac{1}{N} \sum_{i,j,\ell=1}^{N} \pi_{i}^{[1]} p_{ij}^{[1]} f_{ijl} + \frac{1}{N} \sum_{i,k,\ell=1}^{N} \pi_{i}^{[1]} \left( \sum_{j=1}^{N} p_{ij}^{[1]} \right) p_{ik}^{[1]} f_{ikl} \notag\\
	&= \frac{1}{N} \sum_{j,\ell=1}^{N} \pi_{j}^{[1]} \left( \sum_{i=1}^{N} p_{ji}^{[1]} \right) f_{jjl} - \frac{1}{N} \sum_{j,k,\ell=1}^{N} \pi_{j}^{[1]} \left( p^{[1]} \right)_{jk}^{(2)} f_{jkl} \notag\\
	& \quad \cancel{- \frac{1}{N} \sum_{i,j,\ell=1}^{N} \pi_{i}^{[1]} p_{ij}^{[1]} f_{ijl}} + \cancel{\frac{1}{N} \sum_{i,k,\ell=1}^{N} \pi_{i}^{[1]} p_{ik}^{[1]} f_{ikl}} \notag\\
%
%
	&= \frac{1}{N} \sum_{j,\ell=1}^{N} \pi_{j}^{[1]} f_{jjl} - \frac{1}{N} \sum_{i,j,\ell=1}^{N} \pi_{i}^{[1]} \left( p^{[1]} \right)_{ij}^{(2)} f_{ijl} \notag\\
	&= \frac{1}{N} \sum_{i,\ell=1}^{N} \pi_{i}^{[1]} \left[ -\beta_{ii}^{\bm{\xi}[1]} C_{il} + \beta_{il}^{\bm{\xi}[1]} B_{li} + \gamma_{ii}^{\bm{\xi}[1,2]} (r-1) + \eta_{i}^{\bm{\xi}[1]} \right] \notag\\
	& \quad - \frac{1}{N} \sum_{i,j,\ell=1}^{N} \pi_{i}^{[1]} \left( p^{[1]} \right)_{ij}^{(2)} \left[ -\beta_{ij}^{\bm{\xi}[1]} C_{jl} + \beta_{il}^{\bm{\xi}[1]} B_{lj} + \gamma_{ij}^{\bm{\xi}[1,2]} (r-1) + \eta_{i}^{\bm{\xi}[1]} \right],
\end{align}
where
$\left( p^{\left[1\right]} \right)_{ij}^{\left(n\right)}$ is the probability that the random walker in layer 1 moves from replica node $i$ to replica node $j$ in $n$ steps. 
To derive the first equality in Eq.~\eqref{eqn:derivative_rho_cooperation}, we used Eq.~\eqref{eqn:marginalderivation}.
To derive the third equality in Eq.~\eqref{eqn:derivative_rho_cooperation}, we used $\pi_{i}^{\left[1\right]}p_{ij}^{\left[1\right]}=\pi_{j}^{\left[1\right]}p_{ji}^{\left[1\right]}$, $\forall$ $i$, $j$.
Equation~\eqref{eqn:derivative_rho_cooperation} implies that $\left.\frac{\text{d}}{\text{d}\delta}\rho_{\bm{A}}^{\left[1\right]}\left(\bm{\xi}\right)\right|_{\delta=0} > 0$ holds true if and only if
\begin{align}
	&\sum_{i=1}^{N} \pi_{i}^{\left[1\right]} \sum_{\ell=1}^{N} \left[
	-\beta_{ii}^{\bm{\xi}\left[1\right]} C_{il} + \beta_{il}^{\bm{\xi}\left[1\right]} B_{li} 
	+ \gamma_{ii}^{\bm{\xi}\left[1,2\right]} \left( r-1 \right) + \eta_{i}^{\bm{\xi}\left[1\right]}
	\right] 
	> \nonumber \\
	&\sum_{i,j=1}^{N} \pi_{i}^{\left[1\right]} \left( p^{\left[1\right]} \right)_{ij}^{\left(2\right)} \sum_{\ell=1}^{N} \left[
	-\beta_{ij}^{\bm{\xi}\left[1\right]} C_{jl} + \beta_{il}^{\bm{\xi}\left[1\right]} B_{lj} 
	+ \gamma_{ij}^{\bm{\xi}\left[1,2\right]} \left( r-1 \right) + \eta_{i}^{\bm{\xi}\left[1\right]}
	\right].
\label{eq:coop-dB-a}	
\end{align}
By substituting $C_{kl}=c\cdot p_{kl}^{\left[1\right]}$ and $B_{lk}=b\cdot p_{lk} ^{\left[1\right]}$ in Eq.~\eqref{eq:coop-dB-a}, we obtain
\begin{align}
\label{eqn:cooperation_condition}
	&\sum_{i=1}^{N} \pi_{i}^{\left[1\right]} \sum_{\ell=1}^{N}\left[
	-\beta_{ii}^{\bm{\xi}\left[1\right]} c \cdot p_{il}^{\left[1\right]} 
	+ \beta_{il}^{\bm{\xi}\left[1\right]} b \cdot p_{il}^{\left[1\right]} 
	+ \gamma_{ii}^{\bm{\xi}\left[1,2\right]} \left( r-1 \right) 
	+ \eta_{i}^{\bm{\xi}\left[1\right]} 
	\right]
	> \nonumber \\  
	&\sum_{i,j=1}^{N} \pi_{i}^{\left[1\right]} \left( p^{\left[1\right]} \right)_{ij}^{\left(2\right)} 
	\sum_{\ell=1}^{N} \left[
	-\beta_{ij}^{\bm{\xi}\left[1\right]} c \cdot p_{jl}^{\left[1\right]} 
	+ \beta_{il}^{\bm{\xi}\left[1\right]} b \cdot p_{jl}^{\left[1\right]} 
	+ \gamma_{ij}^{\bm{\xi}\left[1,2\right]} \left( r-1 \right) 
	+ \eta_{i}^{\bm{\xi}\left[1\right]}
	\right].
\end{align}
To simplify Eq.~\eqref{eqn:cooperation_condition}, we define
\begin{align}
\theta_{n}^{\bm{\xi}\left[1\right]} :=& \sum_{i,j=1}^{N} \pi_{i}^{\left[1\right]} \left( p^{\left[1\right]} \right)_{ij}^{\left(n\right)} \beta_{ij}^{\bm{\xi}\left[1\right]}
\label{eq:def-theta}
\end{align}
and
\begin{align}
\phi_{n,m}^{\bm{\xi}\left[1,2\right]} :=& \sum_{i,j=1}^{N} \pi_{i}^{\left[1\right]} \left( p^{\left[1,2\right]} \right)_{ij}^{\left(n,m\right)} \gamma_{ij}^{\bm{\xi}\left[1,2\right]},
\label{eq:def-phi}
\end{align}
where $\left( p^{\left[1,2\right]} \right)_{ij}^{\left(n,m\right)}$ is
the probability that the random walker moves from replica node $i$ to replica node $j$ when
the first $n$ steps of the random walk occur in layer 1 and the subsequent $m$ steps occur in layer 2.
By substituting Eqs.~\eqref{eq:def-theta} and \eqref{eq:def-phi} in Eq.~\eqref{eqn:cooperation_condition} and using the fact that
\begin{equation}
\sum_{i,j=1}^{N} \pi_{i}^{\left[1\right]} \left( p^{\left[1\right]} \right)_{ij}^{\left(n\right)} \eta_{i}^{\bm{\xi}\left[1\right]}
%
= \sum_{i=1}^{N} \pi_{i}^{\left[1\right]} \eta_{i}^{\bm{\xi}\left[1\right]}\sum_{j=1}^{N} \left( p^{\left[1\right]} \right)_{ij}^{\left(n\right)} = \sum_{i=1}^{N} \pi_{i}^{\left[1\right]} \eta_{i}^{\bm{\xi}\left[1\right]}=0
\label{eq:pi-eta-sum=0}
\end{equation}
for any $n$, we obtain
\begin{equation}
\label{eqn:cooperation_criteria_reduced-SI}
	c\theta_{2}^{\bm{\xi}\left[1\right]} + b\left(\theta_{1}^{\bm{\xi}\left[1\right]} - \theta_{3}^{\bm{\xi}\left[1\right]}\right) - \left(r-1\right) \phi_{2,0}^{\bm{\xi}\left[1,2\right]} > 0,
\end{equation}
which is
Eq.~\eqref{eqn:cooperation_criteria_reduced}
in the main text. It should be noted that $\eta_{i}^{\bm{\xi}\left[1\right]}$ in Eq.~\eqref{eqn:cooperation_condition} has been canceled out.

Next, we solve the recurrence equations Eqs.~\eqref{eqn:beta_reccurence_soln}, \eqref{eqn:gamma_reccurence}, and \eqref{eqn:eta_recurrence} to determine 
$\theta_{n}^{\bm{\xi}\left[1\right]}$, $n \in \{1, 2, 3\}$ and $\phi_{n,m}^{\bm{\xi}\left[1,2\right]}$ through Eq.~\eqref{eq:def-theta} and \eqref{eq:def-phi}, respectively. The rest of the derivation is identical to section 2.1.1 of the ESM for Ref.~\cite{su2022evolution} because these recurrences were derived in Ref.~\cite{su2022evolution} for any dynamics using the dB rule, except Eq.~\eqref{eqn:eta_recurrence_dB}. Therefore, we omit some details. Note that Eq.~\eqref{eqn:eta_recurrence_dB} contains the recurrence for the $\eta$ term, which is absent in their case.
For $i=j$, Eq.~\eqref{eqn:beta_reccurence_soln} reduces to
\begin{equation}
	\beta_{ii}^{\bm{\xi}\left[1\right]} = \xi_{i}^{\left[1\right]} - \widehat{\xi}^{\left[1\right]} + \frac{1}{N} \sum_{k=1}^{N} p_{ik}^{\left[1\right]} \beta_{kk}^{\bm{\xi}\left[1\right]} + \left(1 - \frac{1}{N}\right) \beta_{ii}^{\bm{\xi}\left[1\right]},
\end{equation}
which gives
\begin{equation}\label{eqn:beta_recurrence_ii_dB}
	\beta_{ii}^{\bm{\xi}\left[1\right]} = N \left(\xi_{i}^{\left[1\right]} - \widehat{\xi}^{\left[1\right]}\right) + \sum_{k=1}^{N} p_{ik}^{\left[1\right]} \beta_{kk}^{\bm{\xi}\left[1\right]}.
\end{equation}
For $i \ne j$, we obtain
\begin{equation}
	\beta_{ij}^{\bm{\xi}\left[1\right]} = \xi_{i}^{\left[1\right]} \xi_{j}^{\left[1\right]} - \widehat{\xi}^{\left[1\right]} + \frac{1}{N} \sum_{k=1}^{N} p_{ik}^{\left[1\right]} \beta_{kj}^{\bm{\xi}\left[1\right]} + \frac{1}{N} \sum_{k=1}^{N} p_{jk}^{\left[1\right]} \beta_{ik}^{\bm{\xi}\left[1\right]} + \left(1 - \frac{2}{N}\right) \beta_{ij}^{\bm{\xi}\left[1\right]},
\end{equation}
which gives
\begin{equation}
\label{eqn:beta_recurrence_ij_dB}
	\beta_{ij}^{\bm{\xi}\left[1\right]} = \frac{N}{2} \left(\xi_{i}^{\left[1\right]} \xi_{j}^{\left[1\right]} - \widehat{\xi}^{\left[1\right]}\right) + \frac{1}{2} \sum_{k=1}^{N} p_{ik}^{\left[1\right]} \beta_{kj}^{\bm{\xi}\left[1\right]} + \frac{1}{2} \sum_{k=1}^{N} p_{jk}^{\left[1\right]} \beta_{ik}^{\bm{\xi}\left[1\right]}.
\end{equation}
For $i, j \in \{1, \ldots, N\}$, Eq.~\eqref{eqn:gamma_reccurence} reduces to
\begin{align}
\gamma_{ij}^{\bm{\xi}\left[1,2\right]} = & \ \xi_{i}^{\left[1\right]} \xi_{j}^{\left[2\right]} - \widehat{\xi}^{\left[1\right]} \widehat{\xi}^{\left[2\right]} + \frac{1}{N^2} \sum_{k_1,k_2=1}^{N} p_{ik_1}^{\left[1\right]} p_{jk_2}^{\left[2\right]} \gamma_{k_1k_2}^{\bm{\xi}\left[1,2\right]} \notag \\
& + \frac{1}{N} \left(1 - \frac{1}{N}\right) \sum_{k_1=1}^{N} p_{ik_1}^{\left[1\right]} \gamma_{k_1j}^{\bm{\xi}\left[1,2\right]} + \frac{1}{N} \left(1 - \frac{1}{N}\right) \sum_{k_2=1}^{N} p_{jk_2}^{\left[2\right]} \gamma_{ik_2}^{\bm{\xi}\left[1,2\right]} \notag \\
& + \left(1 - \frac{1}{N}\right)^2 \gamma_{ij}^{\bm{\xi}\left[1,2\right]},
\end{align}
which gives
\begin{align}
\label{eqn:gamma_recurrence_ij_dB}
\gamma_{ij}^{\bm{\xi}\left[1,2\right]} = & \ \frac{N^2}{2N-1} \left( \xi_{i}^{\left[1\right]} \xi_{j}^{\left[2\right]} - \widehat{\xi}^{\left[1\right]} \widehat{\xi}^{\left[2\right]} \right) + \frac{1}{2N-1} \sum_{k_1,k_2=1}^{N} p_{ik_1}^{\left[1\right]} p_{jk_2}^{\left[2\right]} \gamma_{k_1k_2}^{\bm{\xi}\left[1,2\right]} \notag \\
& + \frac{N-1}{2N-1} \sum_{k_1=1}^{N} p_{ik_1}^{\left[1\right]} \gamma_{k_1j}^{\bm{\xi}\left[1,2\right]} + \frac{N-1}{2N-1} \sum_{k_2=1}^{N} p_{jk_2}^{\left[2\right]} \gamma_{ik_2}^{\bm{\xi}\left[1,2\right]}.
\end{align}
For $i \in \{1, \ldots, N\}$, Eq.~\eqref{eqn:eta_recurrence} reduces to
\begin{equation}\label{eqn:eta_recurrence_dB}
	\eta_{i}^{\bm{\xi}[1]} = \xi_{i}^{\left[1\right]} - \widehat{\xi}^{\left[1\right]} \widehat{\xi}^{\left[2\right]} - \widehat{\xi}^{\left[1\right]} \left(1 - \widehat{\xi}^{\left[2\right]}\right) + \frac{1}{N} \sum_{k=1}^{N} p_{ik}^{\left[1\right]} \eta_{k}^{\bm{\xi}\left[1\right]} + \left(1 - \frac{1}{N}\right) \eta_{i}^{\bm{\xi}\left[1\right]},
\end{equation}
which gives
\begin{equation}\label{eqn:eta_recurrence_dB_reduced}
	\eta_{i}^{\bm{\xi}\left[1\right]} = N \left[\xi_{i}^{\left[1\right]} - \widehat{\xi}^{\left[1\right]} \widehat{\xi}^{\left[2\right]} - \widehat{\xi}^{\left[1\right]} \left(1 - \widehat{\xi}^{\left[2\right]}\right)\right] + \sum_{k=1}^{N} p_{ik}^{\left[1\right]} \eta_{k}^{\bm{\xi}\left[1\right]}.
\end{equation}
By solving the recurrences given by Eqs.~\eqref{eqn:beta_recurrence_ii_dB}, \eqref{eqn:beta_recurrence_ij_dB},  \eqref{eqn:gamma_recurrence_ij_dB}, and \eqref{eqn:eta_recurrence_dB_reduced}, and substituting the solutions in Eqs.~\eqref{eq:def-theta} and \eqref{eq:def-phi} and the obtained $\theta_{n}^{\bm{\xi}\left[1\right]}$ and $\phi_{n,m}^{\bm{\xi}\left[1,2\right]}$ in Eq.~\eqref{eqn:cooperation_criteria_reduced-SI}, we obtain the conditions under which selection favors cooperation.

\subsection{Conditions for the mutant to be selected}\label{sec:conditions-for-selection-in-the-constant-selection-layer}
 
To avoid complication of the equations, in this section, we swap the two layers such that constant selection occurs in layer $1$ and the donation game occurs in layer $2$. Then, Eq.~\eqref{eqn:fixn_beta_gamma} becomes
\begin{align}\label{eqn:fixn_beta_gamma_flipped}
	\left.\frac{\text{d}}{\text{d}\delta}\rho_{\bm{A}}^{\left[1\right]}\left(\bm{\xi}\right)\right|_{\delta=0} = 
	\sum_{i,j,k=1}^{N} \frac{1}{N} \pi_{i}^{[1]} m_{k;ji}^{[1]} 
	\left\{
	\left[\left(\beta_{jk}^{\bm{\xi}[1]} - \beta_{ik}^{\bm{\xi}[1]}\right)\left(r-1\right) + \left(\eta_{j}^{\bm{\xi}[1]} - \eta_{i}^{\bm{\xi}[1]}\right)\right] \right. \notag \\
	\left. + \sum_{\ell=1}^{N} \left[-\left(\gamma_{jk}^{\bm{\xi}[1,2]} - \gamma_{ik}^{\bm{\xi}[1,2]}\right)C_{k\ell} + \left(\gamma_{j\ell}^{\bm{\xi}[1,2]} - \gamma_{i\ell}^{\bm{\xi}[1,2]}\right)B_{\ell k}\right]
	\right\}.
\end{align}
Similarly, Eq.~\eqref{eqn:cooperation_condition} becomes
\begin{align}
	&\sum_{i=1}^{N} \pi_{i}^{\left[1\right]} \sum_{\ell=1}^{N} \left[
		\beta_{ii}^{\bm{\xi}\left[1\right]} \left(r-1\right) + \eta_{i}^{\bm{\xi}\left[1\right]} 
		- \gamma_{ii}^{\bm{\xi}\left[1,2\right]} c p_{k\ell}^{\left[2\right]} + \gamma_{i\ell}^{\bm{\xi}\left[1,2\right]} b p_{i\ell}^{\left[2\right]} 
	\right] > \nonumber \\
	&\sum_{i,j=1}^{N} \pi_{i}^{\left[1\right]} \left( p^{\left[1\right]} \right)_{ij}^{\left(2\right)} \sum_{\ell=1}^{N} \left[
		\beta_{ij}^{\bm{\xi}\left[1\right]} \left(r-1\right) + \eta_{i}^{\bm{\xi}\left[1\right]}
		- \gamma_{ij}^{\bm{\xi}\left[1,2\right]} c p_{j\ell}^{\left[2\right]} +  \gamma_{i\ell}^{\bm{\xi}\left[1,2\right]} b p_{j\ell}^{\left[2\right]} 
	\right].
\label{eq:mutant-selected-1}
\end{align}
By carrying out calculations similar to those in~Ref.~\cite{su2022evolution}, we find that Equation~\eqref{eq:mutant-selected-1} simplifies to
\begin{equation}
\label{eqn:constant_drift_selection_criteria-SI}
	-\left(r-1\right)\theta_{2}^{\bm{\xi}[1]} + c \phi_{2,0}^{\bm{\xi}\left[1,2\right]} +
	b \left(\phi_{0,1}^{\bm{\xi}\left[1,2\right]} - \phi_{2,1}^{\bm{\xi}\left[1,2\right]}\right) > 0.
\end{equation}
Equation~\eqref{eqn:constant_drift_selection_criteria-SI} is equivalent to
Eq.~\eqref{eqn:constant_drift_selection_criteria}
%
%
in the main text; note that Eq.~\eqref{eqn:constant_drift_selection_criteria} is shown in terms of the original layer convention (i.e., the donation game occurs in layer 1, and the constant-selection dynamics occurs in layer 2).

\newpage
\clearpage

\section{Evolution of cooperator and mutant under the dB-Bd rule}\label{sec:SI-C}

In this section, we derive the conditions for the cooperation and mutant type to be favored when updating occurs according to the
death-Birth rule in the game layer (i.e., layer 1) and the Birth-death rule in constant-selection layer (i.e., layer 2), i.e., under the dB-Bd rule.

\subsection{Preliminaries}

For the Bd rule, we obtain
\begin{equation}
	e_{ji}^{\left[2\right]}=\frac{w_{ji}}{\sum_{\ell =1}^{N}w_{j\ell}^{[2]}}\cdot\frac{F_j\left(x\right)}{\sum_{\ell=1}^{N}F_{\ell}\left(x\right)}
	\label{eq:e-Bd}
\end{equation}
for layer 2. By substituting Eq.~\eqref{eq:e-Bd} in Eq.~\eqref{eq:def-m}, we obtain
\begin{align}
	\label{eqn:marginalderivation_Bd}
	m_{k;ji}^{\left[2\right]}=\left. \frac{\partial e_{ji}}{\partial F_k} \right|_{\mathbf{F}=\bm{1}}
	&= \frac{w_{ji}^{[2]}}{\sum_{\ell=1}^{N}w_{j\ell}^{[2]}}	\left. \frac{\partial }{\partial F_k}\frac{F_j\left(x\right)}{\sum_{\ell=1}^{N}F_{\ell}\left(x\right)}\right|_{\mathbf{F=\bm{1}}} \notag\\
	&= \begin{cases}\left.p_{ji}^{[2]}F_{j}\left(x\right) \frac{-1}{\left[\sum_{\ell=1}^{N}F_{\ell}\left(x\right)\right]^2} \right|_{\mathbf{F}=\bm{1}}& \text{if }  j\neq k \notag\\
		\left.p_{ji}^{[2]} \frac{\left[\sum_{\ell =1}^{N}F_{\ell}\left(x\right)\right] - F_{k}\left(x\right)}{\left[\sum_{\ell=1}^{N}w_{\ell i}F_{\ell}\left(x\right)\right]^2} \right|_{\mathbf{F}=\bm{1}} & \text{ if } j=k
	\end{cases} \notag\\
	&= \begin{cases} - \frac{1} {N} p_{ji}^{[2]}& \text{if }  j\neq k\\
		\frac{N-1} {N^2} p_{ji}^{[2]} & \text{ if } j=k
	\end{cases} \notag\\
	&= \frac{1}{N} \left( \delta_{j,k} - \frac{1}{N} \right) p_{ji}^{[2]}.
\end{align}

\subsection{Conditions for the cooperator to be selected}\label{sec:conditions-for-cooperation-dB-Bd}

Because layer 1 uses the dB rule, the condition under which the cooperator is selected remains in the same form, i.e., Eq.~\eqref{eqn:cooperation_criteria_reduced-SI}. Furthermore, $\theta_{n}^{\bm{\xi}\left[1\right]}$, $n \in \{ 1, 2, 3 \}$ remains the same as that under the dB-dB rule. This is because, in Eq.~\eqref{eq:def-theta}, $\left( p^{\left[1\right]} \right)_{ij}^{\left(n\right)}$ only depends on the network structure in layer 1,
$\pi_{i}^{\left[1\right]}$ depends on the network structure in layer 1 and the updating rule used in layer 1, and $\beta_{ij}^{\bm{\xi}[1]}$ only depends on the network structure and updating rule in layer 1. For obtaining $\beta_{ij}^{\bm{\xi}[1]}$, the recurrence equations to be used are Eqs.~\eqref{eqn:beta_recurrence_ii_dB} and \eqref{eqn:beta_recurrence_ij_dB}.

In contrast, the value of $\phi_{2,0}^{\bm{\xi}\left[1,2\right]}$ changes from the case of the dB-dB rule because
$\gamma_{ij}^{\bm{\xi}\left[1,2\right]}$ in Eq.~\eqref{eq:def-phi} depends on the updating rule in layer 2 as well as the network structure of both layers. It should be noted that $\left( p^{\left[1,2\right]} \right)_{ij}^{\left(n,m\right)}$ still only depends on the network structure in both layers and hence is not influenced by the change in the updating rule in layer 2. We derive $\gamma_{ij}^{\bm{\xi}\left[1,2\right]}$ in section~\ref{sec:conditions-for-selection-in-the-constant-selection-layer-dB-Bd} (see the self-consistent equation~\eqref{eqn:gamma21_dB-Bd} for $\gamma_{ij}^{\bm{\xi}[1,2]}$; note that $\gamma_{ij}^{\bm{\xi}[1,2]}=\gamma_{ij}^{\bm{\xi}[2,1]}$).

\subsection{Conditions for the mutant to be selected}\label{sec:conditions-for-selection-in-the-constant-selection-layer-dB-Bd}

Similarly to Eq.~\eqref{eqn:substitution_f_ijk}, in terms of the original layer-label convention (i.e., the donation game occurs in layer 1, and the constant-selection dynamics occurs in layer 2), we set
\begin{equation}
	f_{ijk} \equiv -\gamma_{ij}^{\bm{\xi}\left[2,1\right]}C_{jk}+\gamma_{ik}^{\bm{\xi}\left[2,1\right]}B_{kj}+\beta_{ij}^{\bm{\xi}\left[2\right]}\left(r-1\right)+\eta_{i}^{\bm{\xi}\left[2\right]}.
\end{equation}
Note that the reproductive value under the Bd rule, which is assumed for layer 2 here, is given by~\cite{sood2008voter,masuda2009evolutionary, mcavoy2020social}
\begin{equation}
	\pi_i^{[2]} = \frac{\left( s_i^{[2]} \right)^{-1} }{ \sum_{\ell = 1}^N \left( s_{\ell}^{[2]} \right)^{-1}}.
\end{equation}

Therefore, we obtain
\begin{align}\label{eqn:dB-Bd-constant-selection-layer}
	\left. \frac{\text{d}}{\text{d}\delta} \rho_{\bm{A}}^{[2]} (\bm{\xi}) \right|_{\delta=0} =& \sum_{i,j,k,\ell=1}^N \pi_i^{[2]} m_{k;ji}^{[2]} \left( f_{jk\ell} - f_{ik\ell} \right)\notag\\
	= &\frac{1}{N} \sum_{i,j=1}^{N} \pi_i^{[2]} p_{ji}^{[2]} \sum_{\ell=1}^{N} \left[-\gamma_{ii}^{\bm{\xi}\left[2,1\right]}C_{i\ell}+\gamma_{i\ell}^{\bm{\xi}\left[2,1\right]}B_{\ell i}+\beta_{ii}^{\bm{\xi}\left[2\right]}\left(r-1\right)+\eta_{i}^{\bm{\xi}\left[2\right]} \right] \notag\\
	& - \frac{1}{N} \sum_{i,j=1}^{N} \pi_i^{[2]} p_{ji}^{[2]} \sum_{\ell=1}^{N} \left[-\gamma_{ik}^{\bm{\xi}\left[2,1\right]}C_{k\ell}+\gamma_{i\ell}^{\bm{\xi}\left[2,1\right]}B_{\ell j}+\beta_{ik}^{\bm{\xi}\left[2\right]}\left(r-1\right)+\eta_{i}^{\bm{\xi}\left[2\right]} \right]. 
\end{align}
For detailed derivation of Eq.~\eqref{eqn:dB-Bd-constant-selection-layer}, see Eq.~(SI.60) in Ref.~\cite{su2022evolution}. Equation~\eqref{eqn:dB-Bd-constant-selection-layer} implies that
$ \left. \frac{\text{d}}{\text{d}\delta} \rho_{\bm{A}}^{[2]} (\bm{\xi}) \right|_{\delta=0} > 0$ if and only if
\begin{align}\label{eqn:dB-Bd-selection-criteria-reduced}
	&\sum_{i,j=1}^{N} \pi_i^{[2]} p_{ji}^{[2]} \sum_{\ell=1}^{N}
	\left[-\gamma_{ii}^{\bm{\xi}\left[2,1\right]} cp_{i\ell}^{[1]} + \gamma_{i\ell}^{\bm{\xi}\left[2,1\right]}b p_{i\ell}^{[1]}
	+ \beta_{ii}^{\bm{\xi}\left[2\right]}\left(r-1\right) + \eta_{i}^{\bm{\xi}\left[2\right]} \right] > \nonumber \\
	&\sum_{i,j=1}^{N} \pi_i^{[2]} p_{ji}^{[2]} \sum_{\ell=1}^{N} \left[
	-\gamma_{ij}^{\bm{\xi}\left[2,1\right]} cp_{j\ell}^{[1]} + \gamma_{i\ell}^{\bm{\xi}\left[2,1\right]} bp_{j\ell}^{[1]}
	+ \beta_{ij}^{\bm{\xi}\left[2\right]}\left(r-1\right) + \eta_{i}^{\bm{\xi}\left[2\right]} \right].
\end{align}
In Eq.~\eqref{eqn:dB-Bd-selection-criteria-reduced}, the $\eta_{i}^{\bm{\xi}\left[2\right]}$  terms cancel out, leading to
\begin{align}\label{eqn:dB-Bd-selection-criteria-reduced-2}
	&\sum_{i,j=1}^{N} \pi_i^{[2]} p_{ji}^{[2]} \sum_{\ell=1}^{N}
	\left[-\gamma_{ii}^{\bm{\xi}\left[2,1\right]} cp_{i\ell}^{[1]} + \gamma_{i\ell}^{\bm{\xi}\left[2,1\right]}b p_{i\ell}^{[1]}
	+ \beta_{ii}^{\bm{\xi}\left[2\right]}\left(r-1\right)  \right] > \nonumber \\
	&\sum_{i,j=1}^{N} \pi_i^{[2]} p_{ji}^{[2]} \sum_{\ell=1}^{N} \left[
	-\gamma_{ij}^{\bm{\xi}\left[2,1\right]} cp_{j\ell}^{[1]} + \gamma_{i\ell}^{\bm{\xi}\left[2,1\right]} bp_{j\ell}^{[1]}
	+ \beta_{ij}^{\bm{\xi}\left[2\right]}\left(r-1\right) \right].
\end{align}

To compute Eq.~\eqref{eqn:dB-Bd-selection-criteria-reduced-2}, we follow Ref.~\cite{su2022evolution} to obtain $\beta_{ij}^{\bm{\xi}[2]}$ and $\gamma_{ij}^{\bm{\xi}[21]}$ as solutions of recursive equations as follows.

For $i = j$, we obtain
\begin{align}
	\beta_{ii}^{\bm{\xi}[2]} &= \xi_i^{[2]} - \widehat{\xi}^{[2]} + \sum_{\left(R^{[1]},\alpha^{[1]}\right)} p_{\left(R^{[2]},\alpha^{[2]}\right)}^{\circ} \beta_{\widetilde{\alpha}^{[2]}\left(i\right),\widetilde{\alpha}^{[2]}\left(i\right)}^{\bm{\xi}[2]} \notag \\
	&= \xi_i^{[2]} - \widehat{\xi}^{[2]} + \frac{1}{N} \sum_{k=1}^{N} p_{ki}^{[2]} \beta_{kk}^{\bm{\xi}[2]} + \left( 1 - \frac{1}{N} \sum_{k=1}^{N} p_{ki}^{[2]} \right) \beta_{ii}^{\bm{\xi}[2]}, 
\end{align}
which gives 
\begin{equation}\label{eqn:beta_ii_recurrence_Bd_reduced}
	\beta_{ii}^{\bm{\xi}[2]} = \frac{N \left( \xi_i^{[2]} - \widehat{\xi}_i^{[2]} \right) + \sum_{k=1}^{N} p_{ki}^{[2]} \beta_{kk}^{\bm{\xi}[2]}}{\sum_{k=1}^{N} p_{ki}^{[2]}}. 
\end{equation}
For $i \neq j$, we obtain
\begin{equation}\label{eqn:beta_ij_recurrence_Bd_reduced}
	\beta_{ij}^{\bm{\xi}[2]} = \frac{N \left( \xi_i^{[2]} \xi_j^{[2]} - \widehat{\xi}_i^{[2]}  \right) + \sum_{k=1}^{N} p_{ki}^{[2]} \beta_{kj}^{\bm{\xi}[2]} + \sum_{k=1}^{N} p_{kj}^{[2]} \beta_{ik}^{\bm{\xi}[2]}}{\sum_{k=1}^{N} p_{ki}^{[2]} + \sum_{k=1}^{N} p_{kj}^{[2]}}. 
\end{equation}
Finally, we obtain
\begin{align}\label{eqn:gamma_recurrence_Bd_reduced}
	\gamma_{ij}^{\bm{\xi}[2,1]} &= \xi_i^{[1]} \xi_j^{[2]} -\widehat{\xi}^{[1]} \widehat{\xi}^{[2]} + \sum_{\substack{(R^{[1]},\alpha^{[1]}), \\ (R^{[2]},\alpha^{[2]})}}
	p_{\left(R^{[1]},\alpha^{[1]}\right)}^{\bm{\xi}[1]} p_{\left(R^{[2]},\alpha^{[2]}\right)}^{\bm{\xi}[2]} \gamma_{\widetilde{\alpha}^{[1]}(i),\widetilde{\alpha}^{[2]}(j)}^{\bm{\xi}[2,1]} \notag \\
	&= \xi_i^{[1]} \xi_j^{[2]} - \widehat{\xi}^{[1]} \widehat{\xi}^{[2]} + \frac{1}{N^2} \sum_{k_1,k_2=1}^{N} p_{ik_1 }^{[1]} p_{k_2 j}^{[2]} \gamma_{k_{1}k_{2}}^{\bm{\xi}[2,1]} \notag \\
	&\quad+ \frac{1}{N}\left( 1 - \frac{1}{N} \right) \sum_{k_1=1}^{N} p_{ik_1 }^{[1]} \gamma_{k_{1}j}^{\bm{\xi}[2,1]} \notag \\
	&\quad+ \frac{1}{N}\left( 1 - \frac{1}{N} \sum_{k_2=1}^{N} p_{k_2 j}^{[2]} \right) \sum_{k_2=1}^{N} p_{k_2 j}^{[2]} \gamma_{ik_{2}}^{\bm{\xi}[2,1]} \notag \\
	&\quad + \left( 1 - \frac{1}{N} \right) \left( 1 - \frac{1}{N} \sum_{k_2=1}^{N} p_{k_2 j}^{[2]} \right) \gamma_{ij}^{\bm{\xi}[2,1]},
\end{align}
which gives
\begin{equation}\label{eqn:gamma21_dB-Bd}
	\gamma_{ij}^{\xi[21]} =
	\frac{
		\scalebox{0.8}{$
		N^2 \left( \xi_i^{[1]} \xi_j^{[2]} - \xi_i^{[1]} \xi_j^{[2]} \right) 
		+ \sum_{k_1,k_2=1}^{N} p_{ik_1}^{[1]} p_{k_2 j}^{[2]} \gamma_{k_{1}k_{2}}^{\bm{\xi}[2,1]} 
		+ \left( N - \sum_{k_2=1}^{N} p_{k_2 j}^{[2]} \right) \sum_{k_1=1}^{N} p_{ik_1}^{[1]} \gamma_{k_{1}j}^{\bm{\xi}[2,1]} 
		+ \left( N - 1 \right) \sum_{k_2=1}^{N} p_{k_2 j}^{[2]} \gamma_{ik_{2}}^{\bm{\xi}[2,1]}
		$}
	}{
	N + N \sum_{k_2=1}^{N} p_{k_2 j}^{[2]} - \sum_{k_2=1}^{N} p_{k_2 j}^{[2]}
	}.
\end{equation}

\newpage
\clearpage

\section{\texorpdfstring{Interpretation of Eqs.~\eqref{eqn:cooperation_criteria_reduced} and \eqref{eqn:constant_drift_selection_criteria}}{Interpretation of Eqs. (3) and (4)}}\label{sec:SI-D}

Our main theoretical results, Eqs.~\eqref{eqn:cooperation_criteria_reduced} and \eqref{eqn:constant_drift_selection_criteria} in the main text, which are Eqs.~\eqref{eqn:cooperation_criteria_reduced-SI} and \eqref{eqn:constant_drift_selection_criteria-SI}, are an outcome of detailed mathematical analysis (see 
Section~\ref{sec:application-to-the-death-birth-db-rule}). In this section, we provide a more intuitive explanation of Eqs.~\eqref{eqn:cooperation_criteria_reduced} and \eqref{eqn:constant_drift_selection_criteria}. A similar explanation is provided in
Ref.~\cite{su2022evolution} for two-layer networks in both of which the donation game is played (Fig.\,2 and the associated text in Ref.~\cite{su2022evolution}).
 
Following Ref.~\cite{su2022evolution}, consider an individual $i$ that is chosen to update the strategy in layer 1. Cooperation is favored in layer 1 if a cooperating neighbor of $i$, denoted by $j$, has a larger expected payoff than a uniformly randomly chosen neighbor of $i$ in layer 1, denoted by $\ell$. Individual $j$ receives an average benefit $b\theta_{1}^{\bm{\xi}\left[1\right]}$ from its neighbor, denoted by $k$, in layer 1. Note that $\theta_{1}^{\bm{\xi}\left[1\right]}$ is the probability that $k$ is a cooperator in layer 1.
Individual $j$ pays cost $c\theta_{0}^{\bm{\xi}\left[1\right]}$ because it is a cooperator in layer 1. The payoff to $j$ from layer 2 is equal to $r \phi_{0,0}^{\bm{\xi}\left[1,2\right]} + \left( 1-\phi_{0,0}^{\bm{\xi}\left[1,2\right]} \right)$. Note that
$\phi_{0,0}^{\bm{\xi}\left[1,2\right]}$ is the probability that $j$'s strategy is the same between layers 1 and 2; in our analysis, this means $j$ is a mutant in layer 2 given that $j$ is a cooperator in layer 1. Therefore, the total payoff of $j$ is equal to 
$b \theta_{1}^{\bm{\xi}\left[1\right]} - c\theta_{0}^{\bm{\xi}\left[1\right]}
+ r \phi_{0,0}^{\bm{\xi}\left[1,2\right]} + \left( 1-\phi_{0,0}^{\bm{\xi}\left[1,2\right]} \right)$.

Similarly, individual $\ell$ receives an average benefit
$b\theta_{3}^{\bm{\xi}\left[1\right]}$ from its neighbor, denoted by $k'$, in layer 1. Note that $\theta_{3}^{\bm{\xi}\left[1\right]}$ is the probability that $k'$ is a cooperator in layer 1 because
$k'$ is a three-step neighbor of cooperator $j$ (i.e., $j \to i \to \ell \to k'$).
Individual $\ell$ pays cost $c\theta_{2}^{\bm{\xi}\left[1\right]}$ because $\ell$ is a two-step neighbor of $j$ and hence is a cooperator with probability $\theta_{2}^{\bm{\xi}\left[1\right]}$. The payoff to $\ell$ from layer 2 is equal to $r \phi_{2,0}^{\bm{\xi}\left[1,2\right]} + \left( 1-\phi_{2,0}^{\bm{\xi}\left[1,2\right]} \right)$. Note that
$\phi_{2,0}^{\bm{\xi}\left[1,2\right]}$ is the probability that $\ell$ is a mutant in layer 2 given that $j$ is a cooperator in layer 1. Therefore, the total payoff of $\ell$ is equal to 
$b \theta_{3}^{\bm{\xi}\left[1\right]} - c\theta_{2}^{\bm{\xi}\left[1\right]}
+ r \phi_{2,0}^{\bm{\xi}\left[1,2\right]} + \left( 1-\phi_{2,0}^{\bm{\xi}\left[1,2\right]} \right)$.

Therefore, selection favors cooperation in layer 1 if
\begin{equation}
b \theta_{1}^{\bm{\xi}\left[1\right]} - c\theta_{0}^{\bm{\xi}\left[1\right]}
+ r \phi_{0,0}^{\bm{\xi}\left[1,2\right]} + \left( 1-\phi_{0,0}^{\bm{\xi}\left[1,2\right]} \right)
>
b \theta_{3}^{\bm{\xi}\left[1\right]} - c\theta_{2}^{\bm{\xi}\left[1\right]}
+ r \phi_{2,0}^{\bm{\xi}\left[1,2\right]} + \left( 1-\phi_{2,0}^{\bm{\xi}\left[1,2\right]} \right).
\label{eq:intuitive-1}
\end{equation}
By substituting $n=0$ in Eq.~\eqref{eq:def-theta}, we obtain
\begin{equation}
\theta_{0}^{\bm{\xi}\left[1\right]} = \sum_{i,j=1}^{N} \pi_{i}^{\left[1\right]} \left( p^{\left[1\right]} \right)_{ij}^{\left(0\right)} \beta_{ij}^{\bm{\xi}\left[1\right]}
= \sum_{i=1}^{N} \pi_{i}^{\left[1\right]} \beta_{ii}^{\bm{\xi}\left[1\right]} = 0.
\label{eq:theta_0=0}
\end{equation}
The last equality comes from Eq.~\eqref{eqn:sum_beta_ii=0}.
By substituting $m=n=0$ in Eq.~\eqref{eq:def-phi}, we obtain
\begin{equation}
\phi_{0, 0}^{\bm{\xi}\left[1,2\right]} = \sum_{i,j=1}^{N} \pi_{i}^{\left[1\right]} \left( p^{\left[1,2\right]} \right)_{ij}^{\left(0, 0\right)} \gamma_{ij}^{\bm{\xi}\left[1,2\right]}
= \sum_{i=1}^{N} \pi_{i}^{\left[1\right]} \gamma_{ii}^{\bm{\xi}\left[1,2\right]} = 0.
\label{eq:phi_00=0}
\end{equation}
The last equality comes from Eq.~\eqref{eqn:sum_gamma_ii=0}. By substituting
Eqs.~\eqref{eq:theta_0=0} and \eqref{eq:phi_00=0} in Eq.~\eqref{eq:intuitive-1}, we obtain
\begin{equation}
\label{eqn:cooperation_criteria_reduced-SI-2}
	c\theta_{2}^{\bm{\xi}\left[1\right]} + b\left(\theta_{1}^{\bm{\xi}\left[1\right]} - \theta_{3}^{\bm{\xi}\left[1\right]}\right) - \left(r-1\right) \phi_{2,0}^{\bm{\xi}\left[1,2\right]} > 0,
\end{equation}
which is
Eqs.~\eqref{eqn:cooperation_criteria_reduced} and \eqref{eqn:cooperation_criteria_reduced-SI-2}.

One can derive a parallel explanation of the condition under which the mutant is favored. We assume that individual $i$ updates its type in layer 2, that $j$ is a mutant neighbor of $i$ in layer 2, and that $\ell$ is a random neighbor of $i$ in layer 2. Selection favors the mutant in layer 2 if
\begin{equation}
r \theta_{0}^{\bm{\xi}\left[2\right]} + \left( 1 -\theta_{0}^{\bm{\xi}\left[2\right]} \right)
+ b \phi_{1,0}^{\bm{\xi}\left[1,2\right]} - c \phi_{0,0}^{\bm{\xi}\left[1,2\right]}
>
r \theta_{2}^{\bm{\xi}\left[2\right]} + \left( 1 - \theta_{2}^{\bm{\xi}\left[2\right]} \right)
+ b \phi_{1,2}^{\bm{\xi}\left[1,2\right]} - c \phi_{0,2}^{\bm{\xi}\left[1,2\right]}.
\label{eq:intuitive-2}
\end{equation}
By substituting Eqs.~\eqref{eq:theta_0=0} and \eqref{eq:phi_00=0} in Eq.~\eqref{eq:intuitive-2}, we obtain
\begin{equation}
	-\left(r-1\right)\theta_{2}^{\bm{\xi}[2]} + c \phi_{0,2}^{\bm{\xi}\left[1,2\right]} +
	b \left(\phi_{1,0}^{\bm{\xi}\left[1,2\right]} - \phi_{1,2}^{\bm{\xi}\left[1,2\right]}\right) > 0,
\label{eqn:constant_drift_selection_criteria-SI-2}
\end{equation}
which is Eq.~\eqref{eqn:constant_drift_selection_criteria}.

\newpage

\section{Impact of the network structure and initial condition on promotion of cooperation}\label{sec:SI-E}

\subsection{Impact of the network structure}

To explore what structural properties of networks promote prosociality,
for all the single-layer six-node connected networks, we measured the correlation between each pair of the five indices: $\theta_1^{\bm{\xi}\left[1\right]} - \theta_3^{\bm{\xi}\left[1\right]}$ (which we denote by $\theta_1 - \theta_3$ in this section, as we do in the main text), $n_{\text{init}, \text{C}}$, which is the number of initial conditions out of the six initial conditions (i.e., choice of the single node that initially has the cooperator) for which $(b/c)^* > 0$, mean degree, clustering coefficient, and mean closeness centrality. 
Quantities $\theta_1 - \theta_3$ and $n_{\text{init}, \text{C}}$ are measures of prosociality for a network. The other three indices quantify the network structure.
We regarded the 112 six-node networks as samples based on which we compute the Pearson correlation coefficient. We only consider the dB-dB rule for simplicity.

\begin{figure}
\centering
\includegraphics[width=0.8\textwidth]{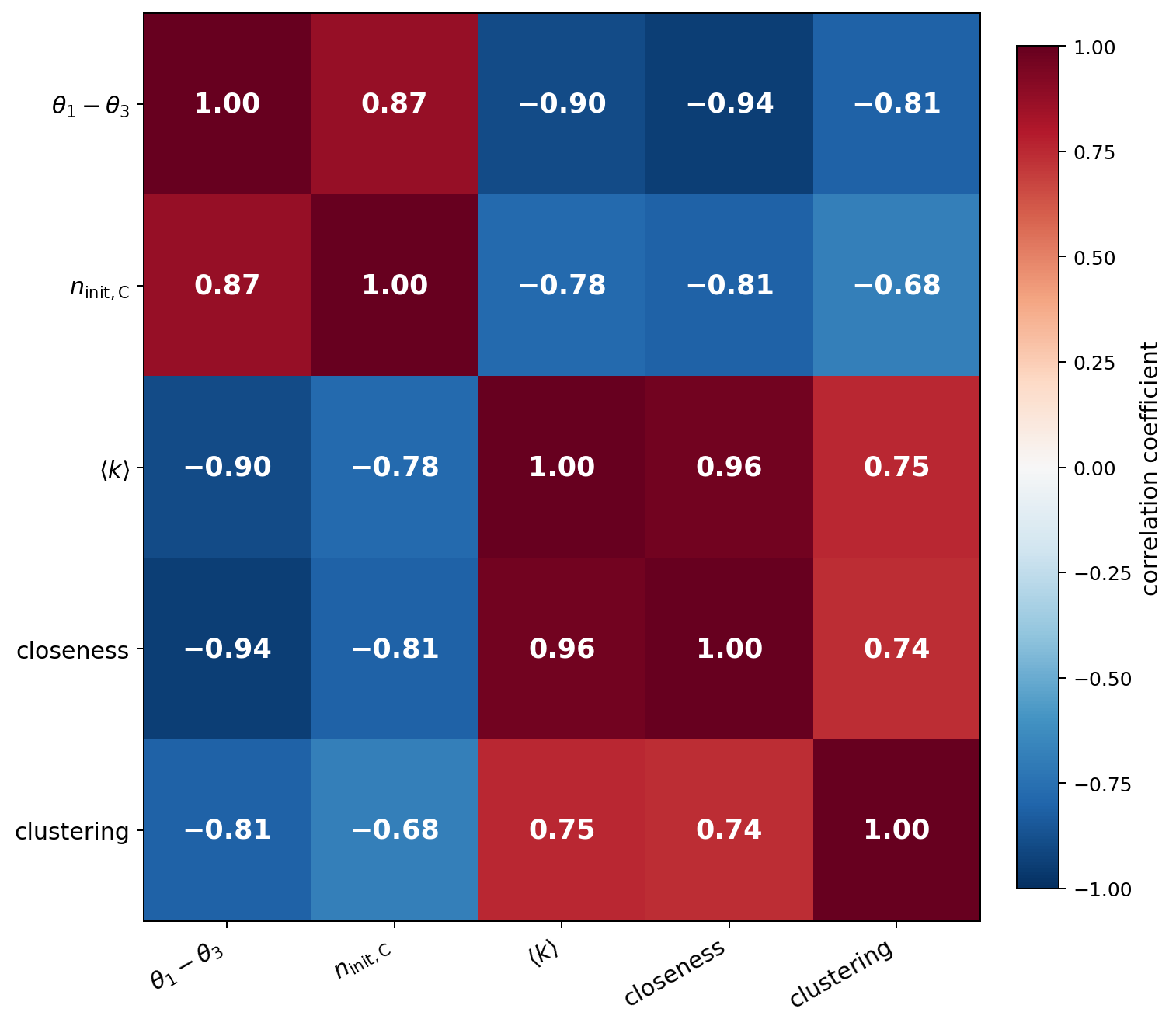}
\caption{Association between prosociality and network structure for one-layer networks with six nodes under the dB-dB rule.
The values shown are the Pearson correlation coefficient between pairs of metrics. Two metrics are related to prosociality
(i.e., $\theta_1 - \theta_3$ and $n_{\text{init}, \text{C}}$). The other three metrics quantify network structure (i.e., mean degree (denoted by $\langle k \rangle$), clustering coefficient, and mean closeness centrality).
}
\label{fig:Pearson-N6-network}
\end{figure}

We show the Pearson correlation coefficients in Fig.~\ref{fig:Pearson-N6-network}. As reported in the main text, 
$\theta_1 - \theta_3$ and $n_{\text{init}, \text{C}}$ are strongly positively correlated. These two quantities are strongly negatively correlated with the three network metrics. Therefore, cooperation is easier for networks with a smaller mean degree, smaller clustering coefficient, and smaller mean closeness. Because the effect of the small mean degree of the network on enhancing fixation of cooperation is established \cite{ohtsuki2006simple, allen2017evolutionary}, we also compute partial correlation coefficients by partialing out the effect of the mean degree. 
The partial correlation with $\theta_1 - \theta_3$ is equal to
$-0.46$ and $-0.65$ for the clustering coefficient and mean closeness, respectively.
The partial correlation with $n_{\text{init}, \text{C}}$ is equal to
$-0.23$ and $-0.35$ for the clustering coefficient and mean closeness, respectively. Therefore, the result that a small clustering coefficient and small closeness centrality tend to promote cooperation is not a byproduct of the effect of the small mean degree.

The result that the small clustering coefficient promotes cooperation contradicts a common notion of spatial reciprocity with which clustering of cooperators in networks enhances cooperation \cite{Nowak1992Nature_spatial, Nowak1993IntJBifuChaos, Nowak2006book}. We do not know the reason for this discrepancy. The discrepancy may be because here we investigate the fixation probability under weak selection, whereas a majority of studies on spatial reciprocity has examined metastable states under strong selection.

Next, we examine the extent to which the coupling with the constant-selection layer changes cooperation.
Under the dB-dB updating rule, we now consider all the $31 \times 112 = 3,584$ ordered pairs of connected six-node networks as layers 1 and 2 of the two-layer network. For layer 1, we only consider networks that yield $(b/c)^* > 0$ for any initial condition. For layer 2, we consider all the connected six-node networks. We examined how the sensitivity of $(b/c)^*$ in layer 1 to the change in the mutant's fitness in layer 2,
quantified by $\left| \text{d}(b/c)^* / \text{d}r \right|$, depends on the network structure of layers 1 and 2. As we did for Fig.~\ref{fig:Pearson-N6-network}, we compute the Pearson correlation coefficient between $\left| \text{d}(b/c)^* / \text{d}r \right|$ and each network metric, regarding each two-layer network as a sample.

We show the results in Table~\ref{tab:Pearson-2-layers}. We find that $\left| \text{d}(b/c)^* / \text{d}r \right|$ is strongly correlated with several metrics of the layer-1 network. In contrast, it is uncorrelated with all of the layer-2 metrics (with the absolute value of the Pearson correlation coefficient $< 0.045$). A large mean degree tends to yield a large mean closeness, small diameter, and large algebraic connectivity. Therefore, a primary implication of Table~\ref{tab:Pearson-2-layers} is that a large mean degree of layer 1, not layer 2, makes $(b/c)^*$ of the two-layer network relatively sensitive to variations in $r$.
%
%

We repeated the same analysis for larger model and empirical two-layer networks in which the layer-1 network allows cooperation (i.e., $(b/c)^* > 0$).
Specifically, we used two-layer ER and BA networks with $N=15$, $30$, $50$, $75$, and $100$ individuals, one instance each, and the LLF network. We excluded the VC7 network because its layer-1 network yields $(b/c)^* < 0$. As a result, we have $11$ two-layer networks, which we regard as samples for computing the Pearson correlation coefficient between $\left| \text{d}(b/c)^* / \text{d}r \right|$ and metrics of network structure.
We show the results in Table~\ref{tab:b/c-vs-init-cnd-larger-networks}. We only show the correlation between $\left| \text{d}(b/c)^* / \text{d}r \right|$ and
the metrics of the layer-1 network, not the layer-2 network, because in the two-layer ER and BA networks, layers 1 and 2 are statistically the same (e.g., the two layers would have similar variances of the degree). We find how properties of network layer 1 affect $\left| \text{d}(b/c)^* / \text{d}r \right|$ is largely similar to the case of networks with $N=6$ individuals, except the effect of the variance of the degree. This discrepancy may be because the degree range of networks with $N=6$ under the constraint that the network has just a single connected component is inherently limited.

\begin{table}
	\centering
\caption{Relationship between $\left| \text{d}(b/c)^* / \text{d}r \right|$ and properties of the two-layer network. Each value shown is the Pearson correlation coefficient between $\left| \text{d}(b/c)^* / \text{d}r \right|$ and each network metric measured for one network layer.
We used the two-layer networks in which the layer-1 network has $(b/c)^* > 0$ regardless of the initial condition. For larger networks, we used the ER and BA networks with $N=15$, $30$, $50$, $75$, and $100$ nodes, and the LLF network. We excluded the VC7 network because its layer-1 network has $(b/c)^* < 0$.}
\newcolumntype{C}{>{\centering\arraybackslash}p{1.4cm}}
\begin{tabular}{|c|C|C|C|}
\hline
network size & \multicolumn{2}{c|}{six-node} & larger \\ \hline
network layer & 1 & 2 & 1 \\
\specialrule{1.5pt}{0pt}{0pt}
mean degree & $0.855$ & $-0.040$ & $0.641$ \\ \hline
maximum degree & $0.284$ & $-0.039$ & $0.481$ \\ \hline
variance of degree & $-0.081$ & $-0.014$ & $0.736$ \\ \hline
mean closeness & $0.831$ & $-0.044$ & $0.750$ \\ \hline
clustering coefficient & $0.286$ & $-0.029$ & $0.834$ \\ \hline
diameter & $-0.532$ & $0.036$ & $-0.532$ \\ \hline
algebraic connectivity & $0.722$ & $-0.034$ & $0.812$ \\ \hline
\end{tabular}
%
\label{tab:Pearson-2-layers}
\end{table}

\subsection{Impact of the initial condition}

In this section, we investigate the impact of the initial condition on the ease of cooperation for single-layer networks.
To this end, we focus on the 32 out of the 112 six-node networks for which all the six initial conditions (i.e., the node on which the single cooperator initially resides, denoted by $v$) yield $(b/c)^* > 0$, such that cooperation occurs for a sufficiently large $b/c$. We removed the ring network, which is one of the 32 networks. This is because all nodes in the ring network are vertex-transitive (i.e., structurally the same due to symmetry), and therefore one cannot ask the impact of the initial condition (i.e., the location of the single initial cooperator) on fixation of cooperation. For each of the remaining 31 networks, we computed the Pearson correlation coefficient between $(b/c)^*$ given the initial condition, $v$, and $v$'s centrality. The node centralities used are the degree, local clustering coefficient, closeness centrality, and betweenness centrality. We confirmed that, except for the local clustering coefficient, for each of the 31 networks, some nodes have different centrality values such that one can compute the Pearson correlation coefficient.
For the local clustering coefficient, all nodes have value $0$ for some networks because they are triangle-free. We excluded these networks because one cannot compute the correlation coefficient. This left us 16 networks for examining the effect of the local clustering coefficient on $(b/c)^*$. We also note that nodes with degree $1$ are assigned with the local clustering coefficient value of $0$ for simplicity.
We include the betweenness centrality in this analysis while we exclude the mean betweenness centrality in Fig.~\ref{fig:Pearson-N6-network}. This is because we consider that it is not appropriate to compare the betweenness centrality across different networks. When computing the correlation coefficient for a given network, we regard each of the six nodes as sample, while we are aware that the computed correlation coefficients are not sufficiently reliable because they are only based on six samples.

\begin{figure}
\centering
\includegraphics[width=0.7\textwidth]{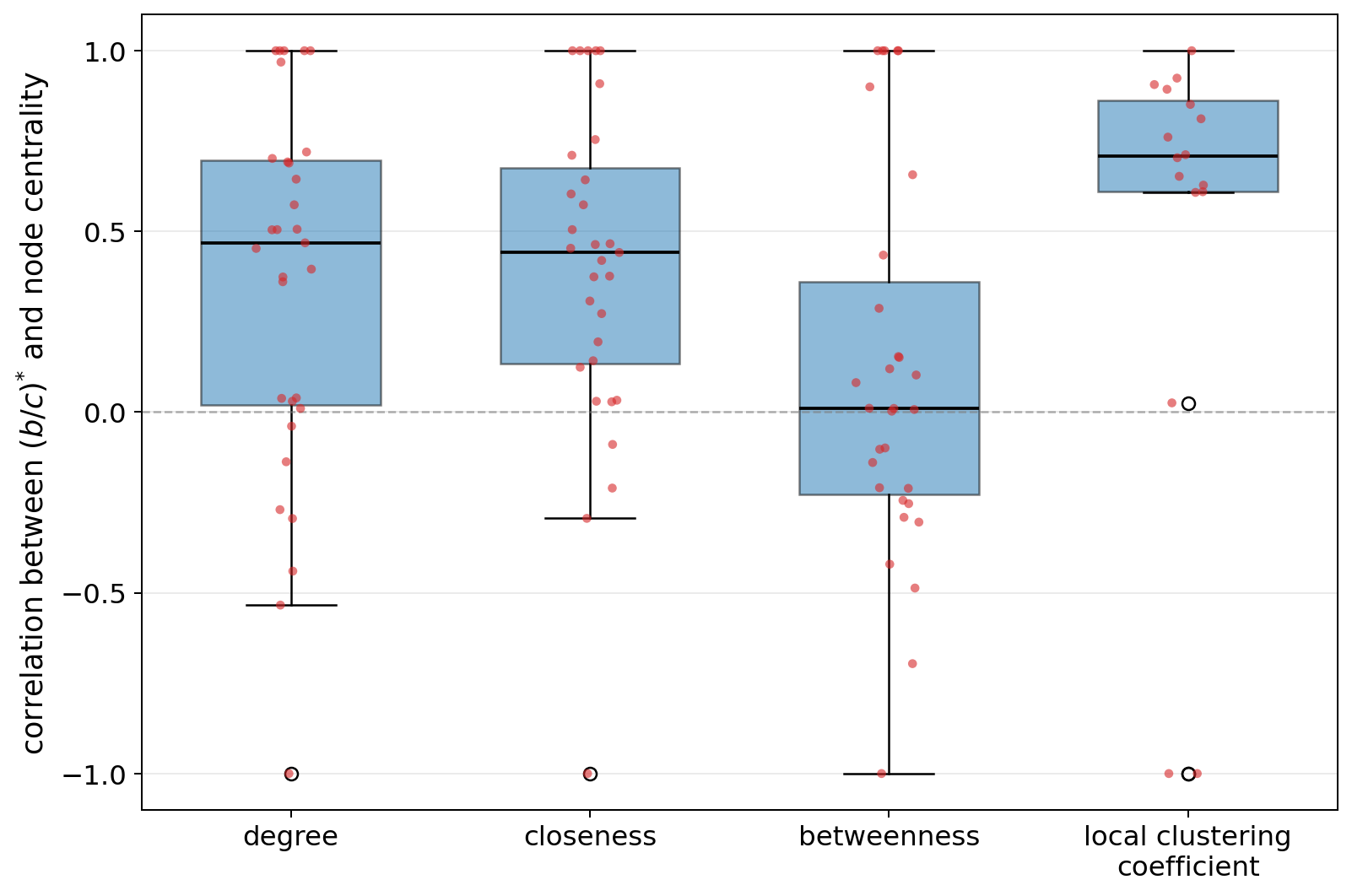}
\caption{Correlation between $(b/c)^*$ and the initial condition for the six-node networks that yield $(b/c)^* > 0$ regardless of the initial condition. A filled dot represents a network, jittered horizontally to avoid overlaps.
An open circle represents an outlier. The vertical axis represents the Pearson correlation coefficient between $(b/c)^*$ and the node's centrality. 
The rectangular boxes span from the first quartile ($Q_1$) to the third quartile ($Q_3$).
The thick horizontal lines represent the median.
The whiskers span from the minimum without outliers ($Q_1-1.5 \times \text{IQR}$) to the maximum without outliers ($Q_3+1.5 \times \text{IQR}$), where IQR $=Q_3-Q_1$.
}
	\label{fig:Pearson-N6-node}
\end{figure}

We show the obtained correlation coefficients in Fig.~\ref{fig:Pearson-N6-node}. We find that the initial cooperator located at a node with a large degree or closeness tends to suppress fixation of cooperation (by raising $(b/c)^*$), albeit modestly. The betweenness centrality is not correlated with $(b/c)^*$. Nodes with a high local clustering coefficients tend to suppress cooperation as well.

We have also investigated the correlation between $(b/c)^*$ and the centralities of $v$ in larger single-layer ER, BA, and empirical networks allowing cooperation (i.e., $(b/c)^* > 0$). We show the results in Table~\ref{tab:b/c-vs-init-cnd-larger-networks}. Although the node's degree in the BA networks is not positively correlated with $(b/c)^*$, the results are largely consistent with those for the networks with $N=6$ nodes shown in Fig.~\ref{fig:Pearson-N6-node}.

\begin{table}
\centering
\caption{Correlation between $(b/c)^*$ and the initial condition for larger networks. Each value shown is the Pearson correlation coefficient between $(b/c)^*$ and the centrality of the node that initially has a cooperator. We denote by ``local clus.'' the local clustering coefficient. We used the single-layer networks that yield $(b/c)^* > 0$ regardless of the initial condition, which excluded layer 1 of the VC7 network. Layer 1 of the LLF network is included. The local clustering coefficient cell is not filled for the ER network with $N=15$ because the network instance we obtained is triangle-free.}
\newcolumntype{C}{>{\centering\arraybackslash}p{1.9cm}}
\begin{tabular}{|c|C|C|C|C|}
\hline
network & degree & closeness & betweenness & local clus. \\
\specialrule{1.5pt}{0pt}{0pt}
ER, $N=15$ & $0.46$ & $0.91$ & $0.87$ & N/A \\ \hline
ER, $N=30$ & $0.46$ & $0.67$ & $0.00$ & $0.25$ \\ \hline
ER, $N=50$ & $0.54$ & $0.78$ & $0.13$ & $0.34$ \\ \hline
ER, $N=75$ & $0.49$ & $0.63$ & $0.16$ & $0.30$ \\ \hline
ER, $N=100$ & $0.55$ & $0.67$ & $0.18$ & $0.31$ \\ \hline
BA, $N=15$ & $0.15$ & $0.30$ & $0.13$ & $0.27$ \\ \hline
BA, $N=30$ & $-0.09$ & $0.19$ & $-0.06$ & $0.63$ \\ \hline
BA, $N=50$ & $-0.11$ & $0.28$ & $-0.06$ & $0.51$ \\ \hline
BA, $N=75$ & $-0.10$ & $0.43$ & $-0.06$ & $0.58$ \\ \hline
BA, $N=100$ & $-0.09$ & $0.38$ & $-0.06$ & $0.40$ \\ \hline
LLF & $0.80$ & $0.77$ & $0.51$ & $-0.17$ \\ \hline
\end{tabular}
\label{tab:b/c-vs-init-cnd-larger-networks}
\end{table}

\newpage
\clearpage

\section{Strategy-type correlation between the two layers during evolutionary dynamics}\label{sec:SI-F}

In this section, we numerically investigate possible correlation between the individual's strategy in the game network layer (i.e., cooperation or defection) and its type in the constant-selection network layer (i.e., resident or mutant) during the evolutionary dynamics. We focus on two-layer networks with $N=6$ of which the layer-1 network fosters cooperation as single-layer network (i.e., $(b/c)^* > 0$). We use five such two-layer networks chosen as follows.
We first impose that the average of $\theta_1^{\bm{\xi}\left[1\right]} - \theta_3^{\bm{\xi}\left[1\right]}$
over the six correlated initial conditions, or the initial conditions in which
the initial cooperator in layer 1 and the initial mutant in layer 2 are the same individual, is larger than
$0.1$. We impose this condition because
$\theta_1^{\bm{\xi}\left[1\right]} - \theta_3^{\bm{\xi}\left[1\right]}$ is the denominator of $\left| \text{d}(b/c)^* / \text{d}r \right|$ and we decide to avoid the cases which involves the division by a quantity near $0$.
Then, among the qualified two-layer networks, we choose the five networks that are the largest in terms of
the average of
$\left| \phi_{2,0}^{\bm{\xi}\left[1,2\right]} \right|$ over the six correlated initial conditions divided by the average of
$\left| \theta_1^{\bm{\xi}\left[1\right]} - \theta_3^{\bm{\xi}\left[1\right]} \right|$ over the six correlated initial conditions.
These five networks make $\left| \text{d}(b/c)^* / \text{d}r \right| = \left|  \phi_{2,0}^{\bm{\xi}\left[1,2\right]} / \left(  \theta_1^{\bm{\xi}\left[1\right]} - \theta_3^{\bm{\xi}\left[1\right]} \right) \right|$ large and therefore
enhance prosociality by relatively a large amount when one couples the donation game network layer with the constant-selection network layer.
It turns out that all the five networks have $\text{d}(b/c)^* / \text{d}r < 0$. Therefore, we set $r=4$ such that $(b/c)^*$ for each of these two-layer networks remains positive and is smaller than $(b/c)^*$ for the one-layer network.
%

For each of the five two-layer networks, we use two initial conditions, each with just one cooperator in layer 1 and one mutant in layer 2. In the first initial condition, the same individual is the cooperator in layer 1 and the mutant in layer 2. As we already mentioned earlier in this section, we call this initial condition correlated. In the second initial condition, the cooperator in layer 1 and the mutant in layer 2 are different individuals. We all this initial condition uncorrelated. Then, we tracked the Pearson correlation coefficient between the state of nodes in layer 1 (by encoding defector and cooperator into $0$ and $1$, respectively) and that of nodes in layer 2 (by encoding resident and mutant into $0$ and $1$, respectively). The correlated initial condition has the correlation coefficient equal to $1$ initially. The interest is whether or not high correlation is sustained during evolutionary dynamics, especially in runs that end up with fixation of cooperation. Likewise, we are interested in whether correlation between the state in layer 1 and layer 2 builds up during evolutionary dynamics starting with the uncorrelated initial condition. We run $5\times 10^5$ simulations with each initial condition (i.e., either the correlated or uncorrelated initial condition) for each of the five two-layer networks. Then, we compute the average correlation coefficient across surviving runs (i.e., runs that have not reached the fixation of cooperation or defection or the fixation of residents and mutants; for computing the correlation coefficient, layer 1 has to have at least one cooperator and one defector, and layer 2 has to have at least one resident and one mutant) at each time point. We use the dB-dB rule and set the selection strength to $\delta = 0.02$. We set $c=1$. We set $b$ such that $b/c$ is $1.5$ times larger than $(b/c)^*$ for the two-layer network with the initial condition used in the panel (i.e., correlated or uncorrelated). This $b$ value is favorable to cooperators, but due to the sufficiently weak selection (i.e., $\delta = 0.02$), the fixation probability of cooperation observed in the following numerical simulations is not much distant from that for the neutral dynamics (i.e., $\delta = 0$).

\begin{figure}
\centering
\includegraphics[width=0.98\textwidth]{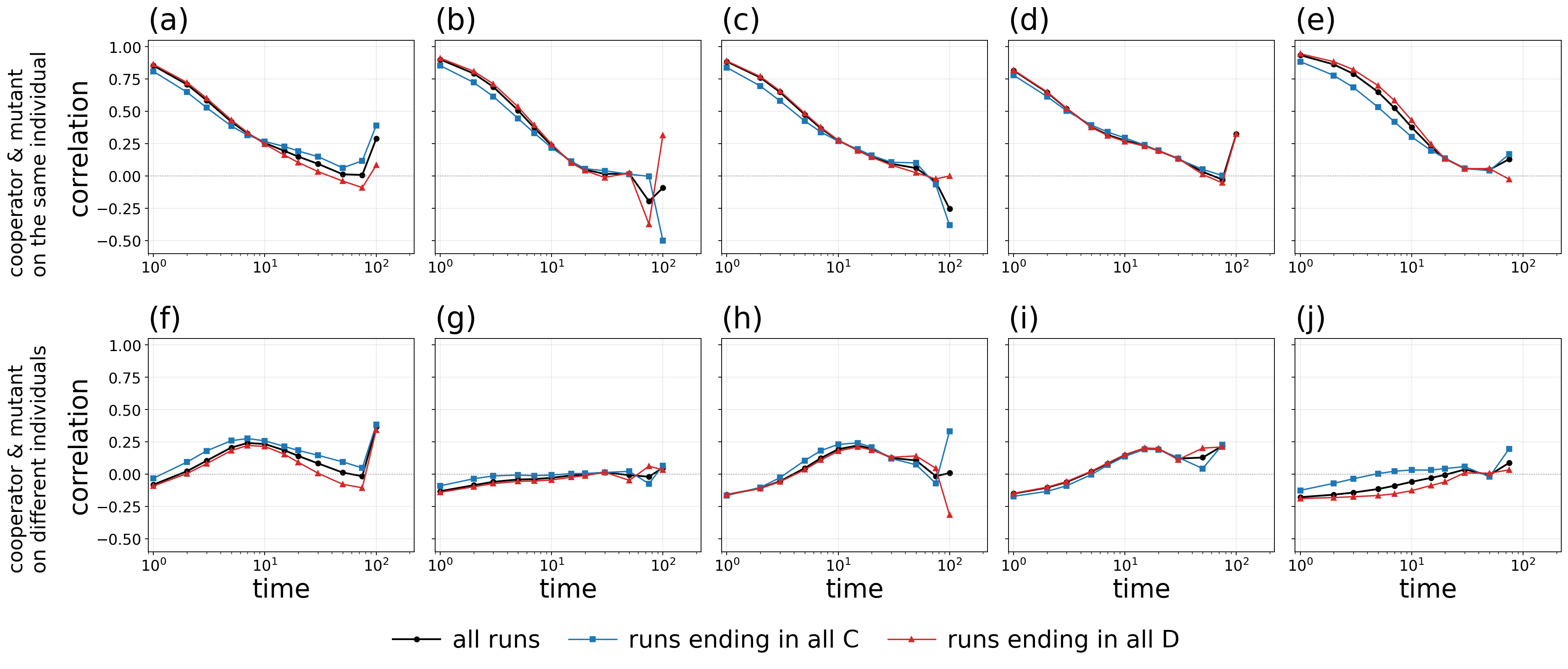}
\caption{Dynamics of the correlation between the states in layer 1 (i.e., defection or cooperation) and those in layer 2 (i.e., resident or mutant). We use five two-layer networks with $N=6$ individuals. (a)--(e): Correlated initial condition. (f)--(j): Uncorrelated initial condition.
Panels (a) and (f), for example, are for the same two-layer network. At each time during the evolutionary dynamics, we show the correlation computed based on all the $5 \times 10^5$ runs, all runs that ended up with fixation of cooperation, and all runs that ended up with fixation of defection, shown by the lines in different colors.
The $(b/c)^*$ value for the layer-1 network is equal to $28.32$ in (a), (b), (e), (f), (g), and (j); the layer-1 network is identical in these three two-layer networks. (Note that we use the same two-layer network in (a) and (f), we use another two-layer network in (b) and (g), and so forth.)
The $(b/c)^*$ value for the layer-1 network is equal to $10.72$ in (c), (d), (h), and (i); the layer-1 network is identical in these two two-layer networks. 
The $(b/c)^*$ value for the two-layer network with $r=4$ is equal to: (a) $6.75$, (b) $8.11$, (c) $2.65$, (d) $1.80$, (e) $7.53$, (f) $26.47$, (g) $31.31$, (h) $10.23$, (i) $11.43$, and (j) $32.45$. Note that $(b/c)^*$ increases for the two-layer network relative to the one-layer network in (g), (i), and (j), making cooperation harder in two-layer than one-layer networks. This is not a contradiction because we have chosen these five two-layer networks to maximize the decrease in $(b/c)^*$ by layer coupling for the correlated initial condition; this does not guarantee that $(b/c)^*$ decreases by layer coupling for the uncorrelated initial condition used in (f)--(j). We set $c=1$ in all panels. We set $b$ to (a) $10.12$, (b) $12.17$, (c) $3.98$, (d) $2.70$, (e) $11.29$, (f) $39.71$, (g) $46.96$, (h) $15.34$, (i) $17.15$, and (j) $48.68$. These $b$ values yield $b/c \approx 1.5 \times (b/c)^*$ for each two-layer network.
The numerically obtained fixation probability for cooperation, which is the fraction of all runs that terminate with the fixation of cooperation, is:
(a) $0.143$, (b) $0.144$, (c) $0.073$, (d) $0.071$, (e) $0.143$, (f) $0.142$, (g) $0.142$, (h) $0.072$, (i) $0.073$, and (j) $0.142$.
}
\label{fig:corr-dyn}
\end{figure}

We show the dynamics of the state correlation for the five two-layer networks and two initial conditions in Fig.~\ref{fig:corr-dyn}. For the correlated initial condition, shown in Fig.~\ref{fig:corr-dyn}(a)--(e), the correlation decays towards zero over time no matter whether the cooperator or defector eventually fixates. The deviation of the correlation from $0$ near the end of the evolutionary dynamics is due to the small number of runs surviving to that time, so we regard it as statistical noise. Therefore, the state-to-state correlation between the two layers does not underlie the enhanced fixation probability for cooperation. In the case of the uncorrelated initial condition, shown in Fig.~\ref{fig:corr-dyn}(f)--(j), strong state-to-state correlation does not build up through evolutionary dynamics, while we see transient Pearson correlation coefficient values up to $\approx 0.25$ in Fig.~\ref{fig:corr-dyn}(f), (h), and (i). However, even in these three networks, the runs in which cooperation fixates do not show notable tendencies. Therefore, although the present numerical simulations are of a small scale and not exhaustive, we suggest that state-to-state correlation is not a mechanism of fixation of cooperation in two-layer networks.

\newpage
\clearpage

\section{Effects of the network structure and initial condition in layer 1 on $\frac{\partial r^*}{\partial b^{\;}}$ and $\frac{\partial r^*}{\partial c^{\;}}$}\label{sec:SI-G}

To explore reasons for striped patterns in
Fig.~\ref{fig:dr}
in the main text, we computed the distribution of $\left( \frac{\partial r^*}{\partial b^{\;}}, \frac{\partial r^*}{\partial c^{\;}} \right)$ for arbitrarily selected four layer-1 networks with $N=6$ individuals.
For each layer-1 network, we first fixed an initial condition in layer 1, i.e., the sole replica node initially hosting a cooperator. Then, for all combinations of one of the four arbitrarily selected layer-2 network, its isomorphism, and initial condition (i.e., the sole replica node in layer 2 that initially hosts the mutant), we computed 
$\left( \frac{\partial r^*}{\partial b^{\;}}, \frac{\partial r^*}{\partial c^{\;}} \right)$. We avoided using all possible networks as the layer-2 network because doing so would make scattergrams too crowded. 

\begin{figure}
\centering
\includegraphics[width=0.9\textwidth]{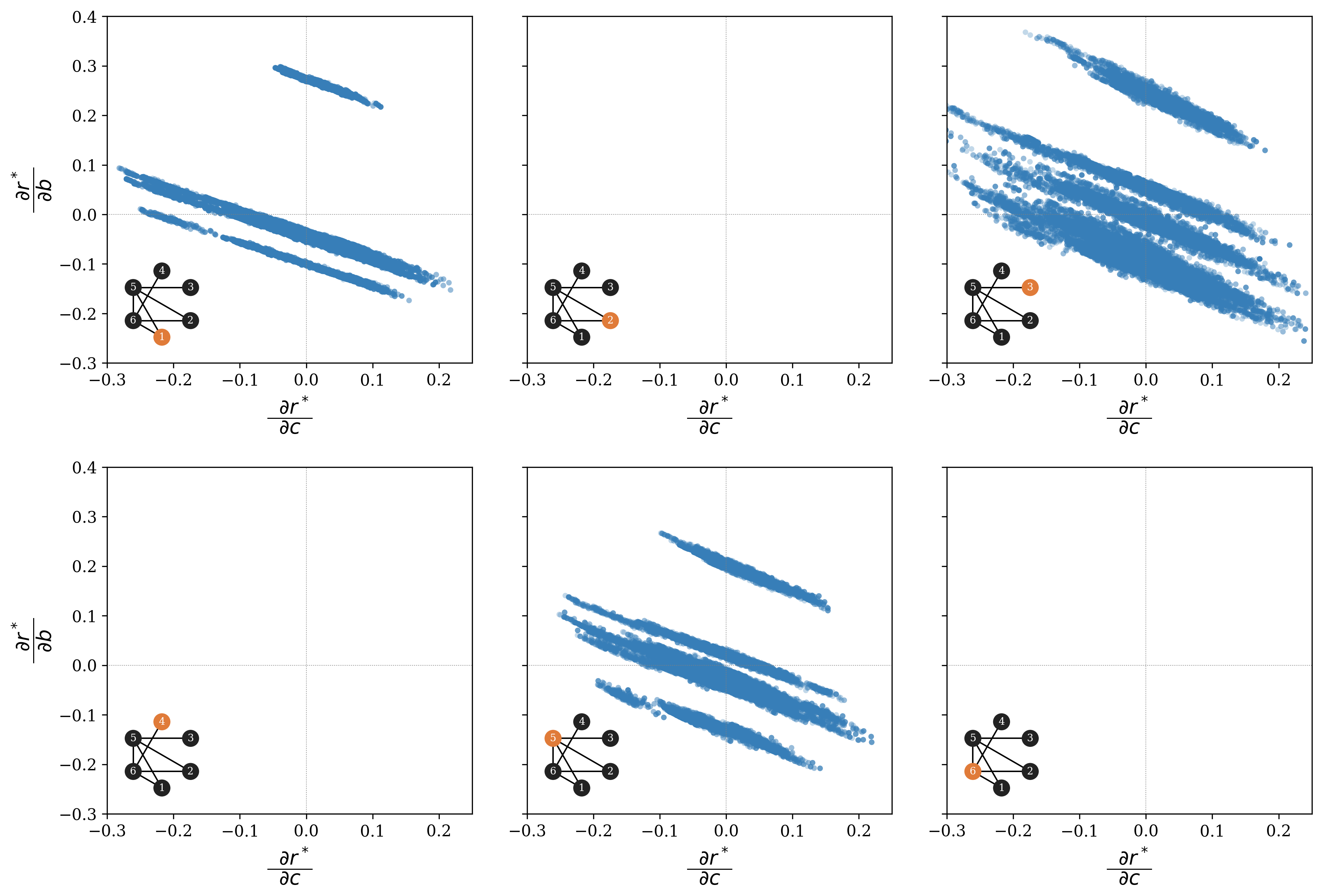}
\includegraphics[width=0.9\textwidth]{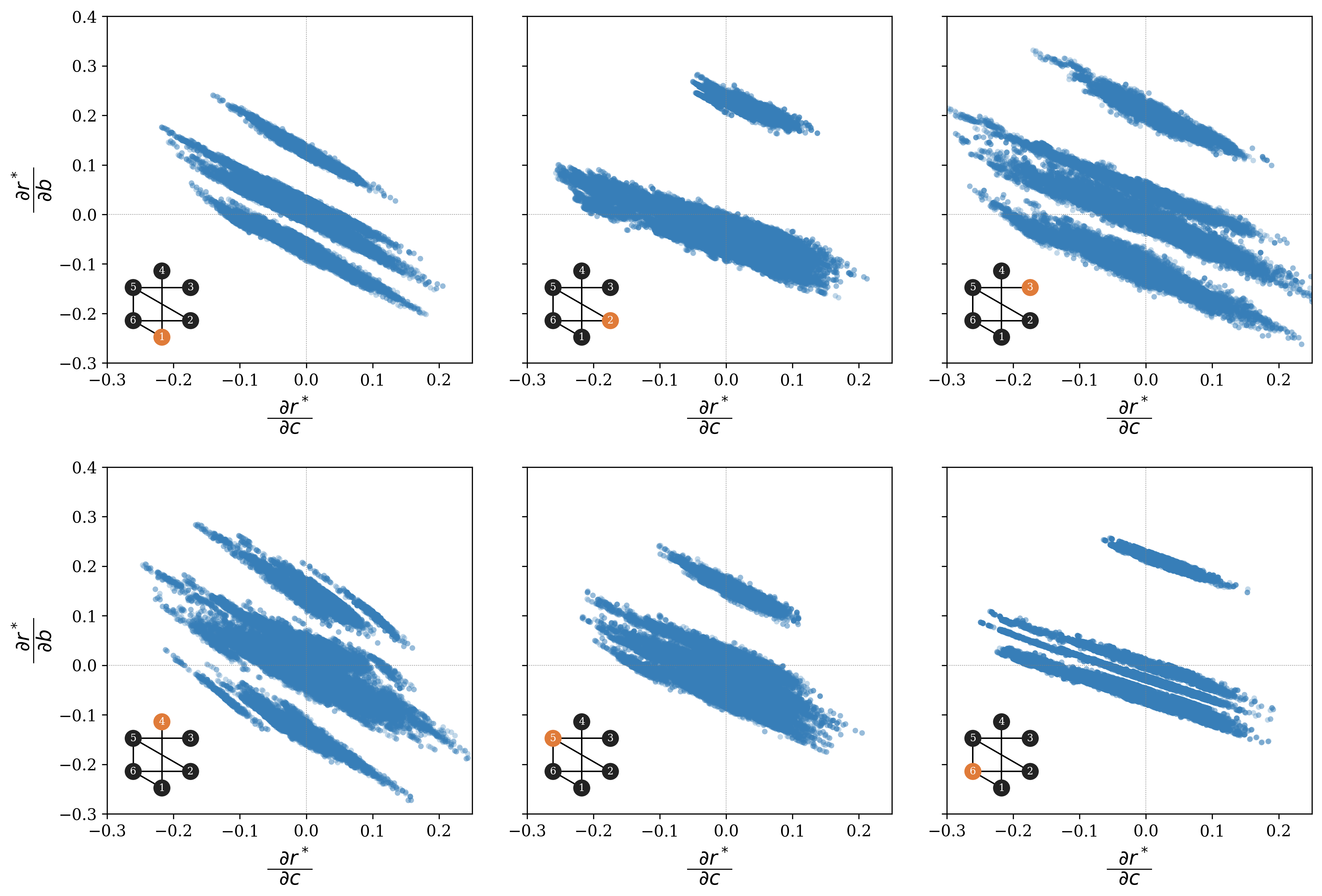}
\caption{Distribution of $\frac{\partial r^*}{\partial b^{\;}}$ and $\frac{\partial r^*}{\partial c^{\;}}$ over layer-2 networks and their initial conditions.
The visualization of the network within each panel shows the layer-1 network and its initial condition used in the panel. Each filled dot represents a pair of the layer-2 network and its initial condition. Dots are not present in four of the 24 panels because these four cases are the same as a pair of the layer-1 network and its initial condition that appears in an earlier panel.
}
	\label{fig:r-sensitivity-SI}
\end{figure}

We show $\left( \frac{\partial r^*}{\partial b^{\;}}, \frac{\partial r^*}{\partial c^{\;}} \right)$ for each pair of the layer-1 network and its initial condition in Fig.~\ref{fig:r-sensitivity-SI}. Each filled dot represents the result for a pair of the layer-2 network and its initial condition. The first six panels show the results for one arbitrarily selected layer-1 network. Each of the six panels corresponds to an initial condition in layer 1. We visualize the layer-1 network and its initial condition within each panel; we recall that the black and orange circles represent the resident and mutant replica nodes, respectively.
The next six panels show the results for a second layer-1 network, and so forth. Some panels are blank because the corresponding initial condition is the same as one that appeared in an earlier panel due to the symmetry of the network. For example, in the top row of Fig.~\ref{fig:r-sensitivity-SI}, nodes 1 and 2 are structurally equivalent. Therefore, placing the initial cooperator on node 1 versus node 2 produces exactly the same result, which is why the second panel does not show any dots.

We find that different layer-1 networks and their initial conditions produce different sets of stripes of the data points in general. Some stripes are only present in some of the panels. Therefore, we conclude that the network structure and initial condition in layer 1 are partial determinants of the stripe patterns observed in
Fig.~\ref{fig:dr}
in the main text.

\clearpage

\begin{center}
\includegraphics[width=0.9\textwidth]{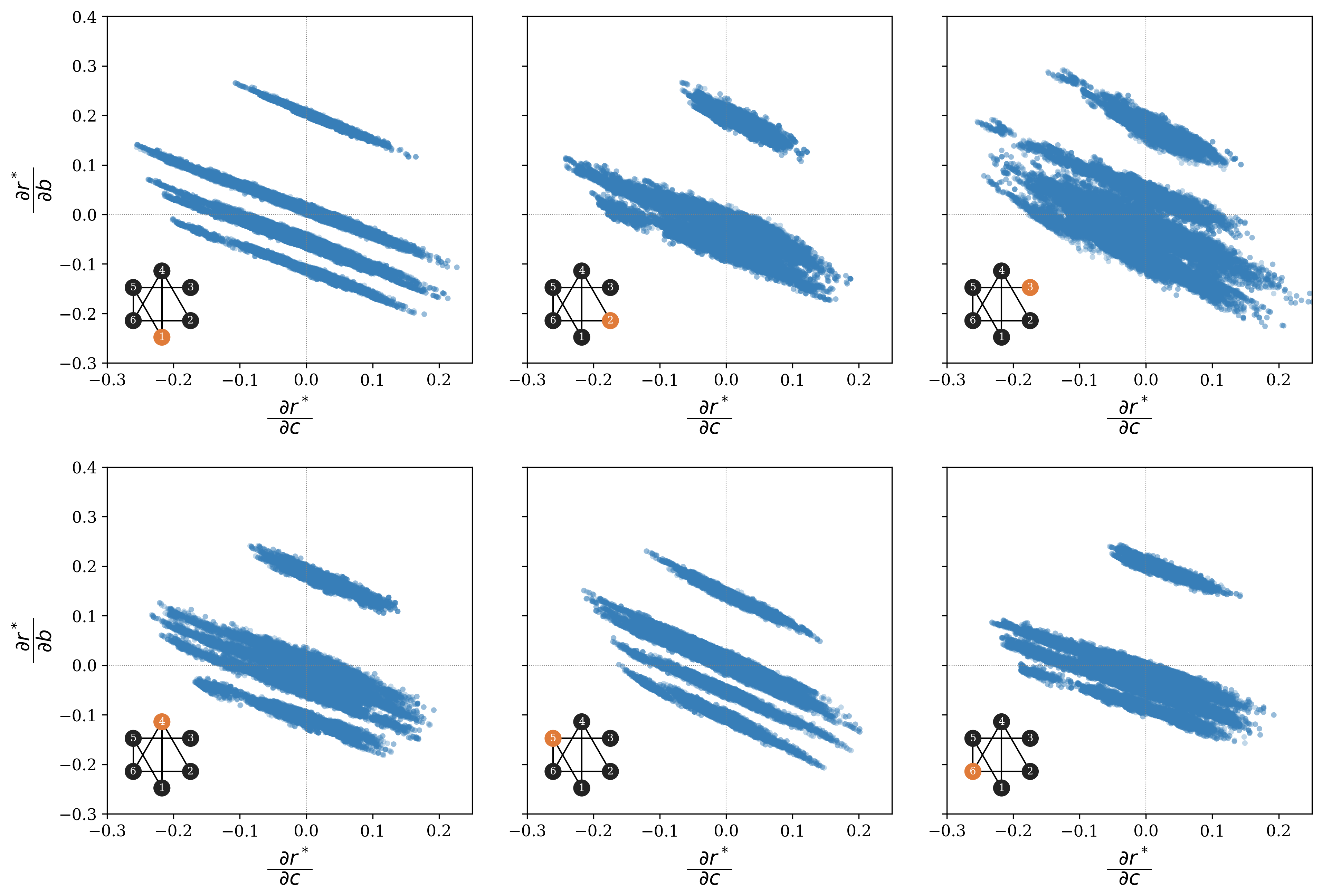}\\
\includegraphics[width=0.9\textwidth]{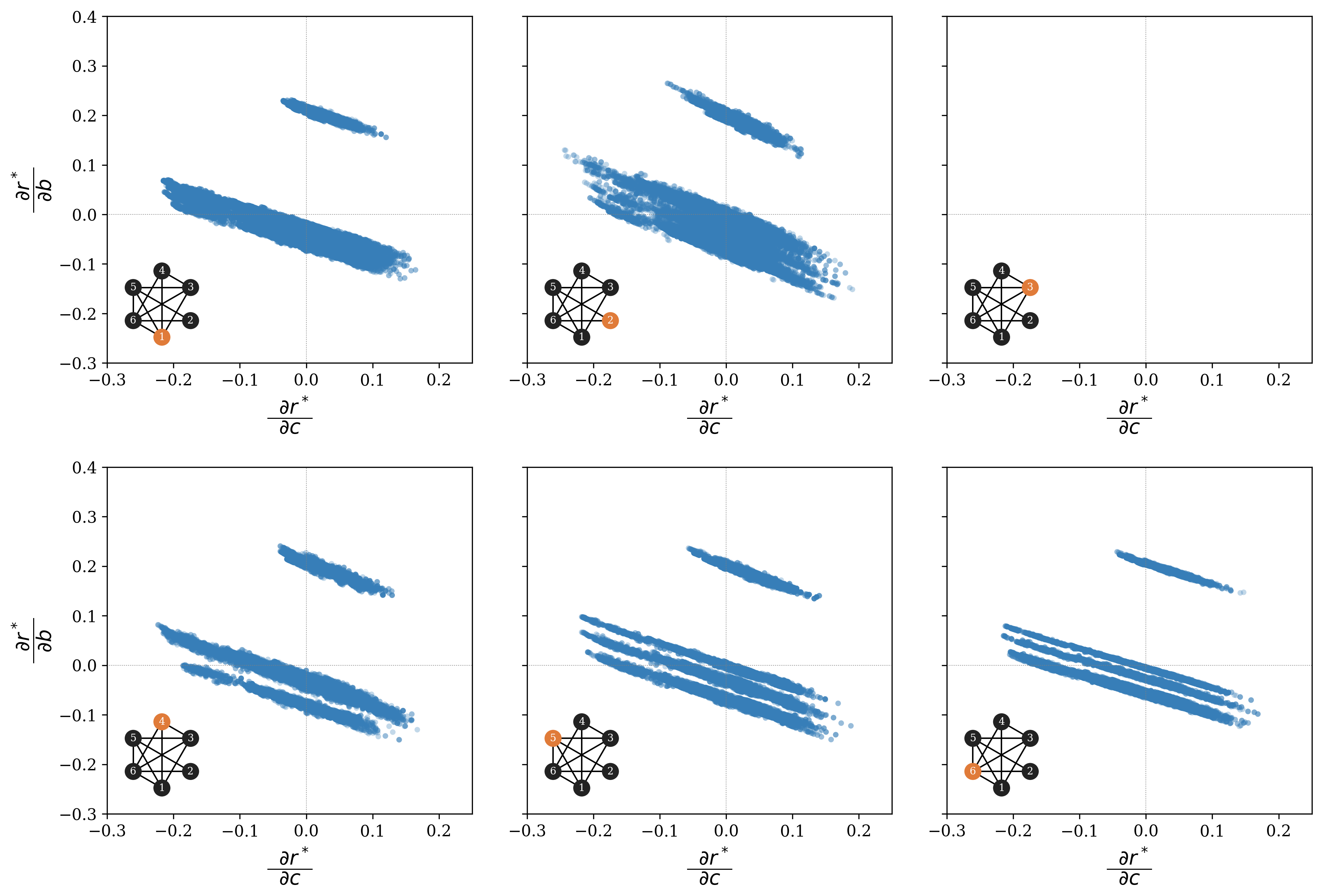}\\
Figure~\ref{fig:r-sensitivity-SI}: (Continued.)
\end{center}

\clearpage

\section{Evolution of the cooperator and mutant on the coupled ring network}\label{sec:SI-H}

We examine the coupled ring network depicted in
Fig.~\ref{fig:examples}A
%
%
in the main text under the dB-dB rule. The expressions for $\theta_i^{\bm{\xi}[1]} $ and $\phi_{m,n}^{\bm{\xi}[1,2]}$ were derived in Ref.~\cite{su2022evolution}. By following their work, we set $\xi_1^{[1]} = 1$, $\xi_j^{[1]} = 0$ for $j\neq 1$, $\xi_i^{[2]} = 1$, and $\xi_j^{[2]} = 0$ for $j \neq i$. Due to the symmetry of the ring network, this setting covers all possible initial conditions although we assumed $i=1$ in
Fig.~\ref{fig:examples}A
%
%
for demonstration. Let $d$ denote the distance between the node with the initial cooperator in layer 1 and that with the initial mutant in layer 2 when we superpose the two identical rings, one from each layer, to regard the superposition as a single ring network. We obtain	
\begin{align}
	\theta_{1}^{\bm{\xi}[1]} &= -\frac{N-1}{2}, \label{eqn:thetaphi_ring_network-1}\\
	\theta_{2}^{\bm{\xi}[1]} &= -\frac{N-2}{2}, \label{eqn:thetaphi_ring_network-2}\\
	\theta_{3}^{\bm{\xi}[1]} &= -\frac{3(N-2)}{4}, \label{eqn:thetaphi_ring_network-3}\\
	\phi_{0,1}^{\bm{\xi}[1,2]} &= - \sum_{\ell=1}^{N-1}
	\frac{\cos\!\left(\dfrac{2\pi \ell d}{N}\right)}{\,2N-1 + \cos\!\left(\dfrac{2\pi \ell}{N}\right)} , \label{eqn:thetaphi_ring_network-4}\\
%
		\phi_{2,0}^{\bm{\xi}[1,2]} &= -2(N-1)\phi_{0,1}^{\bm{\xi}[1,2]} - N\delta_{d,0} + 1 , \label{eqn:thetaphi_ring_network-5}\\
%
	\phi_{2,1}^{\bm{\xi}[1,2]} &= (4N^{2}-6N+3)\phi_{0,1}^{\bm{\xi}[1,2]}
	- \frac{N}{2}\delta_{d,1}
	+ 2N(N-1)\delta_{d,0} - 2N + 3 . \label{eqn:thetaphi_ring_network-6}
\end{align}

By substituting Eqs.~\eqref{eqn:thetaphi_ring_network-1}--\eqref{eqn:thetaphi_ring_network-6} in
Eq.~\eqref{eqn:cooperation_criteria_reduced}
%
%
in the main text, we obtain the condition under which selection favors the cooperator in layer 1 as follows:
\begin{align}
	0
	&< c\theta_{2}^{\bm{\xi}[1]}
	+ b\!\left(\theta_{1}^{\bm{\xi}[1]} - \theta_{3}^{\bm{\xi}[1]}\right)
	- (r-1)\phi_{2,0}^{\bm{\xi}[1,2]} \notag \\
	&= \frac{b}{4}\,(N-4)
	- \frac{c}{2}\,(N-2)
	+ (r-1)\left[2(N-1)\phi_{0,1}^{\bm{\xi}[1,2]} + N\delta_{d,0} - 1\right],
\end{align}
which we can rearrange to
\begin{equation}
\frac{b}{c}> \left(\frac{b}{c}\right)^* \equiv \frac{(N-2)}{2(N-4)} - \frac{4(r-1)}{(N-4)c}\left[2(N-1)\phi_{0,1}^{\bm{\xi}[1,2]} + N\delta_{d,0} - 1\right]
\label{eq:(b/c)^*-ring}
\end{equation}	
assuming $N\ge 4$.
Equation~\eqref{eq:(b/c)^*-ring} suggests that $(b/c)^*$ for the coupled ring network can be moved from the value for the one-layer ring, $(N-2)/\left[2(N-4)\right]$, by choosing $r\neq 1$. Equation~\eqref{eqn:thetaphi_ring_network-4} leads to
\begin{equation}
\lim_{N\to\infty} \phi_{0,1}^{\bm{\xi}[1,2]} = - \int_0^1 \frac{\cos(2\pi dx)}{2} \text{d}x = 0.
\label{eq:phi_01-N-infinity}
\end{equation}
By combining Eqs.~\eqref{eq:(b/c)^*-ring} and \eqref{eq:phi_01-N-infinity}, we obtain
\begin{equation}
\lim_{N\to\infty} \left(\frac{b}{c}\right)^* =
\begin{cases}
\frac{1}{2} & \text{if } d > 0,\\
\frac{1}{2} - \frac{4(r-1)}{c} & \text{if } d = 0.
\end{cases}
\end{equation}
Therefore, on large coupled ring networks, the coupling with the constant-selection layer considerably impacts the emergence of cooperation if and only if the initial cooperator and mutant are on the same individual.

Similarly, by substituting Eqs.~\eqref{eqn:thetaphi_ring_network-1}--\eqref{eqn:thetaphi_ring_network-6} in
%
Eq.~\eqref{eqn:constant_drift_selection_criteria-SI}
, we obtain the condition under which selection favors the mutant in layer 2 as follows:
\begin{align}
	0
	&< - (r-1)\theta_{2}^{\bm{\xi}[1]}
	+ c\,\phi_{2,0}^{\bm{\xi}[1,2]}
	+ b\!\left(\phi_{0,1}^{\bm{\xi}[1,2]} - \phi_{2,1}^{\bm{\xi}[1,2]}\right) \notag \\
	&= \frac{N-2}{2}\,(r-1)
	- 2(N-1)\left[ b(2N-1) + c\right] \phi_{0, 1}^{\bm{\xi}[1,2]}
	- N\left[ 2b(N-1) + c \right]\delta_{d,0}
	+ \frac{bN}{2}\,\delta_{d,1}
	+ c + b(2N-3),
\label{eq:ring-constant-final-1-layer-swapped}
\end{align}
which is equivalent to
\begin{align}
r>&1 + \frac{4(N-1)}{N-2}\left[ b(2N-1) + c \right] \phi_{0, 1}^{\bm{\xi}[1,2]}
	+ \frac{2N}{N-2}\left[ 2b(N-1) + c \right]\delta_{d,0}
	- \frac{bN}{N-2}\,\delta_{d,1}
	- \frac{c}{N-2} + \frac{b(2N-3)}{N-2}.
\label{eq:ring-constant-final-2-layer-swapped}
\end{align}
In terms of the original layer labels, with which layer 1 is the donation game layer and layer 2 is the constant-selection dynamics layer,
Eqs.~\eqref{eq:ring-constant-final-1-layer-swapped} and \eqref{eq:ring-constant-final-2-layer-swapped}
read
\begin{align}
	0
	&< - (r-1)\theta_{2}^{\bm{\xi}[2]}
	+ c\,\phi_{0,2}^{\bm{\xi}[1,2]}
	+ b\!\left(\phi_{1,0}^{\bm{\xi}[1,2]} - \phi_{1,2}^{\bm{\xi}[1,2]}\right) \notag \\
	&= \frac{N-2}{2}\,(r-1)
	- 2(N-1)\left[ b(2N-1) + c\right] \phi_{1, 0}^{\bm{\xi}[1,2]}
	- N\left[ 2b(N-1) + c \right]\delta_{d,0}
	+ \frac{bN}{2}\,\delta_{d,1}
	+ c + b(2N-3)
\end{align}
and
\begin{align}
r>&1 + \frac{4(N-1)}{N-2}\left[ b(2N-1) + c \right] \phi_{1, 0}^{\bm{\xi}[1,2]}
	+ \frac{2N}{N-2}\left[ 2b(N-1) + c \right]\delta_{d,0}
	- \frac{bN}{N-2}\,\delta_{d,1}
	- \frac{c}{N-2} + \frac{b(2N-3)}{N-2},
\end{align}
respectively.

\newpage
\clearpage

\section{Evolution of the cooperator and mutant on the coupled complete graph}\label{sec:SI-I}

In this section, we derive the condition for favoring the cooperator and mutant when both layers are the complete graph with $N$ nodes.
We assume the unweighted complete graphs without self-loops. We also assume the initial condition used in
Fig.~\ref{fig:examples}C
in the main text, in which the individual cooperating in layer 1 coincides with the mutant in layer 2.

\subsection{\texorpdfstring{Derivation of $\theta_n^{\bm{\xi}[1]}$}{Derivation of theta_n^{xi[1]}}}

We first compute
\begin{equation}
\theta_n^{\bm{\xi}[1]}
=\sum_{i=1}^{N}\sum_{j=1}^{N}\pi_i^{[1]}\,\left(p^{[1]}\right)^{(n)}_{ij}
\beta^{\bm{\xi}[1]}_{ij},
\label{eq:theta-replicated}
\end{equation}
for the complete graph. The one-step transition probability matrix of the random walk, $P=\left(p^{[1]}_{ij}\right)$, is given by
\begin{equation}
P=\frac{1}{N-1}\,(J-I),
\end{equation}
where $J$ is matrix with all its entries being $1$, and $I$ is the identity matrix. Matrix $P$ has eigenvalues $1$ and $\lambda :=-\frac{1}{N-1}$ with multiplicities $1$ and $N-1$, respectively. The stationary density of the random walk is given by
\begin{equation}
\pi_i^{[1]}=\frac{1}{N}, \quad i\in \{1, \ldots, N \}.
\label{eq:pi-K_N}
\end{equation}
Using $J^2 = N J$ and the fact that $J$ and $I$ commute, we obtain
\begin{align}
	\label{eq:Pn}
P^n =& \frac{1}{(N-1)^n} \sum_{k=0}^n \binom{n}{k} J^k (-I)^{n-k}\notag\\
=& \left(- \frac{1}{N-1} \right)^n I + \frac{J}{(N-1)^n} \sum_{k=1}^n \binom{n}{k} N^{k-1}(-1)^{n-k} \notag\\
=& \frac{1}{N}J + \lambda^n \left( I- \frac{J}{N}\right),
\end{align}
where $\binom{n}{k}$ is the binomial coefficient. By substituting Eqs.~\eqref{eq:pi-K_N} and \eqref{eq:Pn} in Eq.~\eqref{eq:theta-replicated}, we obtain
\begin{align}
	\theta_n^{\bm{\xi}[1]}
	=& \frac{1}{N}\sum_{i,j=1}^N (P^n)_{ij}\,\beta_{ij}^{\bm{\xi}[1]} \notag \\
	=& \frac{1}{N}\sum_{i,j=1}^N \left[  \frac{1}{N}J_{ij}+\lambda^n\left(\delta_{ij}- \frac{1}{N}J_{ij}\right)\right] \beta_{ij}^{\bm{\xi}[1]} \notag \\
	=& \frac{1}{N^2} \sum_{i,j=1}^N J_{ij}\beta_{ij}^{\bm{\xi}[1]}
	+ \frac{\lambda^n}{N}\left( \sum_{i=1}^N \beta_{ii}^{\bm{\xi}[1]} - \sum_{i,j=1}^N \frac{1}{N}J_{ij}\beta_{ij}^{\bm{\xi}[1]} \right) \notag \\
	=& \frac{1}{N^2} \sum_{i,j=1}^N \beta_{ij}^{\bm{\xi}[1]}
	+ \frac{\lambda^n}{N}\left( \sum_{i=1}^N \beta_{ii}^{\bm{\xi}[1]} - \sum_{i,j=1}^N \frac{1}{N} \beta_{ij}^{\bm{\xi}[1]} \right).
\label{eq:theta-complete-2}
\end{align}
To simplify Eq.~\eqref{eq:theta-complete-2}, we note that substitution of $\pi_i = \frac{1}{N}, \forall i$ in Eq.~\eqref{eqn:sum_beta_ii=0} yields
$\sum_{i=1}^N \beta_{ii}^{\bm{\xi}[1]}=0$. By substituting this equality in Eq.~\eqref{eq:theta-complete-2}, we obtain
\begin{align}
	\theta_n^{\bm{\xi}[1]}
	=& \frac{1}{N^2} \sum_{i,j=1}^N \beta_{ij}^{\bm{\xi}[1]}
	+ \frac{\lambda^n}{N}\left( \sum_{i=1}^N \beta_{ii}^{\bm{\xi}[1]} - \sum_{i,j=1}^N \frac{1}{N} \beta_{ij}^{\bm{\xi}[1]} \right) \notag\\
	=& \frac{1}{N^2} \sum_{i,j = 1; i\neq j}^N \beta_{ij}^{\bm{\xi}[1]}
	+ \frac{\lambda^n}{N}\left( 0 - \sum_{i,j = 1; i\neq j}^N \frac{1}{N} \beta_{ij}^{\bm{\xi}[1]} \right) \notag\\
	=& \frac{1-\lambda^n}{N^2} \sum_{i,j = 1; i\neq j}^N \beta_{ij}^{\bm{\xi}[1]} \notag\\
	\equiv& \frac{1-\lambda^n}{N^2} S_{\mathrm{off}}.
\label{eq:theta-complete-3}
\end{align}

We need to obtain $S_{\mathrm{off}}$ from Eq.~\eqref{eqn:beta_recurrence_ij_dB}. To simplify Eq.~\eqref{eqn:beta_recurrence_ij_dB} for the complete graph, we first note that variables $\xi^{[1]}_i$ and $\xi^{[1]}_j$ are $0$ or $1$ depending on whether a defector or cooperator, respectively, occupies the node.
Because we have only one cooperator in our initial condition, both $\xi^{[1]}_i$ and $\xi^{[1]}_j$, where $i\neq j$, cannot be $1$, leading to $\xi^{[1]}_i\xi^{[1]}_j=0$. The weighted reproductive value $\widehat\xi^{[1]} = 1/N$ comes from the definition of the RV-weighted frequency, Eq.~\eqref{eqn:reproductive-value-1}, and Eq.~\eqref{eqn:reproductive-value-2}. Because all the nodes in the complete graph have the same reproductive value due to symmetry and only one node has a cooperator, the RV-weighted frequency is equal to $1/N$. 
Therefore, we obtain
\begin{equation}
\xi^{[1]}_i\xi^{[1]}_j - \hat\xi^{[1]} = -\frac{1}{N}.
\label{eq:beta_ij-RHS-1-complete-graph}
\end{equation}
By substituting Eq.~\eqref{eq:beta_ij-RHS-1-complete-graph}, $p^{[1]}_{ik}=1/(N-1)$ for $k\neq i$, and $p^{[1]}_{ik}=0$ for $k=i$ in 
Eq.~\eqref{eqn:beta_recurrence_ij_dB}, we obtain
\begin{equation}
	\label{eq:beta-def-2}
	\beta_{ij}^{\bm{\xi}[1]}
	=  -\frac{1}{2}
	+ \frac{1}{2}\sum_{k=1; k\neq i}^{N} \frac{\beta_{kj}^{\bm{\xi}[1]}}{N-1}
	+ \frac{1}{2}\sum_{k=1; k\neq j}^{N} \frac{\beta_{ik}^{\bm{\xi}[1]}}{N-1},
	\qquad (i\neq j).
\end{equation}
By summing both sides of Eq.~\eqref{eq:beta-def-2} over the $i$ and $j$ values with $i\neq j$, we obtain
\begin{equation}
S_{\mathrm{off}}
= -\frac{N(N-1)}{2}
+ \frac{1}{2} \frac{N-2}{N-1}\,S_{\mathrm{off}}
+ \frac{1}{2} \frac{N-2}{N-1}\,S_{\mathrm{off}}.
\label{eq:S_all-calc-1}
\end{equation}
Equation~\eqref{eq:S_all-calc-1} leads to
\begin{equation}
S_{\mathrm{off}} = -\frac{N(N-1)^2}{2}.
\label{eq:Sall}
\end{equation}

Substitution Eq.~\eqref{eq:Sall} in Eq.~\eqref{eq:theta-complete-3} yields
\begin{equation}
	\label{eq:theta-final-K_N}
		\theta_n^{\bm{\xi}[1]}
		= -\,\frac{(N-1)^2}{2N}\,\left[1 - \left(-\frac{1}{N-1}\right)^{\!n}\right],
\end{equation}
where $n \in \{1, 2, \ldots \}$.

\subsection{\texorpdfstring{Derivation of $\phi_{n,m}^{\bm{\xi}[1, 2]}$}{Derivation of \phi_{n,m}^{\xi[1, 2]}}}

We now compute
\begin{equation}
	\phi_{n,m}^{\bm{\xi}[1, 2]}
	:= \sum_{i,j=1}^{N} \pi^{[1]}_i\,\left( p^{[1,2]} \right)^{(n,m)}_{ij}\,\gamma_{ij}^{\bm{\xi}[1, 2]}.
\end{equation}
Because the two layers are both complete graph on $N$ nodes, we obtain
\begin{equation}
\left( p^{[1,2]} \right)^{(n,m)} = P^{\,n}P^{\,m} = P^{\,n+m}
\end{equation}
and
\begin{equation}
	\label{eq:phi-def}
	\phi_{n,m}^{\bm{\xi}[1,2]}
	= \sum_{i,j=1}^{N} \pi^{[1]}_i\,\left( p^{[1,2]} \right)^{(n,m)}_{ij}\,\gamma_{ij}^{\bm{\xi}[1,2]}
	= \frac{1}{N}\sum_{i,j} (P^{\,n+m})_{ij}\,\gamma_{ij}^{\bm{\xi}[1,2]},
\end{equation}
where we used $\pi^{[1]}_i = 1/N$.

To compute $\gamma_{ij}^{\bm{\xi}[1, 2]}$, which we write $\gamma_{ij}$ in the remainder of this section for notational simplicity, we set with $\xi^{[1]}_1=\xi^{[2]}_1=1$ and $\xi^{[L]}_j=0$ for $j\neq1$, $L\in\{1,2\}$. This assumption corresponds to the initial condition that we have assumed, in which the sole initial cooperator in layer 1 is the sole initial mutant in layer 2. We obtain
$\widehat{\xi}^{[1]}=\widehat{\xi}^{[2]}=1/N$
and $\vec{\xi}^{[1]}\vec{\xi}^{[2]\top}-\widehat{\xi}^{[1]}\widehat{\xi}^{[2]}J = U-\frac{1}{N^2}J$, where $\vec{\xi}^{[L]} = (\xi_1^{[L]}, \ldots, \xi_i^{[L]}, \ldots,\xi_N^{[L]})^\top$, $U:=\bm{e}_1 \bm{e}_1^\top$, and vector $\bm{e}_1$ is the standard basis vector, i.e., $\bm{e}_1 = (1,0, \ldots, 0)^\top$. The expression of $\gamma_{ij}$ with $i\neq j$, shown in Eq.~\eqref{eqn:gamma_recurrence_ij_dB}, simplifies to
\begin{equation}
	\label{eq:gamma-eq}
	\gamma_{ij}
	=\frac{N^2}{2N-1}\left(U_{ij}-\frac{1}{N^2}\right)
	+\frac{1}{2N-1}(P\gamma P)_{ij}
	+\frac{N-1}{2N-1}\left((P\gamma)_{ij}+(\gamma P)_{ij}\right).
\end{equation}
We also note that $\sum_{i=1}^N \pi_i\gamma_{ii}=0$ simplifies to
\begin{equation}
\frac{1}{N}\sum_{i=1}^N \gamma_{ii}=0.
\label{eq:gamma_ii-sum=0}
\end{equation}
Using Eqs.~\eqref{eq:Pn}, \eqref{eq:phi-def}, and \eqref{eq:gamma_ii-sum=0}, we obtain
\begin{align}
	\phi_{n,m}
	&=\frac{1}{N}\!\left[\sum_{i=1}^N \left(\frac{1}{N}+\frac{N-1}{N}\lambda^{n+m}\right)\gamma_{ii}
	+\sum_{i, j=1; i\neq j}^N \left(\frac{1}{N}-\frac{1}{N}\lambda^{n+m}\right)\gamma_{ij}\right] \notag\\
	&=\frac{1-\lambda^{n+m}}{N^2} \sum_{i, j=1; i\neq j}^N \gamma_{ij} \notag\\
	&\equiv  \frac{1-\lambda^{n+m}}{N^2} \overline{S}_{\mathrm{off}}.	
\label{eq:phi-SA}
\end{align}

To calculate $\overline{S}_{\mathrm{off}}$, we sum Eq.~\eqref{eq:gamma-eq} over all $i$ and $j$ with $i\neq j$. The sum of the first term on the right-hand side of Eq.~\eqref{eq:gamma-eq} becomes
\begin{equation}
\sum_{i, j = 1; i\neq j}^N \frac{N^2}{2N-1}\left(U_{ij}-\frac{1}{N^2}\right)
=\frac{N^2}{2N-1}\left[ 0-\frac{N(N-1)}{N^2}\right] = -\frac{N(N-1)}{2N-1}.
\label{eq:1st-term-gamma-eq-final}
\end{equation}

The sum of the second term on the right-hand side of Eq.~\eqref{eq:gamma-eq} is written as
\begin{equation}
\sum_{i, j = 1; i\neq j}^N (P\gamma P)_{ij}
=\sum_{k,\ell = 1}^N \gamma_{k\ell}\,T(k,\ell),
\label{eq:2nd-term-gamma-eq}
\end{equation}
where
\begin{equation}
T(k,\ell):=\sum_{i, j = 1; i\neq j}^N p_{ik}p_{j\ell}.
\end{equation}
Because
\begin{equation}
p_{ik} = 
\begin{cases}
0 & \text{if } i=k,\\
\frac{1}{N-1} & \text{if } i\neq k,
\end{cases}
\label{eq:p_{ik}-recap}
\end{equation}
$T(k,\ell)$ counts ordered pairs $(i,j)$ with $i\neq j$, $i\neq k$, $j\neq \ell$ with weight $(N-1)^{-2}$. Among the $N(N-1)$ pairs of $(i, j)$ with $i\neq j$, there are $2(N-1)$ pairs of $(i,j)$ satisfying $i=k$ or $j=\ell$, if $k=\ell$. If $k\neq \ell$, there are $2(N-1)-1$ such $(i, j)$ pairs. Therefore, we obtain
\begin{equation}
\#\{(i,j)\}=\begin{cases}
	(N-2)(N-1) & \text{if } k=\ell,\\[2pt]
	(N-2)(N-1)+1 & \text{if } k\neq\ell,
\end{cases}
\end{equation}
where $\#\{(i,j)\}$ is the number of $(i, j)$ pairs counted in $T(k,\ell)$. Because each $(i,j)$ pair contributes weight $(N-1)^{-2}$, we obtain
\begin{equation}
T(k,\ell)=\begin{cases}
	\dfrac{N-2}{N-1} & \text{if } k=\ell,\\[8pt]
	\dfrac{N^2-3N+3}{(N-1)^2} & \text{if } k\neq\ell.
\end{cases}
\label{eq:T(k,ell)-final}
\end{equation}
By substituting Eq.~\eqref{eq:T(k,ell)-final} in Eq.~\eqref{eq:2nd-term-gamma-eq} and using Eq.~\eqref{eq:gamma_ii-sum=0}, we obtain
\begin{equation}
\sum_{i, j = 1; i\neq j}^N (P\gamma P)_{ij}
=
\frac{N^2-3N+3}{(N-1)^2}\,\overline{S}_{\mathrm{off}}.
\label{eq:2nd-term-gamma-eq-final}
\end{equation}

To calculate the sum of the third term on the right-hand side of Eq.~\eqref{eq:gamma-eq}, we first obtain
\begin{align}
	\sum_{i, j = 1; i\neq j}^N (P\gamma)_{ij}
	=& \sum_{i, j = 1; i\neq j}^N \sum_{k=1}^N p_{ik}\gamma_{kj} \notag\\
	=& \sum_{k,j=1}^N \gamma_{kj}\sum_{i=1; i\neq j}^N p_{ik} \notag\\
	=& \sum_{k,j=1}^N \gamma_{kj}\,(1-p_{jk}) \notag\\
	=& \overline{S}_{\mathrm{off}}-\sum_{k,j=1}^N p_{jk}\gamma_{kj} \notag\\
	=& \overline{S}_{\mathrm{off}}-\frac{\overline{S}_{\mathrm{off}}}{N-1} \notag\\
	=& \frac{N-2}{N-1}\overline{S}_{\mathrm{off}}.
\label{eq:3rd-term-gamma-eq}	
\end{align}
We used Eq.~\eqref{eq:gamma_ii-sum=0} to derive the third last equality. We used Eqs.~\eqref{eq:gamma_ii-sum=0} and \eqref{eq:p_{ik}-recap} to derive the second last equality. By symmetry, the same holds true for $\sum_{i\neq j}(\gamma P)_{ij}$. Therefore, we obtain
\begin{equation}
\sum_{i, j = 1; i\neq j}^N \left((P\gamma)_{ij}+(\gamma P)_{ij}\right)
=\frac{2(N-2)}{N-1}\overline{S}_{\mathrm{off}}.
\label{eq:3rd-term-gamma-eq-final}
\end{equation}

By summing Eq.~\eqref{eq:gamma-eq} over all $i, j \in \{1, \ldots, N\}$ with $i\neq j$ and using Eqs.~\eqref{eq:1st-term-gamma-eq-final}, \eqref{eq:2nd-term-gamma-eq-final}, and \eqref{eq:3rd-term-gamma-eq-final},
we obtain
\begin{equation}
\overline{S}_{\mathrm{off}} = -\frac{N(N-1)}{2N-1}
	+\frac{1}{2N-1}\cdot\frac{N^2-3N+3}{(N-1)^2}\overline{S}_{\mathrm{off}}
	+ \frac{N-1}{2N-1}\cdot\frac{2(N-2)}{N-1}\overline{S}_{\mathrm{off}}.
\label{eq:S_off-and-S_diag-for-gamma}
\end{equation}
Equation~\eqref{eq:S_off-and-S_diag-for-gamma} leads to
\begin{equation}
	\label{eq:Soff-final}
	\overline{S}_{\mathrm{off}}=- \frac{(N-1)^3}{2N-3}.
\end{equation}
By substituting Eq.~\eqref{eq:Soff-final} and $\lambda = - 1/(N-1)$ in Eq.~\eqref{eq:phi-SA}, we obtain
\begin{equation}
	\label{eq:phi-final-K_N}
	\phi_{n,m} = -\,\frac{(N-1)^3}{N^2(2N-3)}\,\left[ 1-\left(-\frac{1}{N-1}\right)^{\,n+m}\right].
\end{equation}

\subsection{Condition for favoring the cooperator and mutant}

By substituting Eqs.~\eqref{eq:theta-final-K_N} and \eqref{eq:phi-final-K_N} in
Eq.~\eqref{eqn:cooperation_criteria_reduced}
%
%
in the main text, we find that spite is favored on the coupled complete graph if
\begin{equation}
\frac{b}{c} < \left( \frac{b}{c} \right)^* = -(N-1) + \frac{r-1}{c} \cdot \frac{2(N-1)^2}{N(2N-3)}.
\label{eq:spite-condition-K_N}
\end{equation}
When $N$ is large, the first term on the right-hand of Eq.~\eqref{eq:spite-condition-K_N} dominates the second term such that the coupling the two complete graphs does not alter the condition for spite substantially. However, the effect is large when $N$ is small.

The condition for the selection of the mutant,
Eq.~\eqref{eqn:constant_drift_selection_criteria}
%
%
in the main text, reads
\begin{equation}
A = -\frac{2(N-1)}{N(2N-3)} \left( \frac{b}{N-1} + c \right).
\end{equation}
This result implies that $r^*$ is substantially different from $1$ if and only if $N$ is small. We also find that the effect of $c$ on $r^*$ is $N-1$ times larger than that of $b$.

\newpage
\clearpage

\section{Evolution of the cooperator and mutant on the coupled star graph}\label{sec:SI-J}

\begin{figure}
\centering
\includegraphics[width=0.5\textwidth]{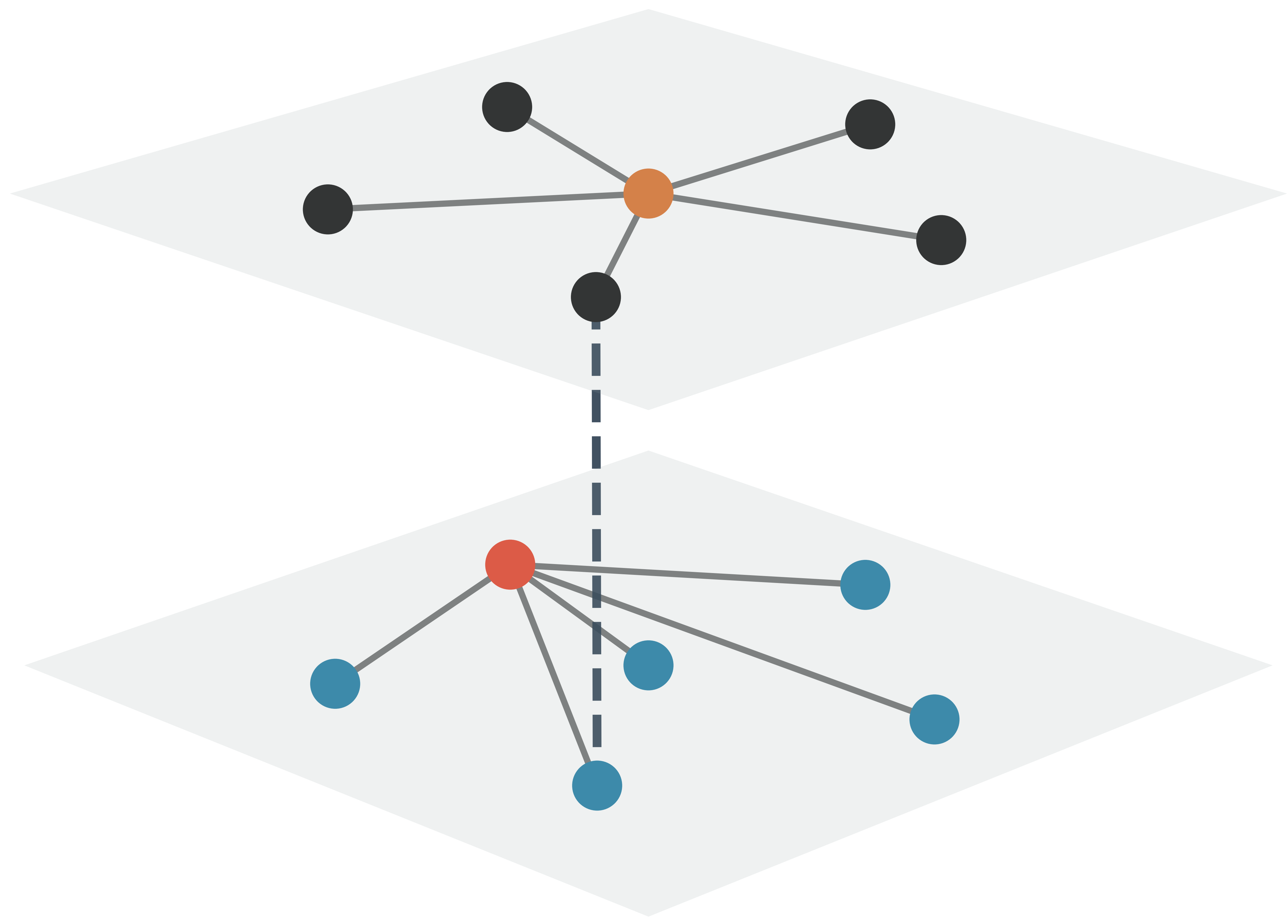}
\caption{Two-layer star graph and the initial condition that we analyze. Note that the hub node in the first layer is not the hub node in layer 2. The node colors are the same as those in
Fig.~\ref{fig:general_rule_multilayer}
in the main text; orange represents the cooperator in layer 1, black represents the defector layer 1, blue represents the resident in layer 2, and red represents the mutant in layer 2. The two replica nodes connected by the dashed line represent the same individual, shown as an example.}
\label{fig:schem-double-star}
\end{figure}

We derive the condition for favoring the cooperator and mutant in two-layer star graphs under the dB-dB rule. See Fig.~\ref{fig:schem-double-star} for the network; we focus on this particular two-layer star graph and initial condition with varying $N$, as in the previous study~\cite{su2022evolution} (see SI Fig.~16 of their paper). The expressions for $\theta_i^{\bm{\xi}[1]} $ and $\phi_{m,n}^{\bm{\xi}[1,2]}$ were derived \cite{su2022evolution} as follows:
\begin{align}
	\theta_{1}^{\bm{\xi}[1]} &= -\frac{2N-3}{2(N-1)}, \label{eq:coupled-star-theta1}\\
	\theta_{2}^{\bm{\xi}[1]} &= -\frac{3(N-2)}{4(N-1)}, \label{eq:coupled-star-theta2}\\
	\theta_{3}^{\bm{\xi}[1]} &= -\frac{2N-3}{2(N-1)}, \label{eq:coupled-star-theta3}\\
	\phi_{0,1}^{\bm{\xi}[1,2]} &= -\frac{N\!\left(8N^{5}-52N^{4}+112N^{3}-107N^{2}+46N-8\right)}
	{8(2N-1)(3N-2)(N-1)^{4}}, \label{eq:coupled-star-phi1}\\ 
	\phi_{2,0}^{\bm{\xi}[1,2]} &= -\frac{N^{2}(N-2)}{2(2N-1)(3N-2)(N-1)^{2}}, \label{eq:coupled-star-phi2}\\ 
	\phi_{2,1}^{\bm{\xi}[1,2]} &= -\frac{N\!\left(8N^{5}-44N^{4}+84N^{3}-79N^{2}+38N-8\right)}
		{8(2N-1)(3N-2)(N-1)^{4}} . \label{eq:coupled-star-phi3}
\end{align}
Note that $\theta_{1}^{\bm{\xi}[1]}=\theta_{3}^{\bm{\xi}[1]}$ because the star graph is a special case of the complete bipartite graph. By substituting Eqs.~\eqref{eq:coupled-star-theta1}--\eqref{eq:coupled-star-phi3} in
Eq.~\eqref{eqn:cooperation_criteria_reduced}
%
%
in the main text, we obtain the condition under which selection favors the cooperator in layer 1 as follows:
\begin{align}
	0
	&< c\,\theta_{2}^{\bm{\xi}[1]}
	+ b\left( \theta_{1}^{\bm{\xi}[1]} - \theta_{3}^{\bm{\xi}[1]}\right)
	- (r-1)\,\phi_{2,0}^{\bm{\xi}[1,2]} \notag \\
	&= -\frac{3c(N-2)}{4(N-1)}
	+ \frac{N^{2}(N-2)(r-1)}{2(2N-1)(3N-2)(N-1)^{2}} \notag \\
	&= \frac{(N-2)\left[ 2N^{2}(r-1)
		- 3c\,(N-1)(2N-1)(3N-2)\right]}
	{4(N-1)^{2}(2N-1)(3N-2)}.
\label{eq:(b/c)^*-two-layer-star}
\end{align}
For $N>2$, Eq.~\eqref{eq:(b/c)^*-two-layer-star} is equivalent to
\begin{equation}
	2N^{2}(r-1) - 3c\,(N-1)(2N-1)(3N-2) > 0.
\end{equation}
Remarkably, the condition does not depend on $b$.

Similarly, by substituting Eqs.~\eqref{eq:coupled-star-theta1}--\eqref{eq:coupled-star-phi3}  in
%
Eq.~\eqref{eqn:constant_drift_selection_criteria-SI}
, we obtain the condition under which selection favors the mutant in the constant-selection layer as follows:
\begin{align}
	0
	&< - (r-1)\,\theta_{2}^{\bm{\xi}[1]}
	+ c\,\phi_{2,0}^{\bm{\xi}[1,2]}
	+ b\left(\phi_{0,1}^{\bm{\xi}[1,2]} - \phi_{2,1}^{\bm{\xi}[1,2]} \right) \notag \\
	&= \frac{3(N-2)(r-1)}{4(N-1)}
	- c\left[
	\frac{N^{2}(N-2)}{2(2N-1)(3N-2)(N-1)^{2}}
	\right] \notag \\
	&\quad
	+ b\left[
	-\frac{N\!\left(8N^{5} - 52N^{4} + 112N^{3} - 107N^{2} + 46N - 8\right)}
	{8(2N-1)(3N-2)(N-1)^{4}} \right. \notag \\
	&\quad + \left. \frac{N\!\left(8N^{5} - 44N^{4} + 84N^{3} - 79N^{2} + 38N - 8\right)}
	{8(2N-1)(3N-2)(N-1)^{4}}
	\right] \notag \\
	&= \frac{(N-2)\left\{
		2N^{2}\left[ b(2N-1) - c(N-1)\right]
		+ 3(N-1)^{2}(2N-1)(3N-2)(r-1)
		\right\}}
	{4(N-1)^{3}(2N-1)(3N-2)}.
\label{eq:r^*-two-layer-star}
\end{align}
For $N>2$, Eq.~\eqref{eq:r^*-two-layer-star} is equivalent to
\begin{equation}
r > 1 - \frac{2N^{2}\left[ b(2N-1) - c(N-1)\right]}{3(N-1)^{2}(2N-1)(3N-2)} .
\end{equation}
We note that the mathematical expression right after the inequality in the first line of
Eq.~\eqref{eq:r^*-two-layer-star} uses the swapped layer labels, with which the constant-selection dynamics occurs in layer 1 and the donation game occurs in layer 2. In terms of the original layer labels, with which the donation game layer occurs in layer 1 and the constant-selection dynamics occurs in layer 2,
the same quantity is expressed as 
$- (r-1)\,\theta_{2}^{\bm{\xi}[2]}
	+ c\,\phi_{0, 2}^{\bm{\xi}[1,2]}
	+ b\left(\phi_{1, 0}^{\bm{\xi}[1,2]} - \phi_{1, 2}^{\bm{\xi}[1,2]} \right)$.

\newpage
\clearpage

\section{Mean and standard deviation of $\left| \frac{\text{d}(b/c)^*}{\text{d}r} \right|$ for two-layer ER and BA networks with $N=15$ individuals}\label{sec:SI-K}

\begin{figure}[b]
\centering
\includegraphics[width=0.9\textwidth]{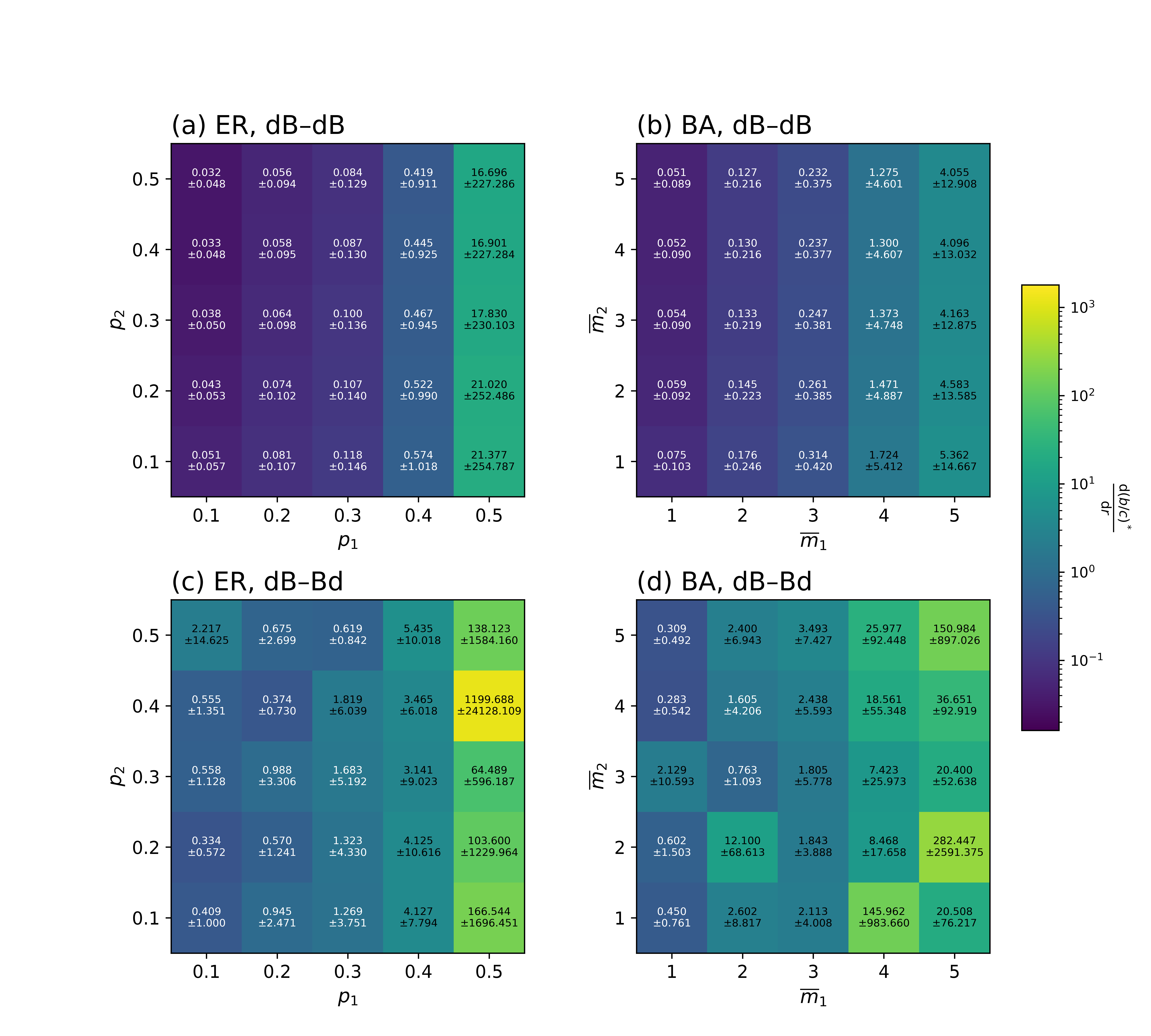}
\caption{Mean, $\mu$, and standard deviation, $\sigma$, of $\left| \text{d}(b/c)^*/\text{d}r \right|$ for two-layer ER and BA networks with $N=15$ individuals. 
(a) ER, dB-dB rule. (b) BA, dB-dB rule. (c) ER, dB-Bd rule. (d) BA, dB-Bd rule. For each ($p_1$, $p_2$) or ($\overline{m}_1$, $\overline{m}_2$), we computed $\mu$ and $\sigma$ on the basis of all the pairs of two-layer network and initial condition for which cooperation can be favored (i.e., $(b/c)^* > 0$ for the layer-1 network).}
\label{fig:mean_std_N=15}
\end{figure}

We showed in
Fig.~\ref{fig:tables_random_graphs}
in the main text the median and the 5th and 95th percentiles of $\left| \text{d}(b/c)^*/\text{d}r \right|$ for two-layer ER and BA networks with $N=15$ individuals under the dB-dB and dB-Bd updating rules. Figure~\ref{fig:mean_std_N=15} shows the mean and standard deviation of $\left| \text{d}(b/c)^*/\text{d}r \right|$ in the corresponding cases. We find that, while the mean does not behave monotonically with respect to the network density parameters (i.e., $p_1$, $p_2$, $\overline{m}_1$, and $\overline{m}_2$), the main trend that $\left| \text{d}(b/c)^*/\text{d}r \right|$ increases as the density of edges in layer 1 (i.e., $p_1$ or $\overline{m}_1$) increases remains the same as in
Fig.~\ref{fig:tables_random_graphs}.

\newpage
\clearpage

\section{Numerical simulations of the stochastic evolutionary dynamics}\label{sec:SI-L}

To validate our theoretical prediction, we carry out direct numerical simulations of the evolutionary dynamics in two-layer networks. For each combination of two-layer network, $b/c$, $r$, updating rule, and strength of selection, we use a particular initial condition, run simulations, and estimate the fixation probability for cooperation as the fraction of runs in which cooperation has fixated.

\begin{figure}
	\centering
	\includegraphics[width=0.95\textwidth]{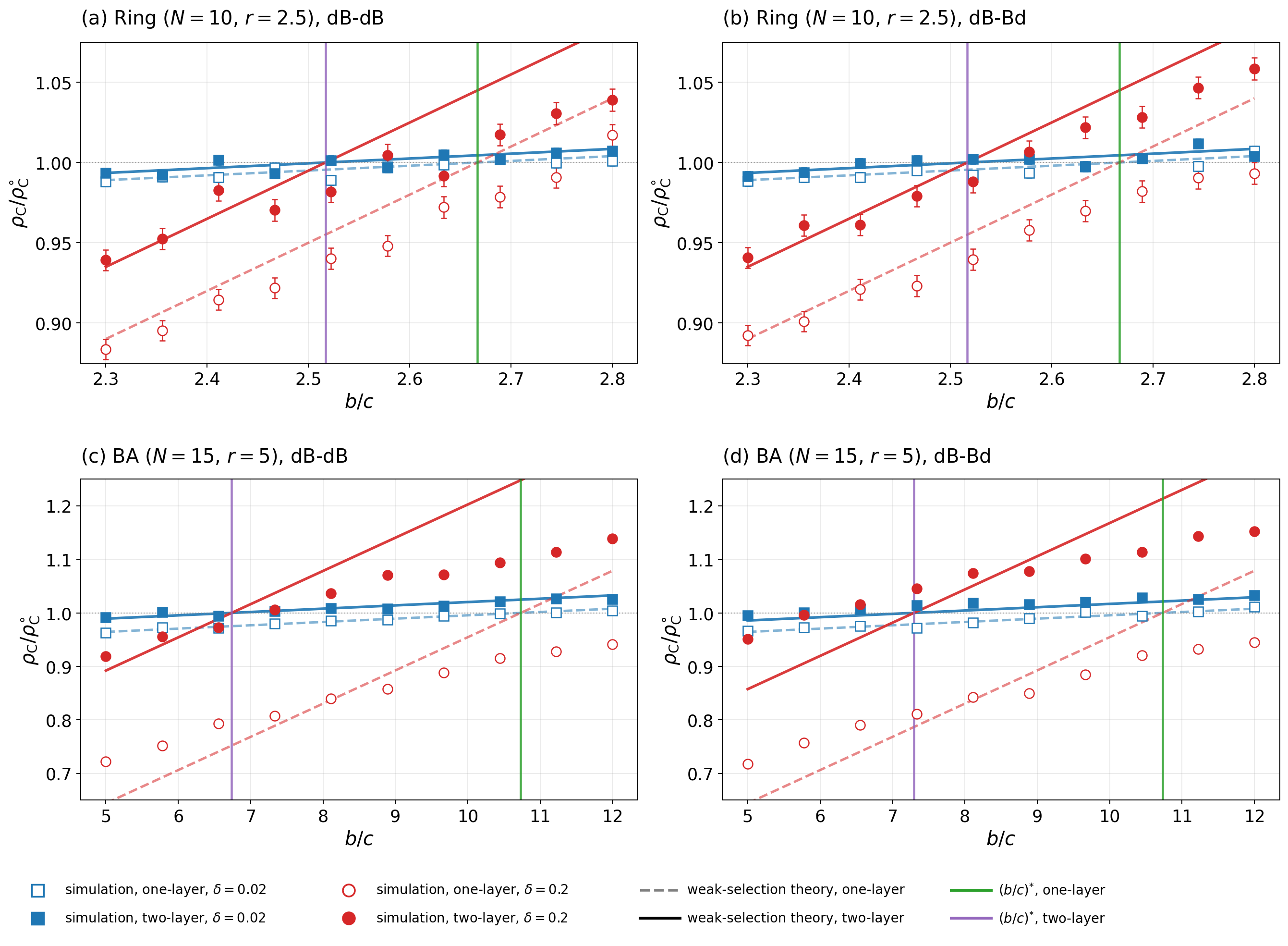}
	\caption{Fixation probability for cooperation compared between the weak selection theory and direct numerical simulations on the ring network with $N=10$ and a BA network with $N=15$. (a) Ring, $N=10$, $r=2.5$, dB-dB. (b) Ring, $N=10$, $r=2.5$, dB-Bd. (c) BA, $N=15$, $r=5$, dB-dB. (d) BA, $N=15$, $r=5$, dB-Bd.
The vertical lines represent the $(b/c)^*$ value from our weak selection theory.
We obtained the theoretical lines for $\delta \in \{ 0.02, 0.2 \}$ by substituting the $\delta$ value in Eq.~\eqref{eqn:cooperation_criteria_reduced} in the main text.
The two-layer ring network is the same as the one used in Fig.~\ref{fig:examples}A in the main text. The results for the one-layer networks are the fixation probability for layer 1 separately considered.
%
%
We set $\overline{m}_1 = \overline{m}_2 = 2$ for the BA network. We use a single instance of two-layer BA network with a fixed initial condition, which is shared between (c) and (d); this is also the case for to the two-layer ER and BA networks used in Figs.~\ref{fig:N=100} and \ref{fig:direct-large}. We used one end node in the initial three-node chain in layer 1 as the location of the initial cooperator in the BA network. In all networks including the BA network, the initial cooperator in layer 1 and the initial mutant in layer 2 are the same individual. We use the same initial conditions in Figs.~\ref{fig:N=100} and \ref{fig:direct-large}.
On the vertical axis, $\rho_{\text{C}}$ is the fixation probability for a single cooperator; $\rho_{\text{C}}^{\circ}$ is the reproductive value of the node (denoted by $v$) on which the single cooperator is initially located (i.e., fixation probability of a neutral mutant placed on $v$). Because we are considering undirected networks, $\rho_{\text{C}}^{\circ}$ is equal to the degree of $v$ divided by the sum of the degree over all the nodes \cite{sood2008voter, allen2017evolutionary}. To numerically compute the fixation probability, we ran $10^6$ and $5\times 10^5$ simulations for the ring and BA networks, respectively, at each $b/c$ value for $\delta = 0.02$. For $\delta = 0.2$, we ran $2\times 10^5$ and $10^5$ simulations for the ring and BA networks, respectively. The number of runs is larger for $\delta = 0.02$ because a weaker selection strength necessitates more runs for accurate estimation of the fixation probability.}
	\label{fig:direct-small}
\end{figure}

We compare the fixation probability between the theory and direct numerical simulations in Fig.~\ref{fig:direct-small} for the two-layer ring network with $N=10$ individuals used in Fig.~\ref{fig:examples}A in the main text and a two-layer BA network with $N=15$ individuals. We set $r=2.5$ and $r=5$ for the ring and BA networks, respectively. We find that the numerically estimated fixation probability closely agrees with the theoretical prediction across $b/c$ values for selection strength $\delta = 0.02$. When $\delta = 0.2$, we observe considerable discrepancies between the theoretical and numerical results, which is expected. For both $\delta$ values, coupling with the constant-selection network layer promotes cooperation relative to the one-layer network.

\begin{figure}
	\centering
	\includegraphics[width=0.7\textwidth]{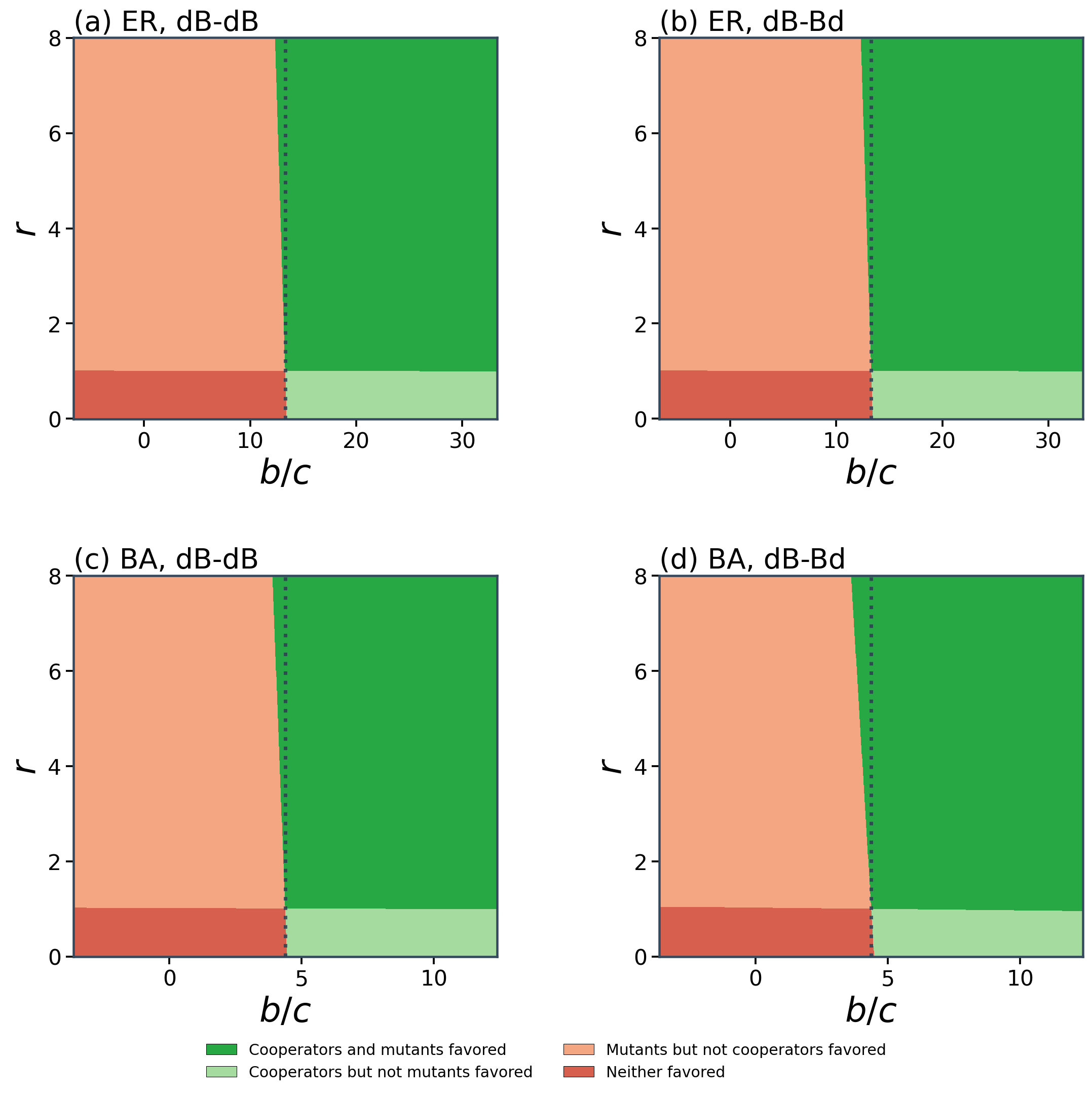}
	\caption{Parameter regions in which the cooperator or mutant is selected in two-layer networks with $N=100$ individuals.
The vertical lines represent the $(b/c)^*$ value for the uncoupled layer-1 network. (a) ER, dB-dB. (b) ER, dB-Bd. (c) BA, dB-dB. (d) BA, dB-Bd. We set $p_1 = p_2 = 0.1$ for the ER network and $\overline{m}_1 = \overline{m}_2 = 2$ for the BA network.}
	\label{fig:N=100}
\end{figure}

\begin{figure}
	\centering
	\includegraphics[width=0.95\textwidth]{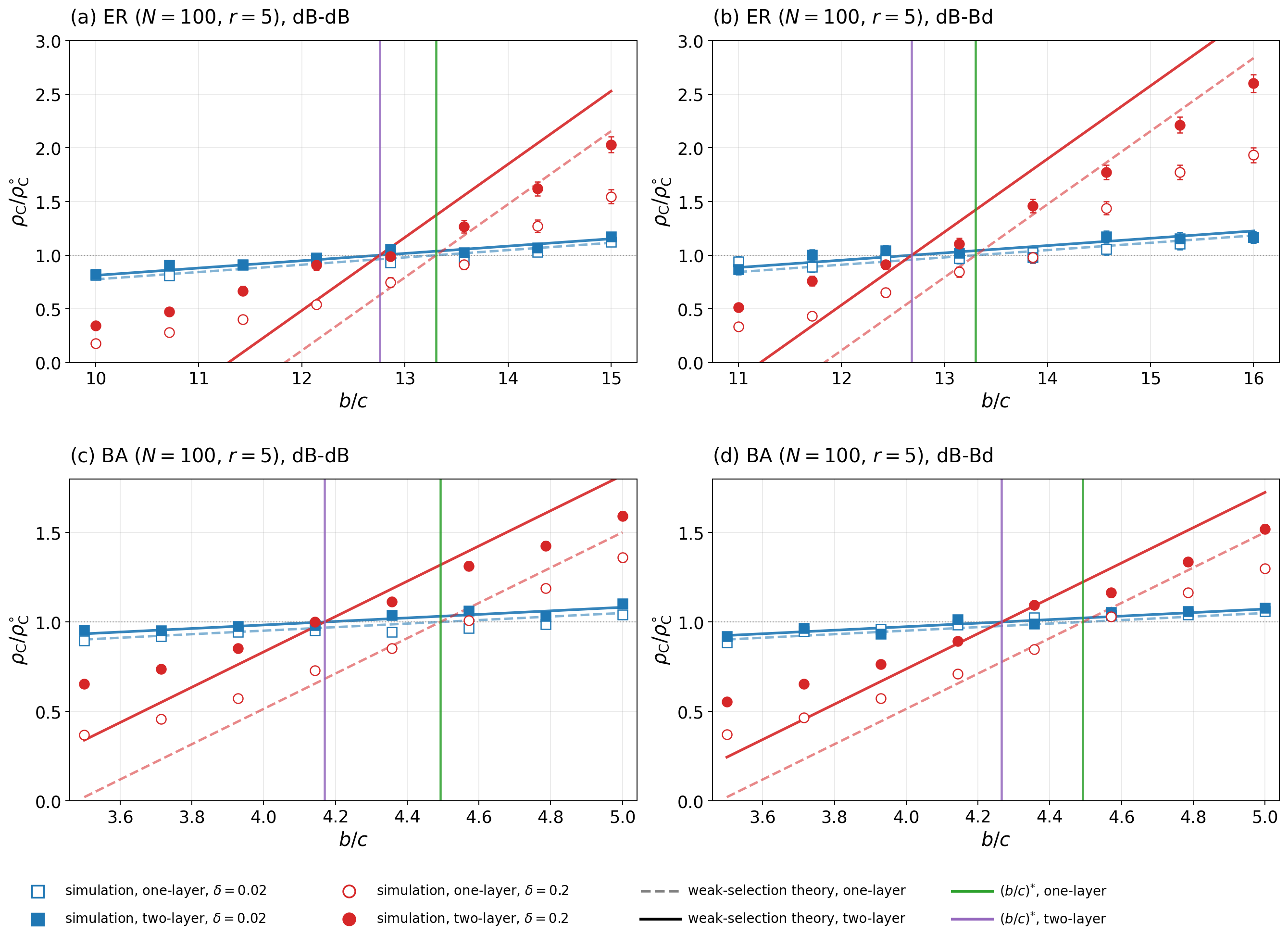}
	\caption{Fixation probability for cooperation compared between the weak selection theory and direct numerical simulations on networks with $N=100$. (a) ER, $N=100$, dB-dB. (b) ER, $N=100$, dB-Bd. (c) BA, $N=100$, dB-dB. (d) BA, $N=100$, dB-Bd. We set $r=5$. See the caption of Fig.~\ref{fig:direct-small} for legends. To numerically compute the fixation probability, we ran $2\times 10^5$ and $3\times 10^5$ simulations for the ER and BA networks, respectively, at each $b/c$ value for $\delta = 0.02$ under the dB-dB rule. These cases yielded noisier evolutionary dynamics such that we used a larger number of runs than in the other cases. In all other cases, we ran $5\times 10^4$ simulations.}
	\label{fig:direct-large}
\end{figure}

In the main text, we have mainly confined ourselves to small networks up to $N=15$ individuals. Therefore, we have also examined the fixation probability for two-layer ER and BA networks with $N=100$ individuals. We first confirmed that the theoretical results for the fixation probability for cooperators and mutants show a typical pattern that small networks show, for both dB-dB and dB-Bd updating rules (Fig.~\ref{fig:N=100}). In particular, $(b/c)^*$ is smaller (i.e., prosociality is promoted) with two coupled network layers if $r>1$ for both networks and both updating rules. We then compared the fixation probability obtained by the weak selection theory and direct numerical simulations. The results, shown in Fig.~\ref{fig:direct-large}, are qualitatively the same as those for smaller networks whose results are shown in Fig.~\ref{fig:direct-small}.

\newpage
\clearpage

\section{``Fixed benefits, fixed costs'' goods scheme}\label{sec:SI-M}\label{sec:fixed-benefits-and-fixed-costs-payoff-scheme}

We analyze the case of the ff goods scheme \cite{mcavoy2020social}. We first derive the conditions under which the cooperator and mutant are favored under the ff goods scheme. Then, we numerically investigate the conditions for the two-layer networks used in the main text.

\subsection{Computation of $(b/c)^*$ and $r^*$}

We recall that, under the pf goods analyzed in the main text, the payoff of the $i$th individual from layer 1 is given by
\begin{equation}
	u_i^{[1]} = -cx_{i}^{\left[1\right]}+\sum_{j=1}^{N} b\frac{w_{ij}^{[1]}}{s_{i}^{[1]}} x_j^{\left[1\right]}.
\label{eq:u_i-pf}	
\end{equation}
Under the ff goods, it changes to
\begin{equation}
	u_i^{[1]} = -cx_{i}^{\left[1\right]}+\sum_{j=1}^{N} b \frac{w_{ij}^{[1]}}{s_{j}^{[1]}} x_j^{\left[1\right]}.
	\label{eq:u_i-ff}	
\end{equation}
The payoff from layer 2, i.e., $u_i^{[2]} = x_i^{\left[2\right]} \left(r-1\right)+1$, remains unchanged.

We rewrite Eq.~\eqref{eq:u_i-ff} as
\begin{align}
	u_k^{\left[1\right]}(\bm{x}) &= \sum_{\ell=1}^{N} \left( -c \cdot p_{k\ell}^{[1]} \cdot x_k^{\left[1\right]} + b \cdot \frac{w_{k\ell}^{[1]}}{s_{\ell}^{[1]}} \cdot x_{\ell}^{\left[1\right]} \right)\notag \\ \notag
	&= \sum_{\ell=1}^{N} \left( -c \cdot p_{k\ell} \cdot x_k^{\left[1\right]} + b \cdot \frac{w_{\ell k}^{[1]}}{s_{\ell}^{[1]}} \cdot x_{\ell}^{\left[1\right]} \right) \\ \notag
	&= \sum_{\ell=1}^{N} \left( -c \cdot p_{k\ell}^{[1]} \cdot x_k^{\left[1\right]} + b \cdot p_{\ell k}^{[1]} \cdot x_{\ell}^{\left[1\right]} \right) \\ 
	&= \sum_{\ell=1}^{N} \left( -C_{k\ell} \cdot x_k^{\left[1\right]} + B_{k\ell } \cdot x_{\ell}^{\left[1\right]} \right).
\end{align}
Then, Eq.~\eqref{eqn:fixn_beta_gamma} becomes 
\begin{align}
	\left.\frac{\text{d}}{\text{d}\delta}\rho_{A}^{\left[1\right]}\left(\bm{\xi}\right)\right|_{\delta=0}= 
	\sum_{i,j,k=1}^{N} \pi_{i}^{[1]} m_{k;ji}^{[1]} 
	\left\{ -\left[\sum_{\ell=1}^{N} \left(\beta _{jk}^{\bm{\xi}[1]}-\beta_{ik}^{\bm{\xi}[1]}\right)C_{k\ell} - \left(\beta_{jl}^{\bm{\xi}[ 1]}-\beta_{il}^{\bm{\xi}[ 1]}\right)B_{ k\ell}\right] \right. \notag \\
	+ \left. \left[\left(\gamma_{jk}^{\bm{\xi}[1,2]}-\gamma_{ik}^{\bm{\xi}[1,2]}\right)\left(r-1\right)+\left(
	\eta_{j}^{\bm{\xi}[1]}-\eta_{i}^{\bm{\xi}[1]}\right)\right] \right\}.
	\label{eq:cooperation-condition-in-rho_A-ff-goods}
\end{align}
Note that the only difference between Eqs.~\eqref{eqn:fixn_beta_gamma} and 
\eqref{eq:cooperation-condition-in-rho_A-ff-goods} is the use of $B_{k\ell}$ instead of $B_{\ell k}$.
Therefore, the condition for the cooperator to be selected in layer 1, i.e.,
\begin{equation}
	\left.\frac{\text{d}}{\text{d}\delta}\rho_{A}^{\left[1\right]}\left(\bm{\xi}\right)\right|_{\delta=0} >0,
\end{equation}
is given by
\begin{align}
\sum_{i,j,k=1}^{N} \pi_{i}^{[1]} m_{k;ji}^{[1]} 
	\left\{- \left[\sum_{\ell=1}^{N} \left(\beta_{jk}^{\bm{\xi}[1]} - \beta_{ik}^{\bm{\xi}[1]}\right)C_{k\ell} - \left(\beta_{jl}^{\bm{\xi}[1]} - \beta_{il}^{\bm{\xi}[1]}\right)B_{k\ell  }\right] \right. \notag \\
	+ \left. \left[\left(\gamma_{jk}^{\bm{\xi}[1,2]} - \gamma_{ik}^{\bm{\xi}[1,2]}\right)\left(r-1\right) + \left(\eta_{j}^{\bm{\xi}[1]} - \eta_{i}^{\bm{\xi}[1]}\right)\right]\right\} > 0.
\label{eq:ff-goods-cnd-1}
\end{align}
By substituting Eq.~\eqref{eqn:marginalderivation}  in Eq.~\eqref{eq:ff-goods-cnd-1}, we obtain
\begin{align}
	&\sum_{i=1}^{N} \pi_{i}^{\left[1\right]} \sum_{\ell=1}^{N} \left[
	-\beta_{ii}^{\bm{\xi}\left[1\right]} C_{i\ell} + \beta_{i\ell}^{\bm{\xi}\left[1\right]} B_{i\ell} 
	+ \gamma_{ii}^{\bm{\xi}\left[1,2\right]} \left( r-1 \right) + \eta_{i}^{\bm{\xi}\left[1\right]}
	\right] > \nonumber \\
	&\sum_{i,j=1}^{N} \pi_{i}^{\left[1\right]} \left( p^{\left[1\right]} \right)_{ij}^{\left(2\right)} \sum_{\ell=1}^{N} \left[
	-\beta_{ij}^{\bm{\xi}\left[1\right]} C_{jl} + \beta_{il}^{\bm{\xi}\left[1\right]} B_{j\ell} 
	+ \gamma_{ij}^{\bm{\xi}\left[1,2\right]} \left( r-1 \right) + \eta_{i}^{\bm{\xi}\left[1\right]}
	\right].
\label{eq:ff-goods-cnd-2}
\end{align}
By substituting $C_{kl}=c p_{kl}^{\left[1\right]}$ and $B_{lk}=b p_{lk} ^{\left[1\right]}$ in Eq.~\eqref{eq:ff-goods-cnd-2}, we obtain
\begin{align}
	&\sum_{i=1}^{N} \pi_{i}^{\left[1\right]} \sum_{\ell=1}^{N} \left[
	-\beta_{ii}^{\bm{\xi}\left[1\right]} c p_{i\ell}^{\left[1\right]} + \beta_{i\ell}^{\bm{\xi}\left[1\right]} b p_{\ell i}^{\left[1\right]} 
	+ \gamma_{ii}^{\bm{\xi}\left[1,2\right]} \left( r-1 \right) + \eta_{i}^{\bm{\xi}\left[1\right]}
	\right] >\nonumber \\
	&\sum_{i,j=1}^{N} \pi_{i}^{\left[1\right]} \left( p^{\left[1\right]} \right)_{ij}^{\left(2\right)} \sum_{\ell=1}^{N} \left[
	-\beta_{ij}^{\bm{\xi}\left[1\right]} c p_{j\ell}^{\left[1\right]} + \beta_{i\ell}^{\bm{\xi}\left[1\right]} b p_{\ell j}^{\left[1\right]} 
	+ \gamma_{ij}^{\bm{\xi}\left[1,2\right]} \left( r-1 \right) + \eta_{i}^{\bm{\xi}\left[1\right]}
	\right].
\label{eqn:cooperation_condition_ff}
\end{align}
By substituting Eqs.~\eqref{eq:def-theta} and \eqref{eq:def-phi} in Eq.~\eqref{eqn:cooperation_condition_ff}, we obtain  
\begin{align}\label{eqn:cooperation_condition_ff_reduced}
	b\left\{\sum_{i=1}^{N} \pi_{i}^{\left[1\right]} \sum_{\ell=1}^{N} \beta_{i\ell}^{\bm{\xi}\left[1\right]}  p_{\ell i}^{\left[1\right]}
	- \sum_{i,j=1}^{N} \pi_{i}^{\left[1\right]} \left( p^{\left[1\right]} \right)_{ij}^{\left(2\right)} \sum_{\ell=1}^{N} 
	\beta_{i\ell}^{\bm{\xi}\left[1\right]} p_{\ell j}^{\left[1\right]} 
	\right\}+	c\theta_{2}^{\bm{\xi}\left[1\right]} - \left(r-1\right) \phi_{2,0}^{\bm{\xi}\left[1,2\right]}  > 0.
\end{align}
Equation~\eqref{eqn:cooperation_condition_ff_reduced} holds true for the game layer (i.e., layer 1) under the dB rule no matter whether the constant-selection layer (i.e., layer 2) obeys the dB or Bd rule. Under the dB-dB rule, the recurrence equations for computing $\beta_{ii}^{\bm{\xi}\left[1\right]},\beta_{ij}^{\bm{\xi}\left[1\right]}$ (with $i \neq j$), $\gamma_{ij}^{\bm{\xi}\left[1,2\right]}$, and $\eta_{i}^{\bm{\xi}\left[1\right]}$ are Eqs.~\eqref{eqn:beta_recurrence_ii_dB}, \eqref{eqn:beta_recurrence_ij_dB}, \eqref{eqn:gamma_recurrence_ij_dB}, and \eqref{eqn:eta_recurrence_dB}, respectively. Under the dB-Bd rule, the recurrence equations are Eqs.~\eqref{eqn:beta_recurrence_ii_dB}, \eqref{eqn:beta_recurrence_ij_dB}, \eqref{eqn:gamma_recurrence_Bd_reduced}, and \eqref{eqn:eta_recurrence_dB}.

The condition under which the mutant is selected in layer 2 (i.e., constant-selection layer), i.e.,
\begin{equation}
	\left.\frac{\text{d}}{\text{d}\delta}\rho_{A}^{\left[2\right]}\left(\bm{\xi}\right)\right|_{\delta=0} >0,
\end{equation}
is given by
\begin{align}\label{eqn:cooperation_condition_ff_constant_selection}
	\sum_{i,j,k=1}^{N} \pi_{i}^{[2]} m_{k; ji}^{[2]} 
	\left\{- \left[\sum_{\ell=1}^{N} \left(\gamma_{jk}^{\bm{\xi}[2,1]} - \gamma_{ik}^{\bm{\xi}[2,1]}\right)C_{k\ell} - \left(\gamma_{jl}^{\bm{\xi}[2,1]} - \gamma_{il}^{\bm{\xi}[2,1]}\right)B_{k\ell  }\right] \right. \notag \\
	+ \left. \left[\left(\beta_{jk}^{\bm{\xi}[2]} - \beta_{ik}^{\bm{\xi}[2]}\right)\left(r-1\right) + \left(\eta_{j}^{\bm{\xi}[1]} - \eta_{i}^{\bm{\xi}[1]}\right)\right]\right\} > 0.
\end{align}
For the dB-dB updating rule, by substituting Eqs.~\eqref{eq:def-theta} and \eqref{eq:def-phi} in Eq.~\eqref{eqn:cooperation_condition_ff_constant_selection}, we obtain  
\begin{align}\label{eqn:cooperation_condition_ff_constant_selection_reduced}
	&\sum_{i=1}^{N} \pi_{i}^{\left[2\right]} \sum_{\ell=1}^{N} \left[
	-\gamma_{ii}^{\bm{\xi}\left[2,1\right]} c p_{i\ell}^{\left[1\right]} + \gamma_{i\ell}^{\bm{\xi}\left[2,1\right]} b p_{\ell i}^{\left[1\right]} 
	+ \beta_{ii}^{\bm{\xi}\left[2\right]} \left( r-1 \right) + \eta_{i}^{\bm{\xi}\left[2\right]}
	\right] >\nonumber \\
	&\sum_{i,j=1}^{N} \pi_{i}^{\left[2\right]} \left( p^{\left[2\right]} \right)_{ij}^{\left(2\right)} \sum_{\ell=1}^{N} \left[
	-\gamma_{ij}^{\bm{\xi}\left[2,1\right]} c p_{j\ell}^{\left[1\right]} + \gamma_{i\ell}^{\bm{\xi}\left[2,1\right]} b p_{\ell j}^{\left[1\right]} 
	+ \beta_{ij}^{\bm{\xi}\left[2\right]} \left( r-1 \right) + \eta_{i}^{\bm{\xi}\left[2\right]}
	\right].
\end{align}
For the dB-Bd updating rule, we similarly obtain  
\begin{align}\label{eqn:dB-Bd-ff-selection-criteria-reduced}
	&\sum_{i,j=1}^{N} \pi_i^{[2]} p_{ji}^{[2]} \sum_{\ell=1}^{N}
	\left[-\gamma_{ii}^{\bm{\xi}\left[2,1\right]} cp_{i\ell}^{[1]} + \gamma_{i\ell}^{\bm{\xi}\left[2,1\right]}b p_{i\ell}^{[1]}
	+ \beta_{ii}^{\bm{\xi}\left[2\right]}\left(r-1\right) + \eta_{i}^{\bm{\xi}\left[2\right]} \right] > \nonumber \\
	&\sum_{i,j=1}^{N} \pi_i^{[2]} p_{ji}^{[2]} \sum_{\ell=1}^{N} \left[
	-\gamma_{ij}^{\bm{\xi}\left[2,1\right]} cp_{j\ell}^{[1]} + \gamma_{i\ell}^{\bm{\xi}\left[2,1\right]} bp_{j\ell}^{[1]}
	+ \beta_{ij}^{\bm{\xi}\left[2\right]}\left(r-1\right) + \eta_{i}^{\bm{\xi}\left[2\right]} \right].
\end{align}
Under the dB-dB rule, the recurrence equations for computing $\beta_{ii}^{\bm{\xi}\left[1\right]}$, $\beta_{ij}^{\bm{\xi}\left[1\right]}$ (with $i\neq j$), $\gamma_{ij}^{\bm{\xi}\left[1,2\right]}$, and $\eta_{i}^{\bm{\xi}\left[1\right]}$ are Eqs.~\eqref{eqn:beta_recurrence_ii_dB}, \eqref{eqn:beta_recurrence_ij_dB}, \eqref{eqn:gamma_recurrence_ij_dB}, and \eqref{eqn:eta_recurrence_dB}, respectively. Under the dB-Bd rule, the recurrence equations are Eqs.~\eqref{eqn:beta_ii_recurrence_Bd_reduced}, \eqref{eqn:beta_ii_recurrence_Bd_reduced}, \eqref{eqn:gamma_recurrence_Bd_reduced}, and \eqref{eqn:eta_recurrence_dB}.

\subsection{Numerical results}

\begin{figure}[t]
\centering
\includegraphics[width=0.9\textwidth]{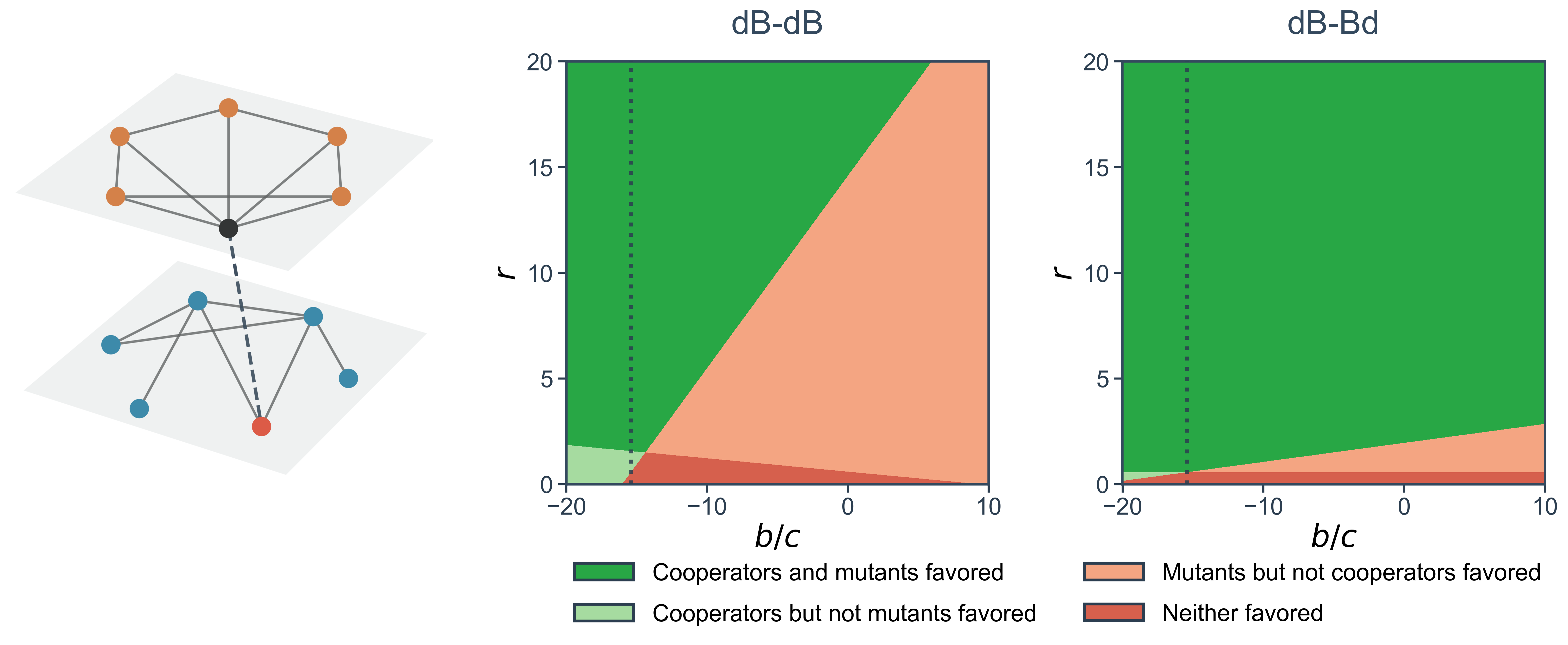}
\caption{Parameter regions in which the cooperator or mutant is favored under the ff good scheme for the two-layer network used in
Fig.~\ref{fig:examples}B
in the main text (replicated in the left panel of this figure). The vertical dotted lines indicate the $(b/c)^*$ value ($= -15.41$) when the two layers are uncoupled.}
\label{fig:examples_ff}
\end{figure}

In Fig.~\ref{fig:examples_ff}, we show the $(b/c)^*$ and $r^*$ values for the two-layer network used in
Fig.~\ref{fig:examples}B
%
%
in the main text, but under the ff goods scheme. For the other three networks used in
Fig.~\ref{fig:examples},
%
%
the $(b/c)^*$ value is the same between the ff and pf goods because the game layer (i.e., layer 1) in these two-layer networks are regular graphs; Eqs.~\eqref{eq:u_i-pf} and \eqref{eq:u_i-ff} are identical for regular graphs. For the network shown in Fig.~\ref{fig:examples_ff},
we obtain qualitatively the same behavior of $(b/c)^*$ as a function of $r$ between the two goods schemes; compare Fig.~\ref{fig:examples_ff}(b) and
Fig.~\ref{fig:examples}B.
%
%
The $r^*$ value as a function of $b/c$ is also similar between the ff and pf goods.
 
\subsection{Random graphs}

For the ff goods scheme, we analyzed the condition under which the cooperator and mutant are selected in two-layer ER and BA networks on $15$ individuals.

\begin{figure}
	\centering
	\includegraphics[width=0.9\textwidth]{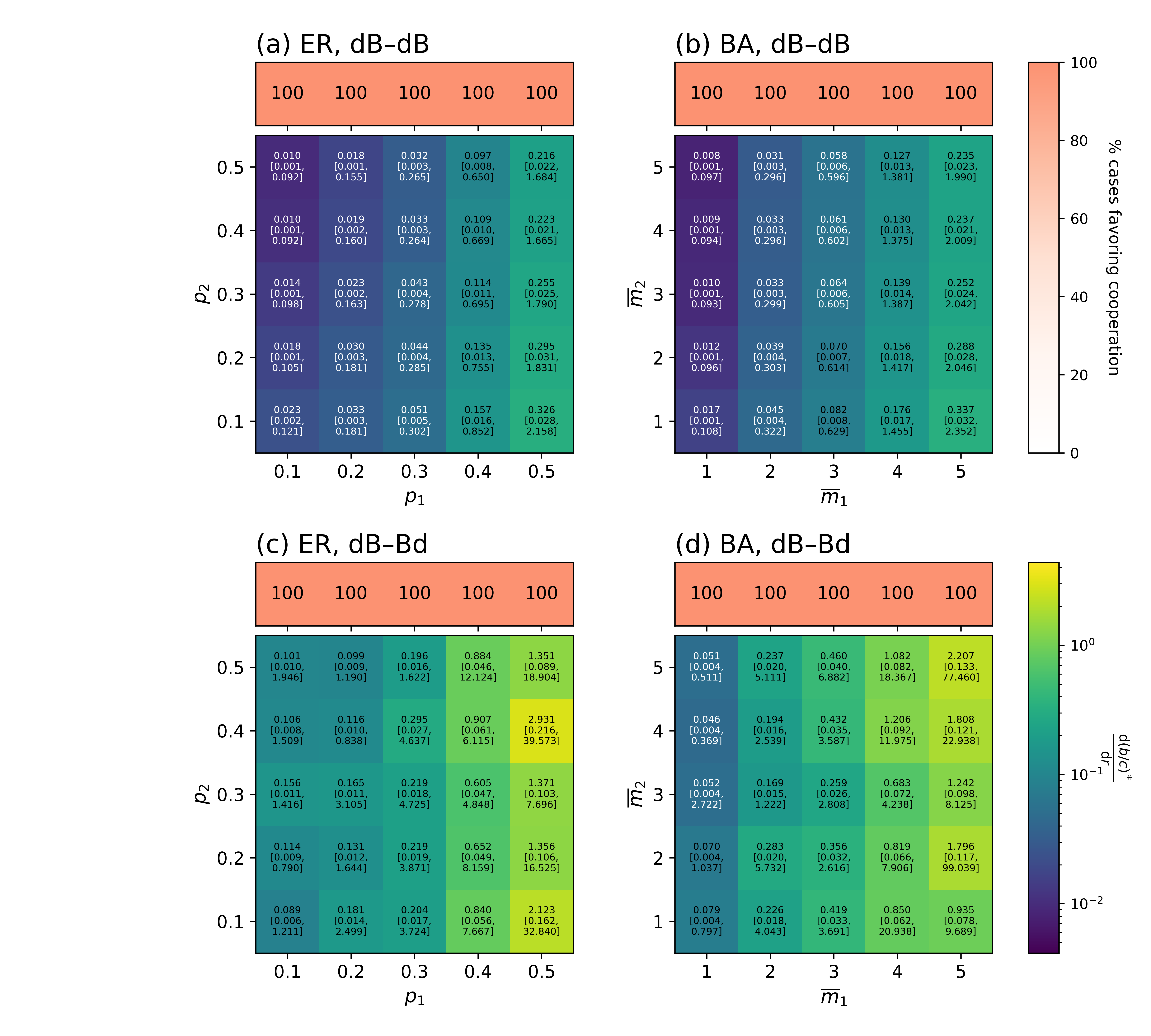}
	\caption{Evolution of cooperation in two-layer ER and BA networks under the ff goods scheme. (a) ER, dB-dB rule. (b) BA, dB-dB rule. (c) ER, dB-Bd rule. (d) BA, dB-Bd rule. The upper part of each panel shows the fraction of pairs of single-layer network and initial condition that yield cooperation when $b/c > (b/c)^*$ for a threshold value $(b/c)^* > 0$. The fraction values are the same between (a) and (c) and between (b) and (d) because whether or not $(b/c)^* > 0$ in layer 1 does not depend on the updating rule used in layer 2. The lower part of each panel shows the median along with the 5th and 95th percentiles (in square brackets) of $\left| \text{d}(b/c)^*/\text{d}r \right|$ for pairs of two-layer network and initial condition.
Each two-layer network is composed of $N=15$ individuals.} 
	\label{fig:tables_random_graphs_ff}
\end{figure}

\begin{figure}[h]
	\centering
	\includegraphics[width=0.8\textwidth]{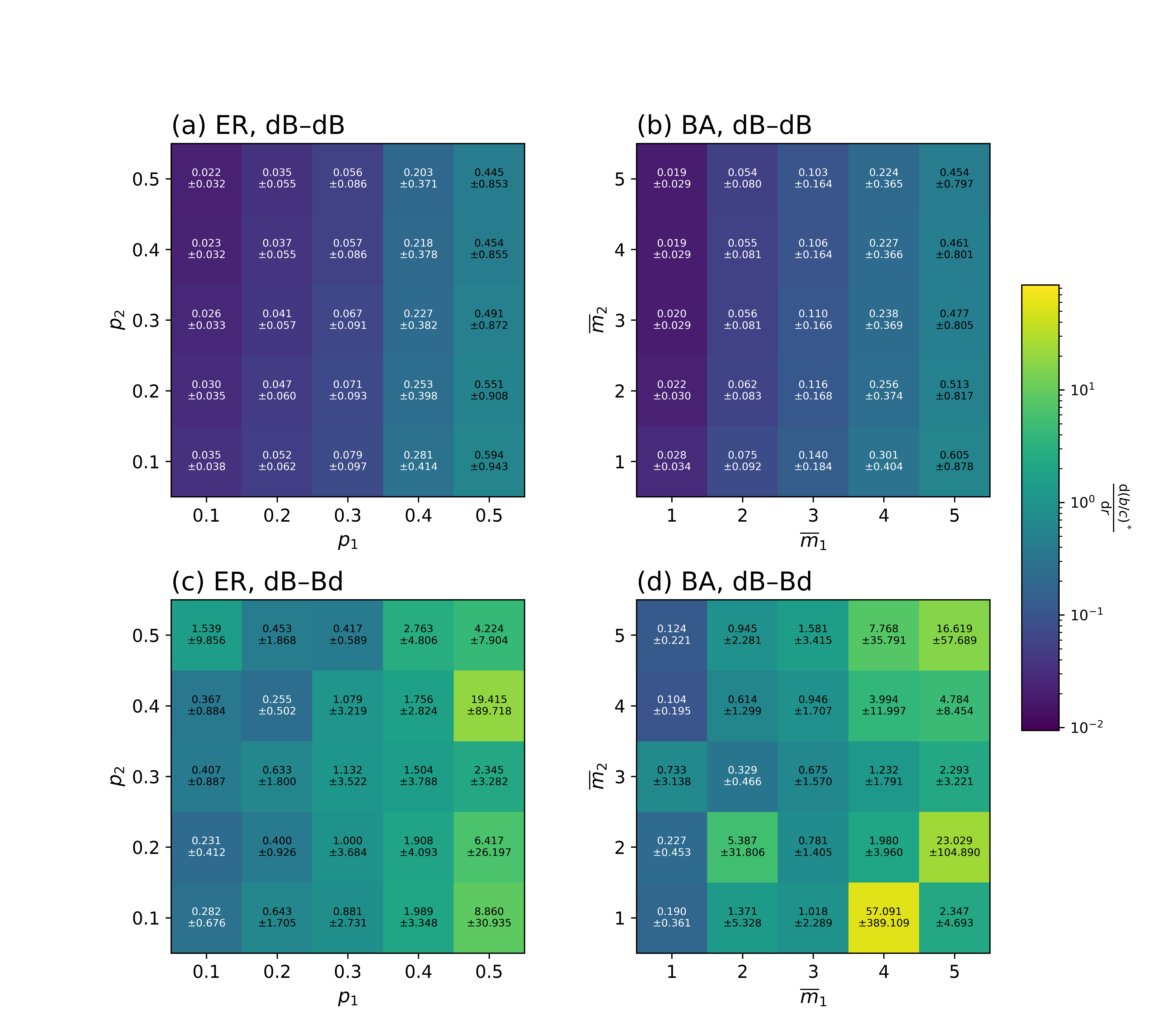}
	\caption{Mean, $\mu$, and standard deviation, $\sigma$, of $\left| \text{d}(b/c)^*/\text{d}r \right|$
for two-layer ER and BA networks with $N=15$ individuals under the ff goods scheme. 
(a) ER, dB-dB rule. (b) BA, dB-dB rule. (c) ER, dB-Bd rule. (d) BA, dB-Bd rule.
See the caption of Fig.~\ref{fig:mean_std_N=15} for details of the computation.}
	\label{fig:mean_std_N=15-ff}
\end{figure}

We show in Fig.~\ref{fig:tables_random_graphs_ff}(a) and (b) the results for the two-layer ER and BA networks, respectively, under the dB-dB updating rule. We find that all the layer-1 networks enable cooperation, not spite (see the upper part of Fig.~\ref{fig:tables_random_graphs_ff}(a) and (b)). The responsiveness of $(b/c)^*$ as the constant selection changes, quantified by $\left| \text{d}(b/c)^*/\text{d}r \right|$, increases as the edge density in layer 1 increases or that in layer 2 decreases (see the lower part of Fig.~\ref{fig:tables_random_graphs_ff}(a) and (b)). These results are similar to those for the pf goods shown in 
Figs.~\ref{fig:tables_random_graphs}A and \ref{fig:tables_random_graphs}B
in the main text. The results are qualitatively the same for the dB-Bd rule, with overall larger $\left| \text{d}(b/c)^*/\text{d}r \right|$ than the case of the dB-dB rule (see Fig.~\ref{fig:tables_random_graphs_ff}(c) and (d)). Finally, we have verified that the results do not substantially change when we switch the measurement of $\left| \text{d}(b/c)^*/\text{d}r \right|$ from the median and percentiles to the mean and standard deviation (see Fig.~\ref{fig:mean_std_N=15-ff}).
 
\newpage
\clearpage

\subsection{Empirical networks}

\begin{figure}[h]
	\centering
	\includegraphics[width=0.7\textwidth]{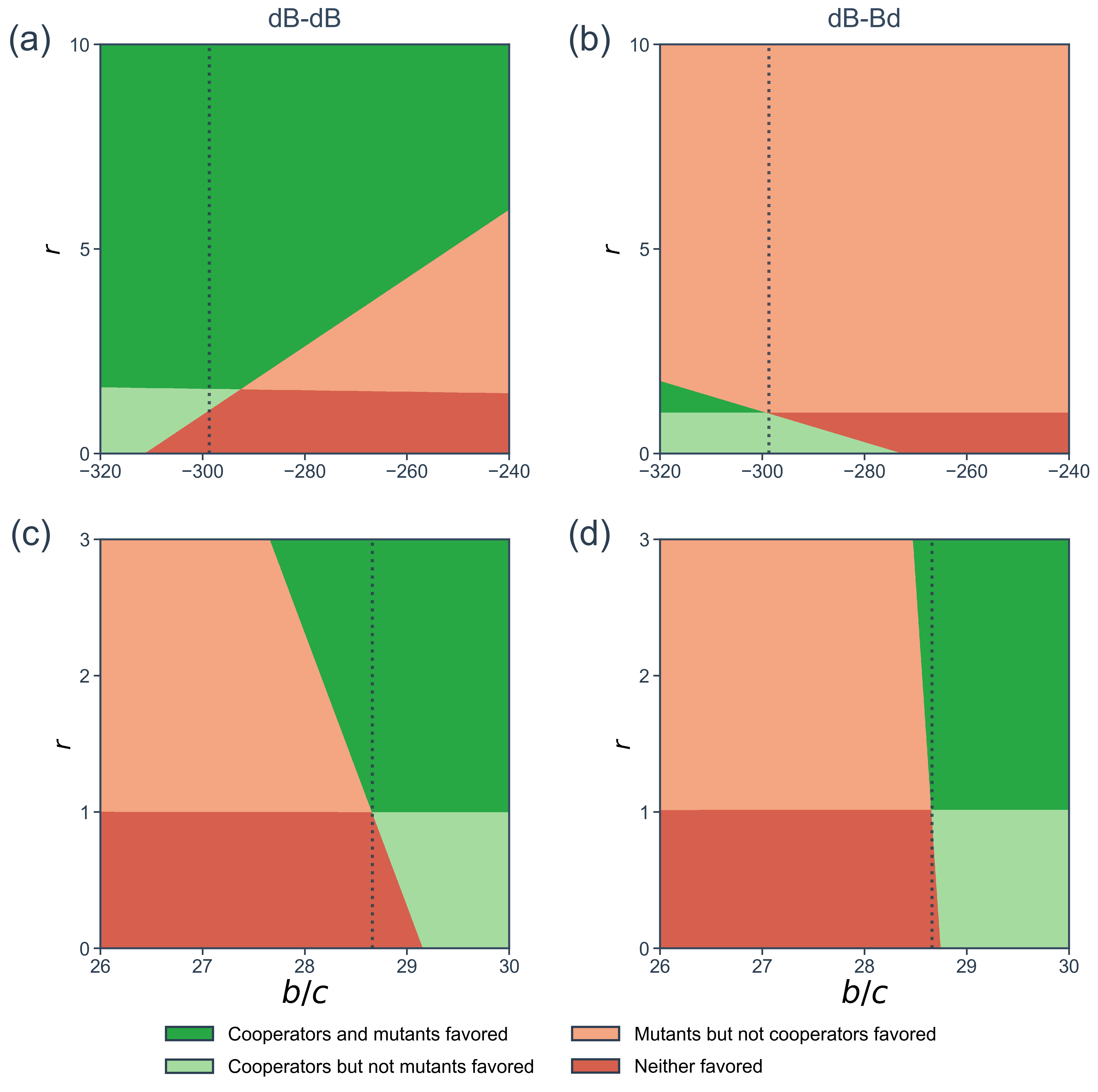}
	\caption{Parameter regions in which the cooperator or mutant is selected in empirical two-layer networks under the ff goods scheme. The vertical lines represent the $(b/c)^*$ value for the uncoupled layer-1 network. (a) VC7, dB-dB. (b) VC7, dB-Bd. (c) LLF, dB-dB. (d) LLF, dB-Bd. The single-layer VC7 and LLF networks yield $(b/c)^*=-298.7$ and $(b/c)^*=28.66$, respectively.}
	\label{fig:empirical_networks_ff}
\end{figure}

We show in Fig.~\ref{fig:empirical_networks_ff} the results for the VC7 and LLF law two-layer networks under the ff goods scheme. The results are qualitatively the same as those under the pf goods scheme. It should be noted that the quantitative difference between the ff and pf goods schemes is large in this case;
for the VC7 network, we have obtained $(b/c)^*=-298.7$, which is far from $(b/c)^*=-89.5$ in the case of the pf goods shown in the main text;
for the LLF law network, we have obtained $(b/c)^*=28.66$ under the ff goods scheme, whereas the main text shows $(b/c)^*=52.4$ under the pf goods scheme.

\newpage
\clearpage

\section{Fixation time}\label{sec:SI-N}\label{sec:fixation-time}

We computed the fixation time for cooperation under the dB-dB rule combined with the pf good scheme. We used the four synthetic two-layer networks used in
Fig.~\ref{fig:examples}
%
%
and the VC7 network. We did not use the LLF network because numerical estimation of the fixation time of cooperation when starting from a single cooperator is computationally demanding. We generated 1,600,000 runs that finished with fixation of cooperation in layer 1 no matter whether the resident or mutant type fixated in layer 2 for the four synthetic networks. We reduced the number of runs to 800,000 for the VC7 network. Then, we recorded the time to fixation of cooperation and averaged it over all the runs. It should be noted that we defined the fixation time as the time at which the cooperator has fixated regardless of whether or not fixation has been attained in layer 2. This choice is to enable fair comparison between the two-layer and one-layer networks; if we wait for the fixation of both layers, then the fixation time would be obviously shorter for one-layer networks than the two-layer networks whose layer-1 network is the same as the one-layer network under comparison. As reference, we also computed the mean fixation time for the one-layer network that is layer 1 of each two-layer network analyzed.

\begin{figure}[t]
\centering
\includegraphics[width=\textwidth]{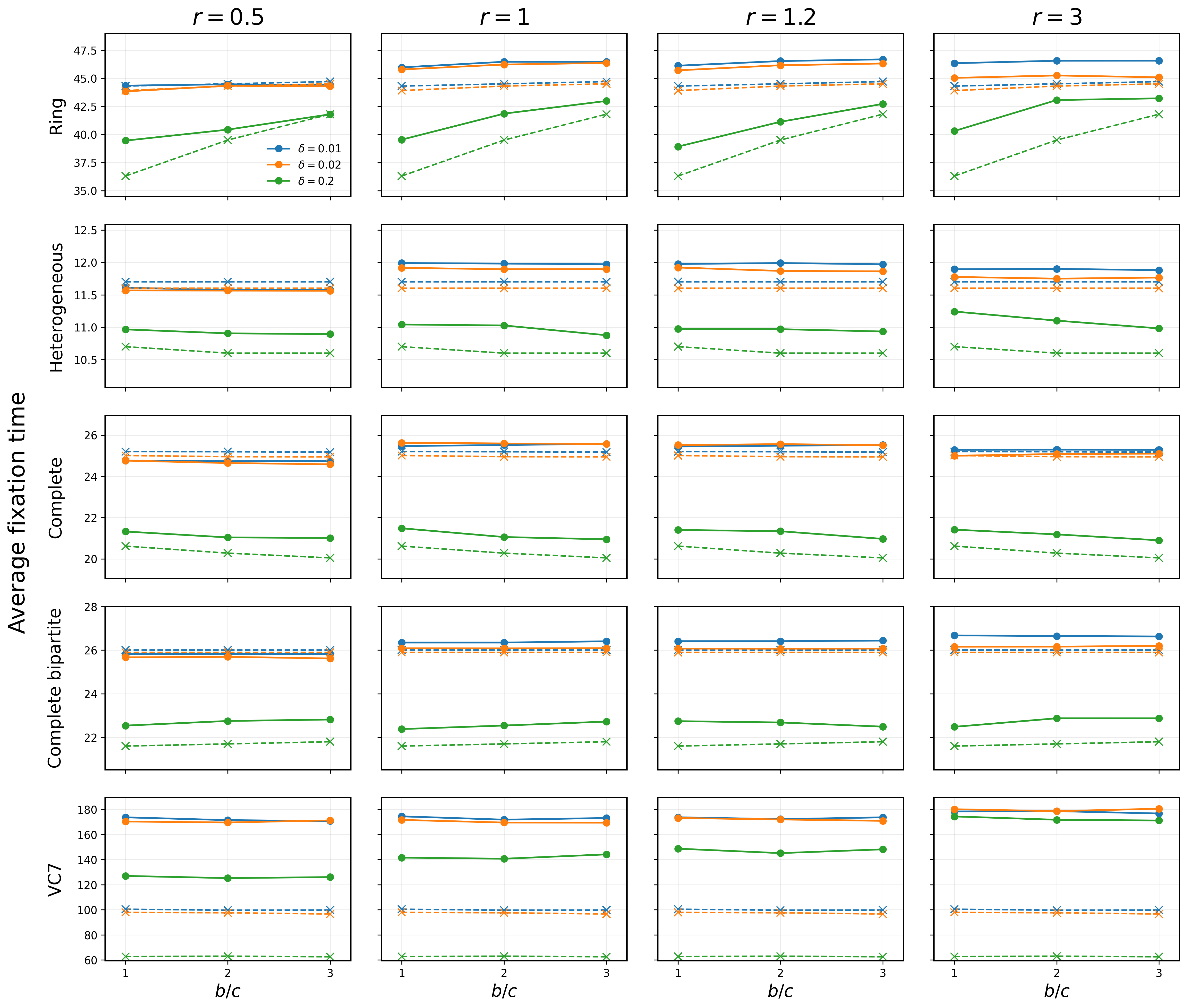}
\caption{Mean fixation time for cooperation in the four synthetic networks and the VC7 network. The ring, heterogeneous, complete, and complete bipartite networks refer to the two-layer networks used in
Fig.~\ref{fig:examples}A,
%
%
(b), (c), and (d), respectively. The solid and dashed lines represent the mean fixation time of cooperation in layer 1 of the two-layer networks and the single-layer counterparts, respectively. We used $r \in \{0.5, 1, 1.2, 3 \}$.}
	\label{fig:fixation_time}
\end{figure}

We show the mean fixation time for cooperation in Fig.~\ref{fig:fixation_time}. We find that the fixation time for the four synthetic two-layer networks is not much larger than that for the corresponding one-layer networks across the parameter values and the choice of two-layer network. For the VC7 network, the two-layer network needs about 1.6--2.8 more time to fixation than one-layer counterparts.

\clearpage


\begin{thebibliography}{10}

\bibitem{Nowak2006Science}
M.~A. Nowak.
\newblock Five rules for the evolution of cooperation.
\newblock {\em Science}, 314:1560--1563, 2006.

\bibitem{Nowak2006book}
M.~A. Nowak.
\newblock {\em Evolutionary Dynamics}.
\newblock Belknap Press of Harvard University Press, Cambridge, MA, 2006.

\bibitem{Sigmund2010book}
K.~Sigmund.
\newblock {\em The Calculus of Selfishness}.
\newblock Princeton University Press, Princeton, NJ, 2010.

\bibitem{Axelrod1984book}
R.~Axelrod.
\newblock {\em Evolution of Cooperation}.
\newblock Basic Books, NY, 1984.

\bibitem{Nowak1992Nature_spatial}
M.~A. Nowak and R.~M. May.
\newblock Evolutionary games and spatial chaos.
\newblock {\em Nature}, 359:826--829, 1992.

\bibitem{Nowak1993IntJBifuChaos}
M.~A. Nowak and R.~M. May.
\newblock The spatial dilemmas of evolution.
\newblock {\em International Journal of bifurcation and chaos}, 3:35--78, 1993.

\bibitem{ohtsuki2006simple}
H.~Ohtsuki, C.~Hauert, E.~Lieberman, and M.~A. Nowak.
\newblock A simple rule for the evolution of cooperation on graphs and social
  networks.
\newblock {\em Nature}, 441:502--505, 2006.

\bibitem{duran2005}
O.~Dur\'{a}n and R.~Mulet.
\newblock Evolutionary prisoner’s dilemma in random graphs.
\newblock {\em Physica D}, 208:257--265, 2005.

\bibitem{Santos2005PhysRevLett}
F.~C. Santos and J.~M. Pacheco.
\newblock Scale-free networks provide a unifying framework for the emergence of
  cooperation.
\newblock {\em Phys. Rev. Lett.}, 95:098104, 2005.

\bibitem{santos2008social}
F.~C. Santos, M.~D. Santos, and J.~M. Pacheco.
\newblock Social diversity promotes the emergence of cooperation in public
  goods games.
\newblock {\em Nature}, 454:213--216, 2008.

\bibitem{li2020evolution}
A.~Li, L.~Zhou, Q.~Su, S.~P. Cornelius, Y.-Y. Liu, L.~Wang, and S.~A. Levin.
\newblock Evolution of cooperation on temporal networks.
\newblock {\em Nature Communications}, 11:2259, 2020.

\bibitem{li2023temporal}
A.~Li, Y.~Meng, L.~Zhou, N.~Masuda, and L.~Wang.
\newblock Temporal networks provide a unifying understanding of the evolution
  of cooperation.
\newblock {\em arXiv preprint arXiv:2309.12686}, 2023.

\bibitem{su2023strategy}
Q.~Su, A.~{McAvoy}, and J.~B. Plotkin.
\newblock Strategy evolution on dynamic networks.
\newblock {\em Nature Computational Science}, 3:763--776, 2023.

\bibitem{meng2025promoting}
Y.~Meng, A.~{McAvoy}, and A.~Li.
\newblock Promoting collective cooperation through temporal interactions.
\newblock {\em Proc. Natl. Acad. Sci. U.S.A.}, 122:e2509575122, 2025.

\bibitem{cardillo2014evolutionary}
A.~Cardillo, G.~Petri, V.~Nicosia, R.~Sinatra, J.~G{\'o}mez-Garde\~{n}es, and
  V.~Latora.
\newblock Evolutionary dynamics of time-resolved social interactions.
\newblock {\em Phys. Rev. E}, 90:052825, 2014.

\bibitem{Easley2010book}
D.~Easley and J.~J.~Kleinberg.
\newblock {\em Networks, Crowds, and Markets}.
\newblock Cambridge University Press, Cambridge, UK, 2010.

\bibitem{Barabasi2016book}
A.~L. Barab\'{a}si.
\newblock {\em Network Science}.
\newblock Cambridge University Press, Cambridge, UK, 2016.

\bibitem{Newman2018book}
M.~E.~J. Newman.
\newblock {\em Networks}.
\newblock Oxford University Press, Oxford, UK, second edition, 2018.

\bibitem{Kivela2014JCompNetw}
M.~Kivel\"{a}, A.~Arenas, M.~Barthelemy, J.~P. Gleeson, Y.~Moreno, and M.~A.
  Porter.
\newblock {Multilayer networks}.
\newblock {\em J. Comp. Netw.}, 2:203--271, 2014.

\bibitem{Boccaletti2014PhysRep}
S.~Boccaletti, G.~Bianconi, R.~Criado, C.~I. {del Genio},
  J.~G\'{o}mez-Garde\~{n}es, M.~Romance, I.~Sendi\~{n}a Nadal, Z.~Wang, and
  M.~Zanin.
\newblock {The structure and dynamics of multilayer networks}.
\newblock {\em Phys. Rep.}, 544:1--122, 2014.

\bibitem{Bianconi2018book}
G.~Bianconi.
\newblock {\em {Multilayer Networks}}.
\newblock Oxford University Press, Oxford, UK, 2018.

\bibitem{de2023more}
M.~{De Domenico}.
\newblock More is different in real-world multilayer networks.
\newblock {\em Nature Physics}, 19:1247--1262, 2023.

\bibitem{Gomezgardenes2012SciRep}
J.~G{\'o}mez-Garde\~{n}es, I.~Reinares, A.~Arenas, and L.~M. Flor{\'\i}a.
\newblock Evolution of cooperation in multiplex networks.
\newblock {\em Sci. Rep.}, 2:620, 2012.

\bibitem{wang2015evolutionary}
Z.~Wang, L.~Wang, A.~Szolnoki, and M.~Perc.
\newblock {Evolutionary games on multilayer networks: A colloquium}.
\newblock {\em European Physical Journal B}, 88:124, 2015.

\bibitem{Jusup2022PhysRep}
M.~Jusup, P.~Holme, K.~Kanazawa, M.~Takayasu, I.~Romi\'c, Z.~Wang,
  S.~Ge\v{c}ek, T.~Lipi\'c, B.~Pobodnik, L.~Wang, W.~Luo, T.~Klanj\v{s}\v{c}ek,
  J.~Fan, S.~Boccaletti, and M.~Perc.
\newblock Social physics.
\newblock {\em Physics Reports}, 948:1--148, 2022.

\bibitem{basak2024evolution}
A.~Basak and S.~Sengupta.
\newblock Evolution of cooperation in multichannel games on multiplex networks.
\newblock {\em PLoS Comput. Biol.}, 20:e1012678, 2024.

\bibitem{zhu2025evolution}
W.~Zhu, X.~Wang, C.~Wang, W.~Xing, L.~Liu, H.~Zheng, J.~Zhao, and S.~Tang.
\newblock Evolution of cooperation and competition in multilayer networks.
\newblock {\em Nonlinear Dynamics}, 113:31619--31635, 2025.

\bibitem{su2022evolution}
Q.~Su, A.~{McAvoy}, Y.~Mori, and J.~B. Plotkin.
\newblock Evolution of prosocial behaviours in multilayer populations.
\newblock {\em Nature Human Behaviour}, 6:338--348, 2022.

\bibitem{santos2014biased}
M.~D. Santos, S.~N. Dorogovtsev, and J.~F.~F. Mendes.
\newblock Biased imitation in coupled evolutionary games in interdependent
  networks.
\newblock {\em Sci. Rep.}, 4:4436, 2014.

\bibitem{wang2020vaccination}
X.~Wang, D.~Jia, S.~Gao, C.~Xia, X.~Li, and Z.~Wang.
\newblock Vaccination behavior by coupling the epidemic spreading with the
  human decision under the game theory.
\newblock {\em Applied Mathematics and Computation}, 380:125232, 2020.

\bibitem{raducha2023evolutionary}
T.~Raducha and M.~{San Miguel}.
\newblock Evolutionary games on multilayer networks: coordination and
  equilibrium selection.
\newblock {\em Sci. Rep.}, 13:11818, 2023.

\bibitem{ohtsuki2007breaking}
H.~Ohtsuki, M.~A Nowak, and J.~M. Pacheco.
\newblock Breaking the symmetry between interaction and replacement in
  evolutionary dynamics on graphs.
\newblock {\em Phys. Rev. Lett.}, 98:108106, 2007.

\bibitem{ohtsuki2007evolutionary}
H.~Ohtsuki, J.~M. Pacheco, and M.~A. Nowak.
\newblock Evolutionary graph theory: Breaking the symmetry between interaction
  and replacement.
\newblock {\em J. Theor. Biol.}, 246:681--694, 2007.

\bibitem{Bramson1981AnnProb}
M.~Bramson and D.~Griffeath.
\newblock {On the Williams-Bjerknes tumour growth model I}.
\newblock {\em Ann. Prob.}, 9:173--185, 1981.

\bibitem{Durrett1988book}
R.~Durrett.
\newblock {\em {Lecture Notes on Particle Systems and Percolation}}.
\newblock Belmont, Wadsworth, CA, 1988.

\bibitem{Antal2006PhysRevLett}
T.~Antal, S.~Redner, and V.~Sood.
\newblock Evolutionary dynamics on degree-heterogeneous graphs.
\newblock {\em Phys. Rev. Lett.}, 96:188104, 2006.

\bibitem{Czaplicka2022ChaosSolitonsFractals}
A.~Czaplicka, C.~Charalambous, R.~Toral, and M.~{San Miguel}.
\newblock Biased-voter model: How persuasive a small group can be?
\newblock {\em Chaos, Solitons and Fractals}, 161:112363, 2022.

\bibitem{mcavoy2020social}
A.~{McAvoy}, B.~Allen, and M.~A. Nowak.
\newblock Social goods dilemmas in heterogeneous societies.
\newblock {\em Nature Human Behaviour}, 4:819--831, 2020.

\bibitem{allen2017evolutionary}
B.~Allen, G.~Lippner, Y.-T. Chen, B.~Fotouhi, N.~Momeni, S.-T. Yau, and M.~A.
  Nowak.
\newblock Evolutionary dynamics on any population structure.
\newblock {\em Nature}, 544:227--230, 2017.

\bibitem{hindersin2015most}
L.~Hindersin and A.~Traulsen.
\newblock Most undirected random graphs are amplifiers of selection for
  birth-death dynamics, but suppressors of selection for death-birth dynamics.
\newblock {\em PLoS Comput. Biol.}, 11:e1004437, 2015.

\bibitem{Masuda2009JTheorBiol}
N.~Masuda.
\newblock Directionality of contact networks suppresses selection pressure in
  evolutionary dynamics.
\newblock {\em J. Theor. Biol.}, 258:323--334, 2009.

\bibitem{shakarian2012review}
P.~Shakarian, P.~Roos, and A.~Johnson.
\newblock A review of evolutionary graph theory with applications to game
  theory.
\newblock {\em Biosystems}, 107:66--80, 2012.

\bibitem{pattni2015evolutionary}
K.~Pattni, M.~Broom, J.~Rycht{\'a}{\v{r}}, and L.~J. Silvers.
\newblock Evolutionary graph theory revisited: when is an evolutionary process
  equivalent to the moran process?
\newblock {\em Proceedings of the Royal Society A}, 471:20150334, 2015.

\bibitem{fu2009evolutionary}
F.~Fu, L.~Wang, M.~A. Nowak, and C.~Hauert.
\newblock Evolutionary dynamics on graphs: Efficient method for weak selection.
\newblock {\em Phys. Rev. E}, 79:046707, 2009.

\bibitem{nowak2010evolutionary}
M.~A. Nowak, C.~E. Tarnita, and T.~Antal.
\newblock Evolutionary dynamics in structured populations.
\newblock {\em Phil. Trans. R. Soc. B}, 365:19--30, 2010.

\bibitem{wang2015universal}
Z.~Wang, S.~Kokubo, M.~Jusup, and J.~Tanimoto.
\newblock Universal scaling for the dilemma strength in evolutionary games.
\newblock {\em Physics of Life Reviews}, 14:1--30, 2015.

\bibitem{su2019evolutionary}
Q.~Su, A.~{McAvoy}, L.~Wang, and M.~A. Nowak.
\newblock Evolutionary dynamics with game transitions.
\newblock {\em Proc. Natl. Acad. Sci. USA}, 116:25398--25404, 2019.

\bibitem{lieberman2005evolutionary}
E.~Lieberman, C.~Hauert, and M.~A. Nowak.
\newblock Evolutionary dynamics on graphs.
\newblock {\em Nature}, 433:312--316, 2005.

\bibitem{vickers1981representing}
M.~Vickers and S.~Chan.
\newblock Representing classroom social structure.
\newblock {\em Victoria Institute of Secondary Education, Melbourne}, 1981.

\bibitem{lazega2001collegial}
E.~Lazega.
\newblock {\em The Collegial Phenomenon}.
\newblock Oxford University Press, Oxford, UK, 2001.

\bibitem{tkadlec2020limits}
J.~Tkadlec, A.~Pavlogiannis, K.~Chatterjee, and M.~A. Nowak.
\newblock Limits on amplifiers of natural selection under death-{B}irth
  updating.
\newblock {\em PLoS Comput. Biol.}, 16:e1007494, 2020.

\bibitem{Svoboda2024PlosComputBiol}
J.~Svoboda, S.~Joshi, J.~Tkadlec, and K.~Chatterjee.
\newblock {Amplifiers of selection for the Moran process with both Birth-death
  and death-Birth updating}.
\newblock {\em PLoS Comput. Biol.}, 20:e1012008, 2024.

\bibitem{altrock2009fixation}
P.~M. Altrock and A.~Traulsen.
\newblock Fixation times in evolutionary games under weak selection.
\newblock {\em New J. Phys.}, 11:013012, 2009.

\bibitem{sui2015speed}
X.~Sui, B.~Wu, and L.~Wang.
\newblock Speed of evolution on graphs.
\newblock {\em Phys. Rev. E}, 92:062124, 2015.

\bibitem{Alvarezrodriguez2021NatHumanBehav}
U.~Alvarez-Rodriguez, F.~Battiston, G.~{Ferraz de Arruda}, Y.~Moreno, M.~Perc,
  and V.~Latora.
\newblock Evolutionary dynamics of higher-order interactions in social
  networks.
\newblock {\em Nature Human Behaviour}, 5:586--595, 2021.

\bibitem{sheng2024strategy}
A.~Sheng, Q.~Su, L.~Wang, and J.~B. Plotkin.
\newblock Strategy evolution on higher-order networks.
\newblock {\em Nature Computational Science}, 4:274--284, 2024.

\bibitem{SzaboToke1998PhysRevE}
G.~Szab\'{o} and C.~T\H{o}ke.
\newblock Evolutionary prisoner's dilemma game on a square lattice.
\newblock {\em Phys. Rev. E}, 58:69--73, 1998.

\bibitem{TraulsenNowakPacheco2006PhysRevE}
A.~Traulsen, M.~A. Nowak, and J.~M. Pacheco.
\newblock Stochastic dynamics of invasion and fixation.
\newblock {\em Phys. Rev. E}, 74:011909, 2006.

\bibitem{Mobilia2003PhysRevLett}
M.~Mobilia.
\newblock {Does a single zealot affect an infinite group of voters?}
\newblock {\em Phys. Rev. Lett.}, 91:028701, 2003.

\bibitem{Mobilia2007JStatMech}
M.~Mobilia, A.~Petersen, and S.~Redner.
\newblock {On the role of zealotry in the voter model}.
\newblock {\em J. Stat. Mech.}, 2007:P08029, 2007.

\bibitem{XieSreenivasan2011PhysRevE}
J.~Xie, S.~Sreenivasan, G.~Korniss, W.~Zhang, C.~Lim, and B.~K. Szymanski.
\newblock {Social consensus through the influence of committed minorities}.
\newblock {\em Phys. Rev. E}, 84:011130, 2011.

\bibitem{Masuda2012SciRep}
N.~Masuda.
\newblock {Evolution of cooperation driven by zealots}.
\newblock {\em Sci. Rep.}, 2:646, 2012.

\bibitem{NakajimaMasuda2015JMathBiol}
Y.~Nakajima and N.~Masuda.
\newblock {Evolutionary dynamics in finite populations with zealots}.
\newblock {\em J. Math. Biol.}, 70:465--484, 2015.

\bibitem{ruodan2023multilayer}
R.~Liu and N.~Masuda.
\newblock Fixation dynamics on multilayer networks.
\newblock {\em SIAM Journal on Applied Mathematics}, 84:2028--2050, 2024.

\bibitem{ShenChu2018EPL}
C.~Shen, C.~Chu, L.~Shi, M.~Jusup, M.~Perc, and Z.~Wang.
\newblock Coevolutionary resolution of the public goods dilemma in
  interdependent structured populations.
\newblock {\em EPL}, 124:48003, 2018.

\bibitem{Barabasi1999Science}
A.~L. Barab\'{a}si and R.~Albert.
\newblock Emergence of scaling in random networks.
\newblock {\em Science}, 286:509--512, 1999.

\bibitem{Bhaumik2026Github}
N.~Masuda.
\newblock Code for: Non-dilemmatic social dynamics promote cooperation in
  multilayer networks.
\newblock
  \url{https://github.com/naokimas/Non-dilemmatic-social-dynamics-promote-cooperation-in-multilayer-network}.
\newblock Accessed: 24 September 2026.

\bibitem{allen2019mathematical}
B.~Allen and A.~{McAvoy}.
\newblock A mathematical formalism for natural selection with arbitrary spatial
  and genetic structure.
\newblock {\em Journal of Mathematical Biology}, 78:1147--1210, 2019.

\bibitem{mcavoy2021fixation}
Alex McAvoy and Benjamin Allen.
\newblock Fixation probabilities in evolutionary dynamics under weak selection.
\newblock {\em Journal of Mathematical Biology}, 82(3):14, 2021.

\bibitem{fisher1999genetical}
R.~A. Fisher.
\newblock {\em The Genetical Theory of Natural Selection: {A} Complete Variorum
  Edition}.
\newblock Oxford University Press, Oxford, UK, 1999.

\bibitem{taylor1990allele}
Peter~D Taylor.
\newblock Allele-frequency change in a class-structured population.
\newblock {\em The American Naturalist}, 135(1):95--106, 1990.

\bibitem{taylor1996inclusive}
Peter~D Taylor.
\newblock Inclusive fitness arguments in genetic models of behaviour.
\newblock {\em Journal of Mathematical Biology}, 34(5):654--674, 1996.

\bibitem{grafen2006theory}
Alan Grafen.
\newblock A theory of fisher's reproductive value.
\newblock {\em Journal of mathematical biology}, 53:15--60, 2006.

\bibitem{maciejewski2014reproductive}
Wes Maciejewski.
\newblock Reproductive value in graph-structured populations.
\newblock {\em Journal of Theoretical Biology}, 340:285--293, 2014.

\bibitem{sood2008voter}
V.~Sood, T.~Antal, and S.~Redner.
\newblock Voter models on heterogeneous networks.
\newblock {\em Physical Review E}, 77:041121, 2008.

\bibitem{masuda2009evolutionary}
Naoki Masuda and Hisashi Ohtsuki.
\newblock Evolutionary dynamics and fixation probabilities in directed
  networks.
\newblock {\em New Journal of Physics}, 11(3):033012, 2009.

\end{thebibliography}
\end{document}